\documentclass[
twocolumn,
reprint,
superscriptaddress,
nofootinbib,
nobibnotes,
amsmath,
amssymb,
aps,
pra,
longbibliography,
]{revtex4-2}
 
\usepackage[latin1]{inputenc}
\usepackage{amsmath}
\usepackage{amsfonts}
\usepackage{hyperref}
\usepackage{amssymb}
\usepackage{graphicx}
\usepackage{hyperref}
\usepackage{colortbl}
\usepackage{subfigure}
\usepackage{svg}

\usepackage[normalem]{ulem}

\usepackage[T1]{fontenc}
\usepackage{mathptmx}
\usepackage{bbold}
\usepackage{scalerel}
\usepackage{xcolor}

\usepackage{mathtools}

\newcommand*\ch[1]{\ensuremath{\mathrm{#1}}}
\newcommand{\hyref}[1]{\hyperref[#1]{\ref{#1}}}
\newcommand{\dd}{\mathrm{d}}

\newcommand{\orange}[1]

\usepackage{titlesec}

\renewcommand{\thesection}{\arabic{section}}

\titleformat{\section}  
  {\fontsize{10}{12}\bfseries} 
  {\thesection} 
  {1em} 
  {\centering} 
  [] 

\begin{document}

\title{Brain-inspired reservoir computing with fluidic iontronic nanochannels}

\author{T. M. Kamsma}
\thanks{These two authors contributed equally to this work}
\affiliation{Institute for Theoretical Physics, Utrecht University,  Princetonplein 5, 3584 CC Utrecht, The Netherlands}
\affiliation{Mathematical Institute, Utrecht University, Budapestlaan 6, 3584 CD Utrecht, The Netherlands}
\author{\normalfont\textsuperscript{,$\dag$}\;J. Kim}
\thanks{These two authors contributed equally to this work}
\affiliation{Department of Mechanical Engineering, Sogang University, 35 Baekbeom-ro (Sinsu-dong), Mapo-gu, Seoul 04107, Republic of Korea}
\author{K. Kim}
\affiliation{Department of Mechanical Engineering, Sogang University, 35 Baekbeom-ro (Sinsu-dong), Mapo-gu, Seoul 04107, Republic of Korea}
\author{W. Q. Boon}
\affiliation{Institute for Theoretical Physics, Utrecht University,  Princetonplein 5, 3584 CC Utrecht, The Netherlands}
\author{C. Spitoni}
\affiliation{Mathematical Institute, Utrecht University, Budapestlaan 6, 3584 CD Utrecht, The Netherlands}
\author{J. Park}
\thanks{Corresponding author}
\affiliation{Department of Mechanical Engineering, Sogang University, 35 Baekbeom-ro (Sinsu-dong), Mapo-gu, Seoul 04107, Republic of Korea}
\author{R. van Roij}
\thanks{Corresponding author}
\affiliation{Institute for Theoretical Physics, Utrecht University,  Princetonplein 5, 3584 CC Utrecht, The Netherlands}

\date{\today}
\begin{abstract}
The brain's remarkable and efficient information processing capability is driving research into brain-inspired (neuromorphic) computing paradigms. Artificial aqueous ion channels are emerging as an exciting platform for neuromorphic computing, representing a departure from conventional solid-state devices by directly mimicking the brain's fluidic ion transport. Supported by a quantitative theoretical model, we present easy to fabricate tapered microchannels that embed a conducting network of fluidic nanochannels between a colloidal structure. Due to transient salt concentration polarisation our devices are volatile memristors (memory resistors) that are remarkably stable. The voltage-driven net salt flux and accumulation, that underpin the concentration polarisation, surprisingly combine into a diffusionlike quadratic dependence of the memory retention time on the channel length, allowing channel design for a specific timescale. We implement our device as a synaptic element for neuromorphic reservoir computing. Individual channels distinguish various time series, that together represent (handwritten) numbers, for subsequent in-silico classification with a simple readout function.  Our results represent a significant step towards realising the promise of fluidic ion channels as a platform to emulate the rich aqueous dynamics of the brain.
\end{abstract}

\maketitle

Neuromorphic computing aims to replicate the information processing of the human brain, which is orders of magnitude more energy efficient than conventional computing devices \cite{mehonic2022brain,schuman2017survey}. This is paramount as the unsustainable trend of energy consumption by computers is growing at an exponential rate, driving investigations into brain-inspired computing paradigms \cite{sangwan2020neuromorphic}. To pursue brain-like information processing, a device structure that goes beyond the conventional von Neumann architecture is necessary \cite{schuman2022opportunities}. To this end, memristors (memory resistors) have emerged as promising artificial analogues to biological synapses that enable brain-inspired circuit architectures \cite{strukov2008missing,chua2013memristor,schuman2017survey,sangwan2020neuromorphic,zhu2020comprehensive}.

Despite the successful implementation of memristors in various conventional platforms, the vast majority of these devices consist (at least partially) of solid-state components, rely on only a single information carrier (usually electrons or holes), and only respond to electric driving forces \cite{sangwan2020neuromorphic,zhu2020comprehensive}. These limitations contrast with the brain's nimble synapses, which can utilize both electrical and chemical signals by relying on transport in an aqueous environment of various ionic and molecular species in parallel \cite{fundNeuroAll}. In light of this disparity, an emerging and exciting approach seeks inspiration not only from the architecture of the brain, but also from its aqueous medium and ionic signal carriers \cite{noy2023nanofluidic}. These so-called iontronic devices employ ions moving in an aqueous environment to carry information, offering the promise of multiple information carriers, chemical regulation, and bio-integrability \cite{han2022iontronics}. Consequently, various iontronic memristors have been presented \cite{powell2011electric,wang2012transmembrane,paulo2023hydrophobically,ramirez2024neuromorphic} that can exhibit synaptic plasticity features \cite{han2023iontronic,ramirez2023synaptical}, and utilize chemical regulation \cite{xiong2023neuromorphic,robin2023long}. Additionally, recent advancements have been made in employing iontronic devices for signaling and computing, with theoretical proposals \cite{robin2021principles,kamsma2023iontronic,kamsma2024advanced}, and demonstrations of traditional truth tables \cite{emmerich2024nanofluidic,sabbagh2023designing,li2023high}, respectively. Despite these prospects, the development of aqueous neuromorphic devices is still in its infancy and neuromorphic computing implementations remain a challenge \cite{han2022iontronics,xie2022perspective,noy2023fluid}.

Here we theoretically propose and experimentally realise the implementation of an aqueous volatile memristor as a synaptic element for iontronic neuromorphic computing. This device is stable over long periods of time, providing reliable and distinct responses to temporal inputs, enabling its use as computing element. Additionally, device fabrication is fast, cost-effective and easy via a soft-lithography process that is almost free-shaping. By constructing a channel of a certain chosen length, made easy by the flexible fabrication process, we can design our channel to feature a specific timescale chosen from a wide range, a desirable property of memristive devices \cite{chicca2020recipe}. The volatile nature, i.e.\ decaying conductance memory when driving forces are removed, with adjustable memory retention times makes our memristor a promising candidate for reservoir computing, a brain-inspired machine learning framework which has drawn attention due to its capability of handling complex time series and sequential tasks \cite{tanaka2019recent,cucchi2021reservoir,cao2022emerging,cucchi2022hands,midya2019reservoir,du2017reservoir}. We implement benchmark protocols of classifying (handwritten) numbers that are encoded as temporal signals. Our aqueous channels process the time series, distinguishing them for subsequent in silico classification with a simple readout function, performing (at least) comparable with more conventional solid-state platforms employing similar protocols \cite{pyo2022non,kim2022implementation,du2017reservoir,midya2019reservoir}. 

\begin{figure*}[ht]
\centering
     \includegraphics[width=1\textwidth]{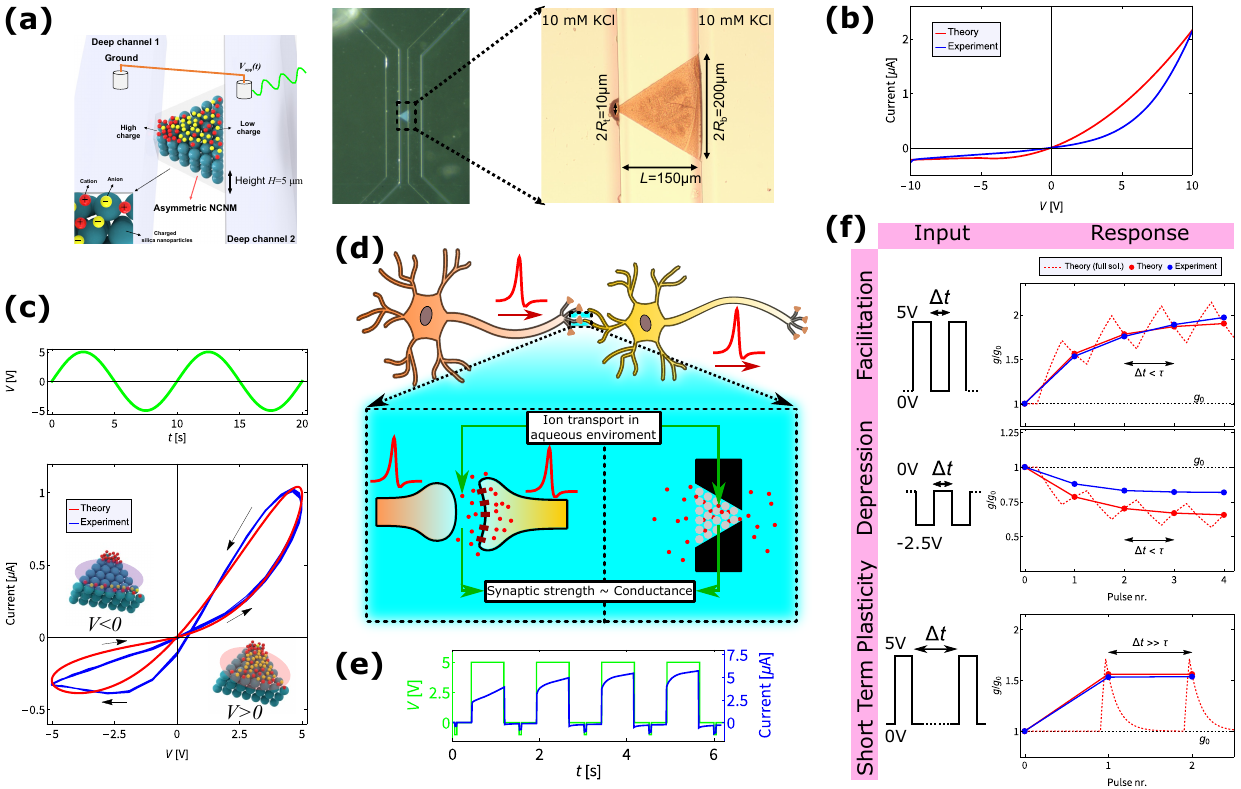}
        \caption{Features and properties of our iontronic memristor through theory and experiment. \textbf{(a)} Schematic (left) and pictures (right) of the device. The channel connects two reservoirs of aqueous KCl electrolyte and incorporates a rigid colloidal structure, forming a network of nanochannels between the colloids. \textbf{(b)} Steady-state $I-V$ curve observed in experiments (blue) and predicted by our theory (red), showing a similar current rectification property. \textbf{(c)} Dynamic $I-V$ curve in response to a sinusoidal voltage over the channel (top, green). The theory (bottom, red) and the experiments (bottom, blue) both exhibit a similar pinched hysteresis loop. \textbf{(d)} Simplified schematic of synaptic signal transmission. An action potential triggers neurotransmitter release (not depicted) from the presynaptic neuron (orange), binding to receptors on the postsynaptic neuron (yellow), potentially inducing ion transport and altering its membrane potential \cite{fundNeuroAll}. The dynamic channel conductance is analogous to the synaptic strength. \textbf{(e)} Current measurements (blue) when four consecutive 5 V pulses and five read pulses (green) are applied. \textbf{(f)} Short-term plasticity features observed in the channel (blue) and predicted by the theory, where we show the full (numerical) solution for $g(t)/g_0$ (red, dashed) and the measurements this would correspond to in the experiment (red, dots). Four consecutive voltage pulses with $\Delta t$ smaller than the channel's memory retention time $\tau$ leads to facilitation (top) and depression (middle) for pulses of $5\text{ V}$ and $-2.5\text{ V}$, respectively. The short term characteristic is clearly visible when $\Delta t\gg\tau$, in this instance no cumulative change in conductance is found (bottom).}
        \label{fig:Channel_Synapse_Hyst_STP}
\end{figure*}

Our iontronic device is understood through a quantitative theoretical model that directly derives from continuum transport equations and identifies an inhomogeneous ionic space charge density, observed between the colloids \cite{choi2016high}, as (a general) main ingredient to induce salt concentration polarisation and consequent ion current rectification. Moreover, our theory elucidates how the \emph{voltage-driven} net salt flux and accumulation, that underpin the (transient) concentration polarisation, surprisingly combine into a \emph{diffusionlike} conductance memory timescale that quadratically depends on the channel length. Consequently, the theory correctly predicts the voltage-dependent (dynamic) conductance, thereby facilitating a great acceleration of the experiments by pointing out the relevant signal voltages, signal timescales, and suitable reservoir computing protocol.

The combination of i) a stable fluidic memristor that can be designed to feature a specific memory timescale, made with ii) an almost free-shaping and cost-effective soft-lithography fabrication process, iii) a theoretical model that quantitatively describes and predicts the device dynamics, and iv) the implementation of an aqueous iontronic device as an element for neuromorphic computing, forms a significant advancement towards developing iontronic devices that can facilitate the wealth of communication pathways harnessed by the brain. 

\section{Fluidic NCNM Memristor: Theory and Experiment}
Our experimental system, as shown in Fig.~\ref{fig:Channel_Synapse_Hyst_STP}(a), consists of a tapered microfluidic channel of uniform height $H=5\text{ }\mu$m and a width that linearly decreases over its length $L=150\text{ }\mu$m from $2R_{\ch{b}}=200\text{ }\mu$m at the broad base to $2R_{\ch{t}}=10\text{ }\mu$m at the narrow tip. The channel, which connects two deep reservoirs containing an aqueous 10 mM KCl electrolyte, is filled with a rigid face-centered cubic (fcc) crystal structure at a near-close-packed volume fraction $\eta\simeq 0.74$ of charged silica spheres with radius $a=100$ nm and approximate surface charge density $\sigma_{\ch{c}}=-0.01$ Cm$^{-2}$. To form the colloidal structure, the colloids, dispersed in a 70\% ethanol solution, are injected into deep channel 1 (as in Fig.~\ref{fig:Channel_Synapse_Hyst_STP}(a)) of $100\text{ }\mu$m height, filling the tapered shallow channel up to the base through capillary action. The fluid halts at the interface between the shallow channel and deep channel 2 due to the Laplace-pressure and consequently evaporates, promoting nanoparticle self-assembly into a close-packed fcc. The device is then dried and prepared for the experiments by filling it with the aqueous electrolyte. The electrolyte in the space between the colloids forms a conducting nanochannel network membrane (NCNM) with pore spaces up to tens of nm in the tetrahedral and octahedral holes of the fcc lattice. This connected porous structure can support an ionic current $I$ driven by a voltage drop $V$ over the length of the channel, defined as tip minus base voltage (i.e.\ $V=-V_{\ch{app}}$). More system parameters and characteristics are laid out in the SI.

Some of us previously showed that these type of colloid-filled microchannels are ionic diodes with excellent ion current rectification (ICR) ratios of up to 55 for the channels on which the devices in this work are based \cite{choi2016high}, and even up to $\sim1600$, the highest reported value at the time, for a more recent version containing two colloidal structures of opposite surface charge \cite{kim2022asymmetric}. The physical phenomenon that underpins the ICR is the strong voltage-dependent salt concentration polarisation, hypothesised to be induced by an experimentally observed inhomogeneous ionic space charge density between the colloidal particles \cite{choi2016high}. Calculations on a standard colloidal Wigner-Seitz cell model, presented in detail in the Supplemental Information (SI), show that a small (essentially invisible) variation of order $\sim1\%$ in the colloidal packing fraction between base and tip of the channel can alter the colloidal surface charge density by several 10\% for a fixed zeta potential. By assuming macroscopic electroneutrality \cite{mani2009propagation,mani2011deionization}, the ionic space charge in the (thin) electric double layers of the charged colloids varies similarly, thereby providing a natural explanation for the hitherto unexplained ionic charge density profile. In order to theoretically investigate the hypothesis that the inhomogeneous space charge density is responsible for the ICR, we employ standard Poisson-Nernst-Planck (PNP) equations for ionic transport to explain that the inhomogeneous ionic space charge density indeed leads to current rectification. Our theoretical framework, based on an efficient slab-averaging approach in tapered channel geometries \cite{boon2021nonlinear,kamsma2023unveiling}, is described below, while the detailed calculations can be found in the SI. 

The PNP equations form an effective theoretical framework to analyse ion transport in charged porous materials \cite{schmuck2015homogenization}. However, the complex three-dimensional geometric structure of the NCNM, with features on length scales varying from the colloidal surface-surface distance all the way up to the channel length, introduces intricate numerical challenges for fully spatially resolved solutions of the PNP equations. To simplify, we consider slab-averages, i.e.\ the average along a cross section \cite{mani2009propagation,zangle2009propagation,jubin2018dramatic,boon2021nonlinear,kamsma2023unveiling}, of the electric potential and the ionic concentrations in the porous structure between the colloids. Although this sacrifices on nanoscale details, it does account for the pinched electric field lines towards the channel tip and for the spatial variation of the ionic charge density. Through this method we reduce the three-dimensional Nernst-Planck equation to a one-dimensional form, providing an expression for the total salt and charge flux through the channel. The divergence of the total salt flux qualitatively shows that the experimentally observed inhomogeneous ionic space charge density forms a source (sink) term of salt, resulting in salt accumulation (depletion) upon a positive (negative) applied voltage $V$. Quantitatively, a divergence-free steady-state condition on the total salt flux provides a differential equation for the voltage-dependent slab-averaged salt concentration profile, which we solve analytically. By viewing the channel as a series of conductive slabs, with the conductance of each slab proportional to the (now known) voltage-dependent salt concentration, we calculate the steady-state channel conductance $g_\infty(V)=I(V)/V$. This describes how an increase (decrease) in salt in the channel at positive (negative) voltages makes the channel more (less) conductive. Our theory thus quantitatively confirms the experimental hypothesis that the ionic space charge distribution results in salt concentration polarisation and hence in current rectification \cite{choi2016high}. Moreover, leveraging the general analytical nature of our theory, we demonstrate that any inhomogeneous ionic space charge density in generic channels (provided they are well-described by slab-averaged PNP equations) is the key ingredient for a source-sink term of salt and thus for current rectification, derived in detail in the SI. Therefore we not only provide a mechanistic insight as to how the space charge leads to current rectification in the channel of present interest, but this understanding could also explain current rectification in channels with other sources of space charge densities and with other geometries \cite{sabbagh2023designing,kim2022asymmetric}. Furthermore, this insight may provide inspiration for future design of devices that exhibit current rectification.

In Fig.~\ref{fig:Channel_Synapse_Hyst_STP}(b) we plot the predicted steady-state current (red) and the experimentally observed current (blue), revealing a similar current rectification. The experimentally observed ICR ratio of 11 is lower than one of our earlier (slightly more complicated) microchannels \cite{choi2016high}, however it is sufficient for this work and we believe it will be straightforward to optimise our channel for higher ratios in the future as higher ratios were already achieved in similar channels \cite{choi2016high,kim2022asymmetric}.

Up to this point, we treated the system in steady-state. When extending our view to the device dynamics we need to consider the time it takes for ions to accumulate into or deplete out of the channel. Utilizing our aforementioned expression for the total salt flux through the channel, we calculate the net flux $\gamma V^{\prime}$ into the channel upon a small applied voltage $V^{\prime}$ and find (see SI for details) that $\gamma\propto D/L$, with $L$ the channel length and $D$ the ionic diffusion coefficient. The contributions to the net flux solely come from the conductive, i.e.\ voltage-driven, flux term in the Nernst-Planck equation. The proportionality to $D/L$ is intuitive as the electric field strength in the channels is proportional to $1/L$ and all flux terms are proportional to the ionic mobilities and hence to $D$. With our expression for the slab-averaged salt concentration profile we also calculate the total change in salt $\alpha V^{\prime}$ upon applying the small voltage $V^{\prime}$, and we find $\alpha\propto L$. This proportionality to $L$ again is intuitive, as the volume of the channel scales with $L$. The ratio $\alpha/\gamma$ between this total change in salt and net flux provides an estimate for the concentration polarisation timescale, given by
\begin{align}\label{eq:ts}
    \tau=\frac{L^2}{4D}\xi,
\end{align}
where $\xi\approx 0.42$ is an involved dimensionless number that depends on the ratio of the channel widths $R_{\ch{t}}/R_{\ch{b}}$ and on the ratio of the internal space charge density at the tip and the base (full expression in SI). Eq.~(\ref{eq:ts}) shows that $\tau$ is a diffusionlike time, which is remarkable as no diffusion terms from the Nernst-Planck equation go directly into the derivation of Eq.~(\ref{eq:ts}). Nevertheless, from the aforementioned intuitive dependencies on $D/L$ and $L$ of the net salt flux and total change in salt, respectively, we surprisingly do retrieve a diffusionlike timescale from a voltage-driven process. Inserting our present system parameters, including $D=1.45\text{ }\mu\text{m}^2\text{ms}^{-1}$, yields a memory timescale of $\tau=1.62$ s. 

With the the steady-state conductance $g_{\infty}(V)$ and the conductance memory timescale $\tau$, we formulate our theory for the time-dependent channel conductance $g(t)$ resulting from a time-dependent voltage $V(t)$. We intuitively expect $g(t)$ to relax towards the instantaneous static conductance $g_{\infty}(V(t))$ on a time scale $\tau$, with a vanishing rate of change $\dd g(t)/\dd t$ once $g(t)=g_{\infty}(V(t))$. We employ a simple equation of motion (which we more rigorously support in the SI) that characterises this relaxation process and write
\begin{align}
	\dfrac{\dd g(t)}{\dd t}=&\frac{g_\infty( V(t))-g(t)}{\tau};\label{eq:cond_ODE}\\
	I(t)=&g(t)V(t),\label{eq:Itd}
\end{align}
where Eq.~(\ref{eq:cond_ODE}) is a form that was also successfully applied in previous studies on various memristors \cite{robin2023long,markin2014analytical,kamsma2023iontronic,kamsma2023unveiling} and Eq.~(\ref{eq:Itd}) is Ohm's law.

To demonstrate the memristive properties of our device we impose a sinusoidal voltage $V(t)$ of amplitude 5V and frequency $f=0.1$ Hz as shown in Fig.~\ref{fig:Channel_Synapse_Hyst_STP}(c, green). The resulting current-voltage ($I-V$) diagram is shown in Fig.~\ref{fig:Channel_Synapse_Hyst_STP}(c, blue), where a clear pinched hysteresis loop is observed, the hallmark of a memristor \cite{chua2014if}. This loop features a pronounced memory effect (i.e.\ more open hysteresis loop) compared to various fluidic memristive devices \cite{robin2023long,xiong2023neuromorphic,brown2022selective,wang2012transmembrane}, allowing for a wide range of comparatively high conductances. Moreover, in Fig.~\ref{fig:Channel_Synapse_Hyst_STP}(c, red) we see that our theory shows good agreement with experimental findings.
\begin{figure}[t!]
\centering
     \includegraphics[width=0.5\textwidth]{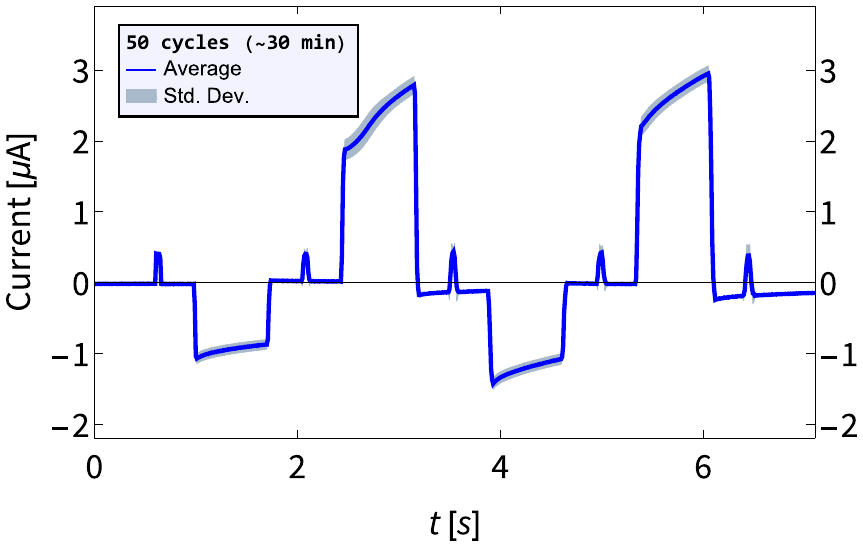}
        \caption{Stability of our NCNM memristors. Averaged current (blue) measured over 50 subsequent voltage pulse train cycles, taking roughly 30 minutes. Each train consists of four write-pulses, of -2 V and 5 V respectively, interspersed with five read-pulses of 1 V. All cycles showed essentially the same behaviour, producing a spread with standard deviation for each current measurement around the average of maximally $\sim7\%$ (blue-gray), demonstrating the device stability.}
        \label{fig:0101stdAvg}
        \vspace{-0.5 cm}
\end{figure}

Memristors are recognised as artificial analogues to synapses, the connections between neurons \cite{chua2013memristor}. In neuronal communication, the change in membrane potential of a postsynaptic neuron, triggered by an influx of ions in response to a signal from a connected presynaptic neuron, is a measure for the synaptic connection strength \cite{fundNeuroAll}. Similarly, our device's measured ion current, also a result of ion flux through a membrane, draws a parallel between the channel's conductance and synaptic strength in neurons, as schematically depicted in Fig.~\ref{fig:Channel_Synapse_Hyst_STP}(d). A crucial aspect of neuronal functioning is short-term plasticity (STP), which allows neurons to adjust their synaptic strength in response to recent input history, being of key importance in information processing \cite{rotman2011short}. STP involves changes in the synaptic strength that decay over timescales ranging from milliseconds to minutes, where an increase of the synaptic strength is called (short-term) facilitation and a decrease (short-term) depression \cite{abbott2004synaptic,deng2011diverse,rotman2011short}. To demonstrate that our fluidic memristor can mimic these aspects of neuronal STP, we apply four consecutive positive and negative ``write-pulses'' of $5\text{ V}$ and $-2.5\text{ V}$, respectively, with a $0.75\text{ s}$ duration, separated by intervals of $\Delta t=0.75\text{ s}<\tau$ smaller than the memory retention time $\tau$. The asymmetric 5 V and -2.5 V voltages helped optimize the conductance response. We measure the channel conductance by applying small and short ``read-pulses'' of -1 V and duration 50 ms after each write-pulse. In Fig.~\ref{fig:Channel_Synapse_Hyst_STP}(e) we show the current measurements (blue) when four 5 V write-pulses and five -1 V read-pulses (green) are applied, with energy consumption of $\sim1-10\text{ }\mu$J for the write-pulses and $\sim10-100\text{ }$nJ for the read-pulses. The read-pulses are converted to the measured channel conductances shown in Fig.~\ref{fig:Channel_Synapse_Hyst_STP}(f). As illustrated in Fig.~\ref{fig:Channel_Synapse_Hyst_STP}(f, blue), our fluidic memristor exhibits both facilitation (top graph) and depression (middle graph), hence replicating the characteristic features of neuronal STP. These results are the average of three devices with two measurements per device, each showing quantitatively similar behaviour. In Fig.~\ref{fig:Channel_Synapse_Hyst_STP}(f, bottom graph) the short-term character of the response is prominently visible when the interval between the pulses is much longer than the typical memory retention time $\tau$ and no cumulative change in conductance is observed. Our experimental findings are mostly in good agreement with Eq.~(\ref{eq:cond_ODE}) shown in Fig.~\ref{fig:Channel_Synapse_Hyst_STP}(f, red), the only notable discrepancy being that the measured strength of depression is weaker than predicted. The overall agreement emphasizes the robustness and predictive power of the theoretical model in quantitatively characterizing the device properties.
\begin{figure*}[t]
\centering
     \includegraphics[width=1\textwidth]{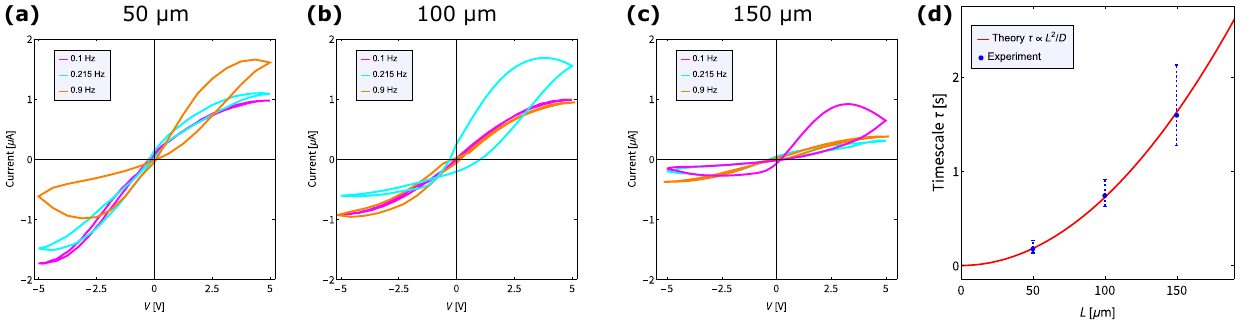}
        \caption{Voltage-driven concentration polarisation occurs over a diffusionlike timescale. Experimental $I-V$ curves for channels of length \textbf{(a)} $50\text{ }\mu$m, \textbf{(b)} $100\text{ }\mu$m, and \textbf{(c)} $150\text{ }\mu$m for a sinusoidal potential with amplitude 5 V and frequencies of 0.1 Hz (magenta), 0.215 Hz (cyan), and 0.9 Hz (orange). Measurements were also conducted for intermediate frequencies of 0.01 Hz, 0.05 Hz, 0.075 Hz, 0.1 Hz, 0.125 Hz, 0.175 Hz, 0.215 Hz, 0.255 Hz, 0.6 Hz, 0.9 Hz, 1.2 Hz, and 2 Hz, shown in the SI. \textbf{(d)} Memory timescale $\tau$ for all channels determined by the frequency $f_{\text{max}}$ for which the enclosed area inside the loop is maximal via the relation $2\pi f_{\text{max}}\tau=1$ \cite{kamsma2024simple}. The measured frequencies around each $f_{\text{max}}$ yield upper and lower bounds, shown here as error bars.}
        \label{fig:TimeScale}
\end{figure*}

To reliably perform neuromorphic reservoir computing our devices need to repeatedly produce the same response to signals. In Fig.~\ref{fig:0101stdAvg}, we demonstrate the stability and reproducibility of our device by applying 50 subsequent cycles of four write-pulses of -2 V and 5 V respectively, taking roughly 30 minutes to complete. The resulting current, averaged over all 50 cycles (blue), shows a narrow spread with a standard deviation (blue-gray) of at most $\sim 7\%$. In the SI, we present additional measurements, including 26 cycles over 4 hours for all 16 different four-pulse voltage trains, reliably finding a similar conductance modulation each cycle with conductance standard deviations of typically a few \% and no more than $\sim10\%$. As fabrication is fast and easy, we normally constructed new devices for new sets of experiments. However, devices were in fact stable enough to be reusable, but did dry out if kept for long with our current way of storing, requiring cleaning the salt residue before using again. The device stability is an important feature that enables it to reliably distinguish a large number of different time series, underpinning the reservoir computing, as we discuss later.
 
The typical timescale $\tau$ over which the conductance memory is retained is an important property of memristive systems and the ability to incorporate a wide range of timescales is desirable \cite{chicca2020recipe}. As per Eq.~(\ref{eq:ts}), we predict that the memory timescale of a device can freely be chosen from a wide range of options by constructing it with the appropriate length $L$ or radii $R_{\ch{t}}/R_{\ch{b}}$, here we focus on the $L$-dependence. The prediction that $\tau$ is determined by a diffusionlike time $\propto L^2/D$, despite the channel being voltage driven, forms a specific, non-trivial and easily falsifiable prediction. To test this we fabricated channels of lengths $50\text{ }\mu\text{m}$, $100\text{ }\mu\text{m}$ and $150\text{ }\mu\text{m}$ and determined for all three at which frequency $f_{\text{max}}$ of an applied sinusoidal voltage the area enclosed in the hysteresis loop in the $I-V$ diagram is maximal. Using the relation $2\pi f_{\text{max}}\tau=1$, we can find an estimate for the timescale $\tau$ \cite{kamsma2024simple}, and check whether $\tau\propto L^2$. The natural relation $2\pi f_{\text{max}}\tau=1$ entails that maximal hysteresis is observed when the voltage changes over the typical memory time $\tau$, i.e.\ providing enough time for the conductance to change, but not enough to reach its steady-state. We mathematically derived this relation for general memristors described by Eqs.~[\ref{eq:cond_ODE}]-[\ref{eq:Itd}] when a sinusoidal voltage is applied \cite{kamsma2024simple}. Additionally, since the equilibrium channel conductance is inversely proportional to the channel length $g_0\propto1/L$ we expect the overall current, and thus the overall loop area, to decrease for longer $L$ at the same voltage. In our experiments we indeed find that channels of different lengths respond to different frequencies, and show overall decreasing conductances for increasing $L$, as can be seen by the various hysteresis loops in Fig.~\ref{fig:TimeScale}(a,b,c). By comparing the different values for $f_{\text{max}}$ we not only confirm that $f_{\text{max}}^{-1}\propto L^2$, but that quantitatively we have good agreement with the theory. Thus by manufacturing channels of different lengths, facile via the flexible fabrication process, a wide range of memory timescales can be achieved. Therefore our device offers a versatility important for tasks that require processing of signals over various timescales \cite{chicca2020recipe}.

\begin{figure*}[ht]
\centering
     \includegraphics[width=1\textwidth]{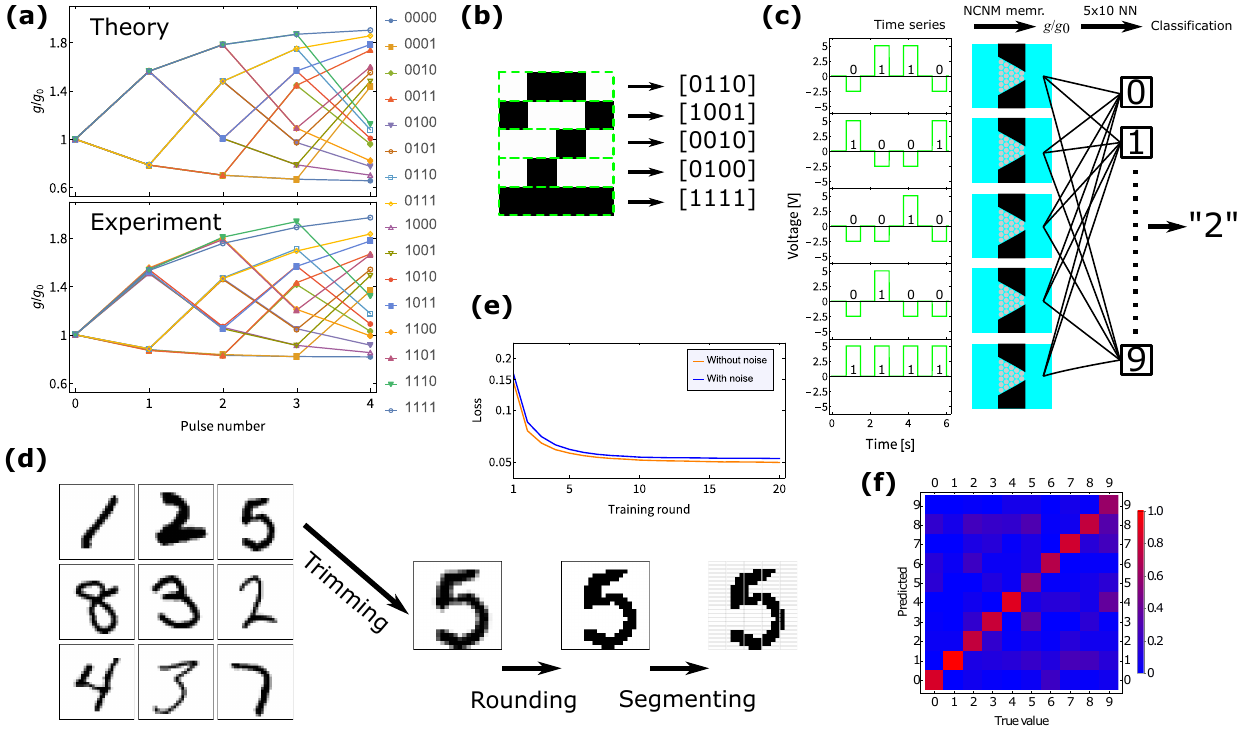}
        \caption{Our iontronic memristor as a reservoir computing element. \textbf{(a)} Theoretical prediction (top) and experimental observation (bottom) of the relative change in conductance as a response to the $2^4=16$ different possible bit-strings, where a ``0'' and ``1'' correspond to pulses of $-2.5\text{ V}$ and $5\text{ V}$, respectively. Three separate devices were used to find the average conductance and typical variation in response to each unique voltage train. \textbf{(b)} Depiction of how a ``2'' can be transformed into 5 distinct bit-strings (other digits are depicted in the SI). \textbf{(c)} Schematic of how the number ``2'' from \textbf{(b)} is translated to 5 voltage trains, yielding 5 conductance values after the fourth (last) pulse. The conductances and variabilities from \textbf{(a)} were then used to train a single-layer fully connected $5\times10$ neural network in silico, that converts the conductances to a classification of the number ``2". \textbf{(d)} Nine examples of handwritten numbers from the MNIST database \cite{deng2012mnist}, where the (originally $28\times 28$ pixel) images are trimmed to $20\times 22$ pixel images, the grayscales are rounded to either white or black pixels and then the image is segmented into 110 bit-strings. \textbf{(e)} The loss function (mean squared loss) during training per training round when the experimentally found noise, which is experimentally quantified using the (device-to-device) variabilities found in our result in \textbf{(a)}, of the devices is not taken into account (orange) and when it is taken into account (blue). \textbf{(f)} The confusion matrix on a test set of 2,000 samples, showing an overall accuracy of 81\%, comparable with recent reported results using more conventional platforms \cite{du2017reservoir,midya2019reservoir}.}
        \label{fig:conductance_bits_MNIST}
\end{figure*}

\section{Aqueous Reservoir Computing}\label{sec:RC}
The short-term memory properties of our fluidic memristor, with a memory retention time that can easily be chosen for each device from a wide range of options as shown in Fig.~\ref{fig:TimeScale}, make it a promising candidate for performing reservoir computing, a brain-inspired framework that leverages a fixed dynamic network, or ``reservoir'', to transform complex temporal input data into an output that can be easily classified \cite{cucchi2022hands}. Unlike traditional neural network approaches, where the full network needs to be trained, in reservoir computing only a comparatively simple read-out function (here a single-layer neural network) that classifies the output of the reservoir requires training \cite{tanaka2019recent}. These properties have driven interest in reservoir computing for analysing a variety of temporal signals \cite{cao2022emerging,tanaka2019recent}. Generally, a device needs two properties for it to be applicable in reservoir computing, (i) a short-term memory and (ii) nonlinear dynamics \cite{cao2022emerging}, which our device satisfies as shown in Fig.~\ref{fig:Channel_Synapse_Hyst_STP}(f).

To demonstrate the reservoir computing capabilities of our fluidic memristor, we carry out an established benchmarking protocol of classifying handwritten numbers using reservoir computing \cite{du2017reservoir,midya2019reservoir}. To build up to this, we first employ a standard method of separating 4-bit strings and show that our memristor already performs remarkably well at this initial task compared to previous results using more conventional platforms \cite{pyo2022non,kim2022implementation}. All $2^4=16$ combinations are translated into a series of voltage pulses with a duration of $0.75\text{ s}$, separated by intervals of $0.75\text{ s}$, where a 0 and a 1 correspond to a voltage of -2.5 V and 5 V, respectively. In theory, applying the voltage pulse trains should yield 16 distinct conductance time traces $g(t)$ shown in Fig.~\ref{fig:conductance_bits_MNIST}(a, top), effectively mapping the 16 possible input patterns onto the 16 different conductance values after the fourth pulse, purely by virtue of the device properties. Excitingly, when we apply this protocol in experiments we find the predicted 16 distinct conductance signatures as shown in Fig.~\ref{fig:conductance_bits_MNIST}(a, bottom), featuring the exact same ordering of conductance values and quantitatively similar changes in conductances, once more highlighting the predictive power of the theory. To obtain Fig.~\ref{fig:conductance_bits_MNIST}(a, bottom), all 16 different voltage pulse trains were repeated twice on three different devices, producing six measurements in total and yielding the average conductances shown in Fig.~\ref{fig:conductance_bits_MNIST}(a, bottom). All 6 runs displayed essentially the same behaviour, producing a spread of conductances after the fourth write-pulse for each pulse train, with standard deviations of around $g/g_0\sim0.06-0.26$ (complete list in SI). Individual device variability is lower, with typically standard deviations of several $g/g_0\sim0.01$ and no more than $0.16$. As we perform an analog computing method, rather than discrete logic, the possible overlap between measured conductances is not an issue and the various time series can reliably encode (handwritten) numbers, as we show next.

To illustrate how the results shown in Fig.~\ref{fig:conductance_bits_MNIST}(a) can be leveraged to classify more complex data inputs with an explanatory example, let us consider the simple single-digit numbers 0-9, represented by black and white $4\times5$ pixel images. By converting a row of 4 pixels to a string of bits by letting a white pixel correspond to a ``0'' and a black pixel to a ``1'', we can encode the entire image with 5 strings of 4 bits, as shown in Fig.~\ref{fig:conductance_bits_MNIST}(b) for the number ``2'' (other digits are shown in the SI). These bit-strings then generate 5 distinct signature outputs, as we saw in Fig.~\ref{fig:conductance_bits_MNIST}(a). A single-layer fully connected $5\times10$ neural network is then trained in silico to classify the 5 measured conductances as numbers. This protocol is schematically illustrated in Fig.~\ref{fig:conductance_bits_MNIST}(c). Other types of simple readout functions could possibly also suffice. We trained our read-out network in silico using the results shown in Fig.~\ref{fig:conductance_bits_MNIST}(a, bottom). To incorporate the (device-to-device) variability, each individual pulse was subject to some noise newly drawn from a normal distribution with mean 0 and standard deviation given by the experimentally determined standard deviation for that specific voltage train. During training, we repeated this process 100 times for each of the numbers 0-9, achieving perfect classification of all 10 digits with noise-free inference measurements. If we also take the noise into account during inference, we still achieve an overall accuracy of 95\%, highlighting the system's robustness against noise. Note that actual training is only performed on a simple and small neural network, that would otherwise not be capable of handling temporal inputs, while the ``hard'' work of separating the time-dependent signals is handled by the internal physics of our fluidic memristor. Ultimately, this successful classification of simple digit images serves as an explanatory proof-of-concept for the broader application of performing complex time-dependent data analysis tasks.

Building on our previous experiments with classifying simple digit images, we go a step further by classifying handwritten numbers from the well-known MNIST database. This database contains a large dataset of $28\times28$ pixel images of handwritten numbers and has become a standard dataset to test and demonstrate the classification capabilities of machine learning methods \cite{deng2012mnist}. We first converted each image into a $20\times22$ pixel black and white image by trimming the edges off and rounding the grayscales to either black or white pixels, as depicted in Fig.~\ref{fig:conductance_bits_MNIST}(d). Each image was then sectioned into pixel rows of four pixels, which can be encoded with a voltage pulse train as outlined before, leading to a total of 110 conductance states per image. The readout function consists of a single-layer fully connected $110\times 10$ neural network, trained on a dataset of 20,000 samples. The training incorporated the conductance response and device-to-device noise using the results shown in Fig.~\ref{fig:conductance_bits_MNIST}(a), with the noise taken into account like we did for the classification $4\times 5$ pixel digits. The noise hardly had any effect on overall accuracy, as can be seen by the nearly identical decrease of the loss function during training in Fig.~\ref{fig:conductance_bits_MNIST}(e). Via this rudimentary straightforward protocol, we achieved an accuracy of 81\% on a test set of 2,000 samples, comparable with earlier reported accuracies of 83\% and 85.6\% resulting from the same protocol using solid-state memristors \cite{midya2019reservoir,du2017reservoir}. In Fig.~\ref{fig:conductance_bits_MNIST}(f) the classification is schematically depicted in a confusion matrix, where we see how often each combination of true and predicted numbers occurred in the test set. 

Our successful implementation of an aqueous iontronic device as a synaptic element for reservoir computing with a performance that is (at least) on par with more traditional platforms \cite{pyo2022non,kim2022implementation,du2017reservoir,midya2019reservoir} is a promising demonstration of the potential that our fluidic platform offers for brain-inspired computing. The functionality extracted from our simple devices is remarkable, obviating the need for complicated circuits to distinguish the various time series of interest here, instead relying on the stable conductance modulation of individual devices. Consequently, as we advance towards circuits integrating multiple coupled NCNM devices, we anticipate that the robust individual device properties demonstrated here will enable the realization of expanded functionalities with relatively simple circuits.

\section{Discussion and Conclusion}
We implemented a fluidic iontronic volatile memristor as a synaptic element for neuromorphic reservoir computing, while the device relies on the same aqueous electrolyte medium and ionic signal carriers as biological neurons. Our memristor consists of a tapered microchannel that features a conducting network of nanochannels embedded in a rigid colloidal structure, forming a nanochannel network membrane (NCNM). Device fabrication is fast, cost-effective, and easy via an almost free-shaping soft-lithography process. The trait that underpins the conductance memory effect of the channel is its steady-state diode behaviour, for which NCNM devices have shown excellent performance \cite{choi2016high,kim2022asymmetric}, translating into a wide range of achievable conductances. Additionally, our device exhibited stable and reliable (dynamic) conductance modulation, enabling its use as a computing element. Moreover, the quadratic dependence of the memory timescale on the channel length offers a straightforward method to design a channel to feature a specific timescale, a sought after feature for advancing neuromorphic computing capabilities \cite{chicca2020recipe}.

Our memristor is inspired and supported by a comprehensive theory directly derived from the underlying physical equations of diffusive and electric continuum ion transport. We experimentally quantitatively verified the predictions of our theory on multiple occasions, amongst which the specific and surprising prediction that the memory retention time of the channel depends on the channel diffusion time, despite the channel being constantly voltage-driven. The theory exclusively relies on physical parameters, such as channel dimensions and ion concentrations, and enabled streamlined experimentation by pinpointing the relevant signal timescales, signal voltages, and suitable reservoir computing protocol. Additionally, we identify an inhomogeneous charge \emph{density} as the key ingredient for iontronic channels to exhibit current rectification  (provided they are well-described by slab-averaged PNP equations). Consequently, our theory paves the way for targeted advancements in iontronic circuits and facilitates efficient exploration of their diverse applications.

For future prospects, a next step is the integration of multiple devices, where the flexible fabrication methods do offer a clear path towards circuits that couple multiple channels. Additionally, optimising the device to exhibit strong conductance modulation for lower voltages would be of interest to bring electric potentials found in nature into the scope of possible inputs and reduce the energy consumption for conductance modulation. From a theoretical perspective, the understanding of the (origin of the) inhomogeneous space charge and the surface conductance is still somewhat limited. These contain (physical) parameters that are now partially chosen from a reasonable physical regime to yield good agreement, but do not directly follow from underlying physical equations. We also assume that the inhomogeneous ionic space charge distribution is constant, while it might well be voltage-dependent. Lastly, our theoretical model treats the complex porous structure in terms of slab-averages, thereby possibly missing out on detailed features. These constraints of the theoretical model could account for some of the discrepancies between theory and experiment, which is notable in the steady-state current in Fig.~\ref{fig:Channel_Synapse_Hyst_STP}(b) and the decrease in conductance in Fig.~\ref{fig:Channel_Synapse_Hyst_STP}(f). For the purposes of this work our current approach is sufficient, however, a more in-depth study could offer a more profound understanding into the interesting features of the channel.

In conclusion, in order to narrow the gap between the promise of aqueous iontronic neuromorphic computation and its implementation, our work demonstrates the capabilities of a fluidic memristor by employing it as an artificial synapse for carrying out neuromorphic reservoir computing. Temporal signals, in the form of voltage pulse trains, that together represent (handwritten) numbers were distinguished by individual channels for subsequent in silico classification with a simple readout function, demonstrating (at least) comparable performance to more conventional solid-state platforms \cite{pyo2022non,kim2022implementation,du2017reservoir,midya2019reservoir}. Additionally, the device is fabricated with a cost-effective easy soft-lithography process. The achieved computing properties are inspired and supported by a quantitative predictive theoretical model of the device dynamics. Consequently, our work establishes a solid foundation, both theoretically and experimentally, for future investigations into fluidic memristive systems and their application in aqueous neuromorphic computing architectures, paving the way for computing systems that more closely resemble the brains fascinating aqueous processes.
\\\\\\

\begin{acknowledgments}
This work is part of the D-ITP consortium, a program of the Netherlands Organisation for Scientific Research (NWO) that is funded by the Dutch Ministry of Education, Culture and Science (OCW). This work is also supported by a National Research Foundation of Korea (NRF) grant funded by the Korean government (MSIP) (2020R1A2C2009093) and by the Korea Environment Industry \& Technology Institute (KEITI) through its Ecological Imitation-based Environmental Pollution Management Technology Development Project funded by the Korea Ministry of Environment (MOE) (2019002790007).
\end{acknowledgments}
\vspace{0.5 cm}
\section*{Author contributions}
T.M.K.\ conceptualized the work and developed the theory; J.K.\ carried out the experiments; K.K. contributed to carrying out the experiments; W.Q.B.\ contributed to developing the theory; C.S., J.P.\ and R.v.R.\ supervised the research. All authors discussed the results and contributed to the manuscript.

\section*{Methods}
The fabrication of microchannel and formation of the NCNM for the fluidic memristor is similar to previously reported methods \cite{choi2016high,kim2022asymmetric} and is described in the SI in detail. A master for multi-layered channels (target heights are 5 $\mu$m for shallow channel and 100 $\mu$m for deep) was created using a multi-step UV exposure with negative photoresist (PR, SU-8 2005, 3050, Microchem Co., USA). After surface treatment of the master with (3,3,3-trifluoropropyl)silane (452807, Sigma-Aldrich, USA) for easy separation, Polydimethylsiloxane (PDMS, Sylgard, Dow Corning Korea Ltd., Korea) was poured and cured by heating. The detached PDMS device was bonded with a slide glass. The formation of NCNM was formed by a self-assembly of homogeneous nanoparticles with negative surface charge in the desired shallow channel using Laplace pressure to halt the solvent at the base and evaporation of solvent. A close-packed fcc was formed by the growth of the ordered lattice induced by the evaporation.


\begin{thebibliography}{52}%
\makeatletter
\providecommand \@ifxundefined [1]{%
 \@ifx{#1\undefined}
}%
\providecommand \@ifnum [1]{%
 \ifnum #1\expandafter \@firstoftwo
 \else \expandafter \@secondoftwo
 \fi
}%
\providecommand \@ifx [1]{%
 \ifx #1\expandafter \@firstoftwo
 \else \expandafter \@secondoftwo
 \fi
}%
\providecommand \natexlab [1]{#1}%
\providecommand \enquote  [1]{``#1''}%
\providecommand \bibnamefont  [1]{#1}%
\providecommand \bibfnamefont [1]{#1}%
\providecommand \citenamefont [1]{#1}%
\providecommand \href@noop [0]{\@secondoftwo}%
\providecommand \href [0]{\begingroup \@sanitize@url \@href}%
\providecommand \@href[1]{\@@startlink{#1}\@@href}%
\providecommand \@@href[1]{\endgroup#1\@@endlink}%
\providecommand \@sanitize@url [0]{\catcode `\\12\catcode `\$12\catcode
  `\&12\catcode `\#12\catcode `\^12\catcode `\_12\catcode `\%12\relax}%
\providecommand \@@startlink[1]{}%
\providecommand \@@endlink[0]{}%
\providecommand \url  [0]{\begingroup\@sanitize@url \@url }%
\providecommand \@url [1]{\endgroup\@href {#1}{\urlprefix }}%
\providecommand \urlprefix  [0]{URL }%
\providecommand \Eprint [0]{\href }%
\providecommand \doibase [0]{https://doi.org/}%
\providecommand \selectlanguage [0]{\@gobble}%
\providecommand \bibinfo  [0]{\@secondoftwo}%
\providecommand \bibfield  [0]{\@secondoftwo}%
\providecommand \translation [1]{[#1]}%
\providecommand \BibitemOpen [0]{}%
\providecommand \bibitemStop [0]{}%
\providecommand \bibitemNoStop [0]{.\EOS\space}%
\providecommand \EOS [0]{\spacefactor3000\relax}%
\providecommand \BibitemShut  [1]{\csname bibitem#1\endcsname}%
\let\auto@bib@innerbib\@empty
\bibitem [{\citenamefont {Mehonic}\ and\ \citenamefont
  {Kenyon}(2022)}]{mehonic2022brain}%
  \BibitemOpen
  \bibfield  {author} {\bibinfo {author} {\bibfnamefont {A.}~\bibnamefont
  {Mehonic}}\ and\ \bibinfo {author} {\bibfnamefont {A.~J.}\ \bibnamefont
  {Kenyon}},\ }\bibfield  {title} {\bibinfo {title} {Brain-inspired computing
  needs a master plan},\ }\href@noop {} {\bibfield  {journal} {\bibinfo
  {journal} {Nature}\ }\textbf {\bibinfo {volume} {604}},\ \bibinfo {pages}
  {255} (\bibinfo {year} {2022})}\BibitemShut {NoStop}%
\bibitem [{\citenamefont {Schuman}\ \emph {et~al.}(2017)\citenamefont
  {Schuman}, \citenamefont {Potok}, \citenamefont {Patton}, \citenamefont
  {Birdwell}, \citenamefont {Dean}, \citenamefont {Rose},\ and\ \citenamefont
  {Plank}}]{schuman2017survey}%
  \BibitemOpen
  \bibfield  {author} {\bibinfo {author} {\bibfnamefont {C.~D.}\ \bibnamefont
  {Schuman}}, \bibinfo {author} {\bibfnamefont {T.~E.}\ \bibnamefont {Potok}},
  \bibinfo {author} {\bibfnamefont {R.~M.}\ \bibnamefont {Patton}}, \bibinfo
  {author} {\bibfnamefont {J.~D.}\ \bibnamefont {Birdwell}}, \bibinfo {author}
  {\bibfnamefont {M.~E.}\ \bibnamefont {Dean}}, \bibinfo {author}
  {\bibfnamefont {G.~S.}\ \bibnamefont {Rose}},\ and\ \bibinfo {author}
  {\bibfnamefont {J.~S.}\ \bibnamefont {Plank}},\ }\bibfield  {title} {\bibinfo
  {title} {A survey of neuromorphic computing and neural networks in
  hardware},\ }\href@noop {} {\bibfield  {journal} {\bibinfo  {journal} {arXiv
  preprint arXiv:1705.06963}\ } (\bibinfo {year} {2017})}\BibitemShut {NoStop}%
\bibitem [{\citenamefont {Sangwan}\ and\ \citenamefont
  {Hersam}(2020)}]{sangwan2020neuromorphic}%
  \BibitemOpen
  \bibfield  {author} {\bibinfo {author} {\bibfnamefont {V.~K.}\ \bibnamefont
  {Sangwan}}\ and\ \bibinfo {author} {\bibfnamefont {M.~C.}\ \bibnamefont
  {Hersam}},\ }\bibfield  {title} {\bibinfo {title} {Neuromorphic
  nanoelectronic materials},\ }\href@noop {} {\bibfield  {journal} {\bibinfo
  {journal} {Nature Nanotechnology}\ }\textbf {\bibinfo {volume} {15}},\
  \bibinfo {pages} {517} (\bibinfo {year} {2020})}\BibitemShut {NoStop}%
\bibitem [{\citenamefont {Schuman}\ \emph {et~al.}(2022)\citenamefont
  {Schuman}, \citenamefont {Kulkarni}, \citenamefont {Parsa}, \citenamefont
  {Mitchell}, \citenamefont {Date},\ and\ \citenamefont
  {Kay}}]{schuman2022opportunities}%
  \BibitemOpen
  \bibfield  {author} {\bibinfo {author} {\bibfnamefont {C.~D.}\ \bibnamefont
  {Schuman}}, \bibinfo {author} {\bibfnamefont {S.~R.}\ \bibnamefont
  {Kulkarni}}, \bibinfo {author} {\bibfnamefont {M.}~\bibnamefont {Parsa}},
  \bibinfo {author} {\bibfnamefont {J.~P.}\ \bibnamefont {Mitchell}}, \bibinfo
  {author} {\bibfnamefont {P.}~\bibnamefont {Date}},\ and\ \bibinfo {author}
  {\bibfnamefont {B.}~\bibnamefont {Kay}},\ }\bibfield  {title} {\bibinfo
  {title} {Opportunities for neuromorphic computing algorithms and
  applications},\ }\href@noop {} {\bibfield  {journal} {\bibinfo  {journal}
  {Nature Computational Science}\ }\textbf {\bibinfo {volume} {2}},\ \bibinfo
  {pages} {10} (\bibinfo {year} {2022})}\BibitemShut {NoStop}%
\bibitem [{\citenamefont {Strukov}\ \emph {et~al.}(2008)\citenamefont
  {Strukov}, \citenamefont {Snider}, \citenamefont {Stewart},\ and\
  \citenamefont {Williams}}]{strukov2008missing}%
  \BibitemOpen
  \bibfield  {author} {\bibinfo {author} {\bibfnamefont {D.~B.}\ \bibnamefont
  {Strukov}}, \bibinfo {author} {\bibfnamefont {G.~S.}\ \bibnamefont {Snider}},
  \bibinfo {author} {\bibfnamefont {D.~R.}\ \bibnamefont {Stewart}},\ and\
  \bibinfo {author} {\bibfnamefont {R.~S.}\ \bibnamefont {Williams}},\
  }\bibfield  {title} {\bibinfo {title} {The missing memristor found},\
  }\href@noop {} {\bibfield  {journal} {\bibinfo  {journal} {Nature}\ }\textbf
  {\bibinfo {volume} {453}},\ \bibinfo {pages} {80} (\bibinfo {year}
  {2008})}\BibitemShut {NoStop}%
\bibitem [{\citenamefont {Chua}(2013)}]{chua2013memristor}%
  \BibitemOpen
  \bibfield  {author} {\bibinfo {author} {\bibfnamefont {L.}~\bibnamefont
  {Chua}},\ }\bibfield  {title} {\bibinfo {title} {Memristor, hodgkin-huxley,
  and edge of chaos},\ }\href@noop {} {\bibfield  {journal} {\bibinfo
  {journal} {Nanotechnology}\ }\textbf {\bibinfo {volume} {24}},\ \bibinfo
  {pages} {383001} (\bibinfo {year} {2013})}\BibitemShut {NoStop}%
\bibitem [{\citenamefont {Zhu}\ \emph {et~al.}(2020)\citenamefont {Zhu},
  \citenamefont {Zhang}, \citenamefont {Yang},\ and\ \citenamefont
  {Huang}}]{zhu2020comprehensive}%
  \BibitemOpen
  \bibfield  {author} {\bibinfo {author} {\bibfnamefont {J.}~\bibnamefont
  {Zhu}}, \bibinfo {author} {\bibfnamefont {T.}~\bibnamefont {Zhang}}, \bibinfo
  {author} {\bibfnamefont {Y.}~\bibnamefont {Yang}},\ and\ \bibinfo {author}
  {\bibfnamefont {R.}~\bibnamefont {Huang}},\ }\bibfield  {title} {\bibinfo
  {title} {A comprehensive review on emerging artificial neuromorphic
  devices},\ }\href@noop {} {\bibfield  {journal} {\bibinfo  {journal} {Applied
  Physics Reviews}\ }\textbf {\bibinfo {volume} {7}},\ \bibinfo {pages}
  {011312} (\bibinfo {year} {2020})}\BibitemShut {NoStop}%
\bibitem [{\citenamefont {{L. Squire, D. Berg, F. Bloom, S. du Lac, A. Ghosh,
  N. Spitzer}}(2008)}]{fundNeuroAll}%
  \BibitemOpen
  \bibfield  {author} {\bibinfo {author} {\bibnamefont {{L. Squire, D. Berg, F.
  Bloom, S. du Lac, A. Ghosh, N. Spitzer}}},\ }\href@noop {} {\emph {\bibinfo
  {title} {Fundamental Neuroscience}}},\ \bibinfo {edition} {3rd}\ ed.\
  (\bibinfo  {publisher} {Academic Press},\ \bibinfo {year} {2008})\BibitemShut
  {NoStop}%
\bibitem [{\citenamefont {Noy}\ and\ \citenamefont
  {Darling}(2023)}]{noy2023nanofluidic}%
  \BibitemOpen
  \bibfield  {author} {\bibinfo {author} {\bibfnamefont {A.}~\bibnamefont
  {Noy}}\ and\ \bibinfo {author} {\bibfnamefont {S.~B.}\ \bibnamefont
  {Darling}},\ }\bibfield  {title} {\bibinfo {title} {Nanofluidic computing
  makes a splash},\ }\href@noop {} {\bibfield  {journal} {\bibinfo  {journal}
  {Science}\ }\textbf {\bibinfo {volume} {379}},\ \bibinfo {pages} {143}
  (\bibinfo {year} {2023})}\BibitemShut {NoStop}%
\bibitem [{\citenamefont {Han}\ \emph {et~al.}(2022)\citenamefont {Han},
  \citenamefont {Oh},\ and\ \citenamefont {Chung}}]{han2022iontronics}%
  \BibitemOpen
  \bibfield  {author} {\bibinfo {author} {\bibfnamefont {S.~H.}\ \bibnamefont
  {Han}}, \bibinfo {author} {\bibfnamefont {M.-A.}\ \bibnamefont {Oh}},\ and\
  \bibinfo {author} {\bibfnamefont {T.~D.}\ \bibnamefont {Chung}},\ }\bibfield
  {title} {\bibinfo {title} {Iontronics: Aqueous ion-based engineering for
  bioinspired functionalities and applications},\ }\href@noop {} {\bibfield
  {journal} {\bibinfo  {journal} {Chemical Physics Reviews}\ }\textbf {\bibinfo
  {volume} {3}},\ \bibinfo {pages} {031302} (\bibinfo {year}
  {2022})}\BibitemShut {NoStop}%
\bibitem [{\citenamefont {Powell}\ \emph {et~al.}(2011)\citenamefont {Powell},
  \citenamefont {Cleary}, \citenamefont {Davenport}, \citenamefont {Shea},\
  and\ \citenamefont {Siwy}}]{powell2011electric}%
  \BibitemOpen
  \bibfield  {author} {\bibinfo {author} {\bibfnamefont {M.~R.}\ \bibnamefont
  {Powell}}, \bibinfo {author} {\bibfnamefont {L.}~\bibnamefont {Cleary}},
  \bibinfo {author} {\bibfnamefont {M.}~\bibnamefont {Davenport}}, \bibinfo
  {author} {\bibfnamefont {K.~J.}\ \bibnamefont {Shea}},\ and\ \bibinfo
  {author} {\bibfnamefont {Z.~S.}\ \bibnamefont {Siwy}},\ }\bibfield  {title}
  {\bibinfo {title} {Electric-field-induced wetting and dewetting in single
  hydrophobic nanopores},\ }\href@noop {} {\bibfield  {journal} {\bibinfo
  {journal} {Nature nanotechnology}\ }\textbf {\bibinfo {volume} {6}},\
  \bibinfo {pages} {798} (\bibinfo {year} {2011})}\BibitemShut {NoStop}%
\bibitem [{\citenamefont {Wang}\ \emph {et~al.}(2012)\citenamefont {Wang},
  \citenamefont {Kvetny}, \citenamefont {Liu}, \citenamefont {Brown},
  \citenamefont {Li},\ and\ \citenamefont {Wang}}]{wang2012transmembrane}%
  \BibitemOpen
  \bibfield  {author} {\bibinfo {author} {\bibfnamefont {D.}~\bibnamefont
  {Wang}}, \bibinfo {author} {\bibfnamefont {M.}~\bibnamefont {Kvetny}},
  \bibinfo {author} {\bibfnamefont {J.}~\bibnamefont {Liu}}, \bibinfo {author}
  {\bibfnamefont {W.}~\bibnamefont {Brown}}, \bibinfo {author} {\bibfnamefont
  {Y.}~\bibnamefont {Li}},\ and\ \bibinfo {author} {\bibfnamefont
  {G.}~\bibnamefont {Wang}},\ }\bibfield  {title} {\bibinfo {title}
  {Transmembrane potential across single conical nanopores and resulting
  memristive and memcapacitive ion transport},\ }\href@noop {} {\bibfield
  {journal} {\bibinfo  {journal} {Journal of the American Chemical Society}\
  }\textbf {\bibinfo {volume} {134}},\ \bibinfo {pages} {3651} (\bibinfo {year}
  {2012})}\BibitemShut {NoStop}%
\bibitem [{\citenamefont {Paulo}\ \emph {et~al.}(2023)\citenamefont {Paulo},
  \citenamefont {Sun}, \citenamefont {Di~Muccio}, \citenamefont {Gubbiotti},
  \citenamefont {Morozzo~della Rocca}, \citenamefont {Geng}, \citenamefont
  {Maglia}, \citenamefont {Chinappi},\ and\ \citenamefont
  {Giacomello}}]{paulo2023hydrophobically}%
  \BibitemOpen
  \bibfield  {author} {\bibinfo {author} {\bibfnamefont {G.}~\bibnamefont
  {Paulo}}, \bibinfo {author} {\bibfnamefont {K.}~\bibnamefont {Sun}}, \bibinfo
  {author} {\bibfnamefont {G.}~\bibnamefont {Di~Muccio}}, \bibinfo {author}
  {\bibfnamefont {A.}~\bibnamefont {Gubbiotti}}, \bibinfo {author}
  {\bibfnamefont {B.}~\bibnamefont {Morozzo~della Rocca}}, \bibinfo {author}
  {\bibfnamefont {J.}~\bibnamefont {Geng}}, \bibinfo {author} {\bibfnamefont
  {G.}~\bibnamefont {Maglia}}, \bibinfo {author} {\bibfnamefont
  {M.}~\bibnamefont {Chinappi}},\ and\ \bibinfo {author} {\bibfnamefont
  {A.}~\bibnamefont {Giacomello}},\ }\bibfield  {title} {\bibinfo {title}
  {Hydrophobically gated memristive nanopores for neuromorphic applications},\
  }\href@noop {} {\bibfield  {journal} {\bibinfo  {journal} {Nature
  Communications}\ }\textbf {\bibinfo {volume} {14}},\ \bibinfo {pages} {8390}
  (\bibinfo {year} {2023})}\BibitemShut {NoStop}%
\bibitem [{\citenamefont {Ramirez}\ \emph {et~al.}(2024)\citenamefont
  {Ramirez}, \citenamefont {Portillo}, \citenamefont {Cervera}, \citenamefont
  {Nasir}, \citenamefont {Ali}, \citenamefont {Ensinger},\ and\ \citenamefont
  {Mafe}}]{ramirez2024neuromorphic}%
  \BibitemOpen
  \bibfield  {author} {\bibinfo {author} {\bibfnamefont {P.}~\bibnamefont
  {Ramirez}}, \bibinfo {author} {\bibfnamefont {S.}~\bibnamefont {Portillo}},
  \bibinfo {author} {\bibfnamefont {J.}~\bibnamefont {Cervera}}, \bibinfo
  {author} {\bibfnamefont {S.}~\bibnamefont {Nasir}}, \bibinfo {author}
  {\bibfnamefont {M.}~\bibnamefont {Ali}}, \bibinfo {author} {\bibfnamefont
  {W.}~\bibnamefont {Ensinger}},\ and\ \bibinfo {author} {\bibfnamefont
  {S.}~\bibnamefont {Mafe}},\ }\bibfield  {title} {\bibinfo {title}
  {Neuromorphic responses of nanofluidic memristors in symmetric and asymmetric
  ionic solutions},\ }\href@noop {} {\bibfield  {journal} {\bibinfo  {journal}
  {The Journal of Chemical Physics}\ }\textbf {\bibinfo {volume} {160}}
  (\bibinfo {year} {2024})}\BibitemShut {NoStop}%
\bibitem [{\citenamefont {Han}\ \emph {et~al.}(2023)\citenamefont {Han},
  \citenamefont {Kim}, \citenamefont {Oh},\ and\ \citenamefont
  {Chung}}]{han2023iontronic}%
  \BibitemOpen
  \bibfield  {author} {\bibinfo {author} {\bibfnamefont {S.~H.}\ \bibnamefont
  {Han}}, \bibinfo {author} {\bibfnamefont {S.~I.}\ \bibnamefont {Kim}},
  \bibinfo {author} {\bibfnamefont {M.-A.}\ \bibnamefont {Oh}},\ and\ \bibinfo
  {author} {\bibfnamefont {T.~D.}\ \bibnamefont {Chung}},\ }\bibfield  {title}
  {\bibinfo {title} {Iontronic analog of synaptic plasticity: Hydrogel-based
  ionic diode with chemical precipitation and dissolution},\ }\href@noop {}
  {\bibfield  {journal} {\bibinfo  {journal} {Proceedings of the National
  Academy of Sciences}\ }\textbf {\bibinfo {volume} {120}},\ \bibinfo {pages}
  {e2211442120} (\bibinfo {year} {2023})}\BibitemShut {NoStop}%
\bibitem [{\citenamefont {Ramirez}\ \emph {et~al.}(2023)\citenamefont
  {Ramirez}, \citenamefont {G{\'o}mez}, \citenamefont {Cervera}, \citenamefont
  {Mafe},\ and\ \citenamefont {Bisquert}}]{ramirez2023synaptical}%
  \BibitemOpen
  \bibfield  {author} {\bibinfo {author} {\bibfnamefont {P.}~\bibnamefont
  {Ramirez}}, \bibinfo {author} {\bibfnamefont {V.}~\bibnamefont {G{\'o}mez}},
  \bibinfo {author} {\bibfnamefont {J.}~\bibnamefont {Cervera}}, \bibinfo
  {author} {\bibfnamefont {S.}~\bibnamefont {Mafe}},\ and\ \bibinfo {author}
  {\bibfnamefont {J.}~\bibnamefont {Bisquert}},\ }\bibfield  {title} {\bibinfo
  {title} {Synaptical tunability of multipore nanofluidic memristors},\
  }\href@noop {} {\bibfield  {journal} {\bibinfo  {journal} {The Journal of
  Physical Chemistry Letters}\ }\textbf {\bibinfo {volume} {14}},\ \bibinfo
  {pages} {10930} (\bibinfo {year} {2023})}\BibitemShut {NoStop}%
\bibitem [{\citenamefont {Xiong}\ \emph {et~al.}(2023)\citenamefont {Xiong},
  \citenamefont {Li}, \citenamefont {He}, \citenamefont {Xie}, \citenamefont
  {Zong}, \citenamefont {Jiang}, \citenamefont {Ma}, \citenamefont {Wu},
  \citenamefont {Fei}, \citenamefont {Yu} \emph
  {et~al.}}]{xiong2023neuromorphic}%
  \BibitemOpen
  \bibfield  {author} {\bibinfo {author} {\bibfnamefont {T.}~\bibnamefont
  {Xiong}}, \bibinfo {author} {\bibfnamefont {C.}~\bibnamefont {Li}}, \bibinfo
  {author} {\bibfnamefont {X.}~\bibnamefont {He}}, \bibinfo {author}
  {\bibfnamefont {B.}~\bibnamefont {Xie}}, \bibinfo {author} {\bibfnamefont
  {J.}~\bibnamefont {Zong}}, \bibinfo {author} {\bibfnamefont {Y.}~\bibnamefont
  {Jiang}}, \bibinfo {author} {\bibfnamefont {W.}~\bibnamefont {Ma}}, \bibinfo
  {author} {\bibfnamefont {F.}~\bibnamefont {Wu}}, \bibinfo {author}
  {\bibfnamefont {J.}~\bibnamefont {Fei}}, \bibinfo {author} {\bibfnamefont
  {P.}~\bibnamefont {Yu}}, \emph {et~al.},\ }\bibfield  {title} {\bibinfo
  {title} {Neuromorphic functions with a polyelectrolyte-confined fluidic
  memristor},\ }\href@noop {} {\bibfield  {journal} {\bibinfo  {journal}
  {Science}\ }\textbf {\bibinfo {volume} {379}},\ \bibinfo {pages} {156}
  (\bibinfo {year} {2023})}\BibitemShut {NoStop}%
\bibitem [{\citenamefont {Robin}\ \emph {et~al.}(2023)\citenamefont {Robin},
  \citenamefont {Emmerich}, \citenamefont {Ismail}, \citenamefont {Nigu{\`e}s},
  \citenamefont {You}, \citenamefont {Nam}, \citenamefont {Keerthi},
  \citenamefont {Siria}, \citenamefont {Geim}, \citenamefont {Radha} \emph
  {et~al.}}]{robin2023long}%
  \BibitemOpen
  \bibfield  {author} {\bibinfo {author} {\bibfnamefont {P.}~\bibnamefont
  {Robin}}, \bibinfo {author} {\bibfnamefont {T.}~\bibnamefont {Emmerich}},
  \bibinfo {author} {\bibfnamefont {A.}~\bibnamefont {Ismail}}, \bibinfo
  {author} {\bibfnamefont {A.}~\bibnamefont {Nigu{\`e}s}}, \bibinfo {author}
  {\bibfnamefont {Y.}~\bibnamefont {You}}, \bibinfo {author} {\bibfnamefont
  {G.-H.}\ \bibnamefont {Nam}}, \bibinfo {author} {\bibfnamefont
  {A.}~\bibnamefont {Keerthi}}, \bibinfo {author} {\bibfnamefont
  {A.}~\bibnamefont {Siria}}, \bibinfo {author} {\bibfnamefont
  {A.}~\bibnamefont {Geim}}, \bibinfo {author} {\bibfnamefont {B.}~\bibnamefont
  {Radha}}, \emph {et~al.},\ }\bibfield  {title} {\bibinfo {title} {Long-term
  memory and synapse-like dynamics in two-dimensional nanofluidic channels},\
  }\href@noop {} {\bibfield  {journal} {\bibinfo  {journal} {Science}\ }\textbf
  {\bibinfo {volume} {379}},\ \bibinfo {pages} {161} (\bibinfo {year}
  {2023})}\BibitemShut {NoStop}%
\bibitem [{\citenamefont {Robin}\ \emph {et~al.}(2021)\citenamefont {Robin},
  \citenamefont {Kavokine},\ and\ \citenamefont
  {Bocquet}}]{robin2021principles}%
  \BibitemOpen
  \bibfield  {author} {\bibinfo {author} {\bibfnamefont {P.}~\bibnamefont
  {Robin}}, \bibinfo {author} {\bibfnamefont {N.}~\bibnamefont {Kavokine}},\
  and\ \bibinfo {author} {\bibfnamefont {L.}~\bibnamefont {Bocquet}},\
  }\bibfield  {title} {\bibinfo {title} {Modeling of emergent memory and
  voltage spiking in ionic transport through angstrom-scale slits},\
  }\href@noop {} {\bibfield  {journal} {\bibinfo  {journal} {Science}\ }\textbf
  {\bibinfo {volume} {373}},\ \bibinfo {pages} {687} (\bibinfo {year}
  {2021})}\BibitemShut {NoStop}%
\bibitem [{\citenamefont {Kamsma}\ \emph
  {et~al.}(2023{\natexlab{a}})\citenamefont {Kamsma}, \citenamefont {Boon},
  \citenamefont {ter Rele}, \citenamefont {Spitoni},\ and\ \citenamefont {van
  Roij}}]{kamsma2023iontronic}%
  \BibitemOpen
  \bibfield  {author} {\bibinfo {author} {\bibfnamefont {T.~M.}\ \bibnamefont
  {Kamsma}}, \bibinfo {author} {\bibfnamefont {W.~Q.}\ \bibnamefont {Boon}},
  \bibinfo {author} {\bibfnamefont {T.}~\bibnamefont {ter Rele}}, \bibinfo
  {author} {\bibfnamefont {C.}~\bibnamefont {Spitoni}},\ and\ \bibinfo {author}
  {\bibfnamefont {R.}~\bibnamefont {van Roij}},\ }\bibfield  {title} {\bibinfo
  {title} {Iontronic neuromorphic signaling with conical microfluidic
  memristors},\ }\href {https://doi.org/10.1103/PhysRevLett.130.268401}
  {\bibfield  {journal} {\bibinfo  {journal} {Phys. Rev. Lett.}\ }\textbf
  {\bibinfo {volume} {130}},\ \bibinfo {pages} {268401} (\bibinfo {year}
  {2023}{\natexlab{a}})}\BibitemShut {NoStop}%
\bibitem [{\citenamefont {Kamsma}\ \emph
  {et~al.}(2024{\natexlab{a}})\citenamefont {Kamsma}, \citenamefont {Rossing},
  \citenamefont {Spitoni},\ and\ \citenamefont {van
  Roij}}]{kamsma2024advanced}%
  \BibitemOpen
  \bibfield  {author} {\bibinfo {author} {\bibfnamefont {T.~M.}\ \bibnamefont
  {Kamsma}}, \bibinfo {author} {\bibfnamefont {E.~A.}\ \bibnamefont {Rossing}},
  \bibinfo {author} {\bibfnamefont {C.}~\bibnamefont {Spitoni}},\ and\ \bibinfo
  {author} {\bibfnamefont {R.}~\bibnamefont {van Roij}},\ }\bibfield  {title}
  {\bibinfo {title} {Advanced iontronic spiking modes with multiscale diffusive
  dynamics in a fluidic circuit},\ }\href@noop {} {\bibfield  {journal}
  {\bibinfo  {journal} {arXiv preprint arXiv:2401.14921}\ } (\bibinfo {year}
  {2024}{\natexlab{a}})}\BibitemShut {NoStop}%
\bibitem [{\citenamefont {Emmerich}\ \emph {et~al.}(2024)\citenamefont
  {Emmerich}, \citenamefont {Teng}, \citenamefont {Ronceray}, \citenamefont
  {Lopriore}, \citenamefont {Chiesa}, \citenamefont {Chernev}, \citenamefont
  {Artemov}, \citenamefont {Di~Ventra}, \citenamefont {Kis},\ and\
  \citenamefont {Radenovic}}]{emmerich2024nanofluidic}%
  \BibitemOpen
  \bibfield  {author} {\bibinfo {author} {\bibfnamefont {T.}~\bibnamefont
  {Emmerich}}, \bibinfo {author} {\bibfnamefont {Y.}~\bibnamefont {Teng}},
  \bibinfo {author} {\bibfnamefont {N.}~\bibnamefont {Ronceray}}, \bibinfo
  {author} {\bibfnamefont {E.}~\bibnamefont {Lopriore}}, \bibinfo {author}
  {\bibfnamefont {R.}~\bibnamefont {Chiesa}}, \bibinfo {author} {\bibfnamefont
  {A.}~\bibnamefont {Chernev}}, \bibinfo {author} {\bibfnamefont
  {V.}~\bibnamefont {Artemov}}, \bibinfo {author} {\bibfnamefont
  {M.}~\bibnamefont {Di~Ventra}}, \bibinfo {author} {\bibfnamefont
  {A.}~\bibnamefont {Kis}},\ and\ \bibinfo {author} {\bibfnamefont
  {A.}~\bibnamefont {Radenovic}},\ }\bibfield  {title} {\bibinfo {title}
  {Nanofluidic logic with mechano--ionic memristive switches},\ }\href@noop {}
  {\bibfield  {journal} {\bibinfo  {journal} {Nature Electronics}\ ,\ \bibinfo
  {pages} {1}} (\bibinfo {year} {2024})}\BibitemShut {NoStop}%
\bibitem [{\citenamefont {Sabbagh}\ \emph {et~al.}(2023)\citenamefont
  {Sabbagh}, \citenamefont {Fraiman}, \citenamefont {Fish},\ and\ \citenamefont
  {Yossifon}}]{sabbagh2023designing}%
  \BibitemOpen
  \bibfield  {author} {\bibinfo {author} {\bibfnamefont {B.}~\bibnamefont
  {Sabbagh}}, \bibinfo {author} {\bibfnamefont {N.~E.}\ \bibnamefont
  {Fraiman}}, \bibinfo {author} {\bibfnamefont {A.}~\bibnamefont {Fish}},\ and\
  \bibinfo {author} {\bibfnamefont {G.}~\bibnamefont {Yossifon}},\ }\bibfield
  {title} {\bibinfo {title} {Designing with iontronic logic gates-from a single
  polyelectrolyte diode to an integrated ionic circuit},\ }\href
  {https://doi.org/10.1021/acsami.3c00062} {\bibfield  {journal} {\bibinfo
  {journal} {ACS Applied Materials \& Interfaces}\ }\textbf {\bibinfo {volume}
  {15}},\ \bibinfo {pages} {23361} (\bibinfo {year} {2023})},\ \bibinfo {note}
  {pMID: 37068481},\ \Eprint
  {https://arxiv.org/abs/https://doi.org/10.1021/acsami.3c00062}
  {https://doi.org/10.1021/acsami.3c00062} \BibitemShut {NoStop}%
\bibitem [{\citenamefont {Li}\ \emph {et~al.}(2023)\citenamefont {Li},
  \citenamefont {Li}, \citenamefont {Zhang}, \citenamefont {Hu},\ and\
  \citenamefont {Li}}]{li2023high}%
  \BibitemOpen
  \bibfield  {author} {\bibinfo {author} {\bibfnamefont {J.}~\bibnamefont
  {Li}}, \bibinfo {author} {\bibfnamefont {M.}~\bibnamefont {Li}}, \bibinfo
  {author} {\bibfnamefont {K.}~\bibnamefont {Zhang}}, \bibinfo {author}
  {\bibfnamefont {L.}~\bibnamefont {Hu}},\ and\ \bibinfo {author}
  {\bibfnamefont {D.}~\bibnamefont {Li}},\ }\bibfield  {title} {\bibinfo
  {title} {High-performance integrated iontronic circuits based on single
  nano/microchannels},\ }\href@noop {} {\bibfield  {journal} {\bibinfo
  {journal} {Small}\ ,\ \bibinfo {pages} {2208079}} (\bibinfo {year}
  {2023})}\BibitemShut {NoStop}%
\bibitem [{\citenamefont {Xie}\ \emph {et~al.}(2022)\citenamefont {Xie},
  \citenamefont {Xiong}, \citenamefont {Li}, \citenamefont {Gao}, \citenamefont
  {Zong}, \citenamefont {Liu},\ and\ \citenamefont {Yu}}]{xie2022perspective}%
  \BibitemOpen
  \bibfield  {author} {\bibinfo {author} {\bibfnamefont {B.}~\bibnamefont
  {Xie}}, \bibinfo {author} {\bibfnamefont {T.}~\bibnamefont {Xiong}}, \bibinfo
  {author} {\bibfnamefont {W.}~\bibnamefont {Li}}, \bibinfo {author}
  {\bibfnamefont {T.}~\bibnamefont {Gao}}, \bibinfo {author} {\bibfnamefont
  {J.}~\bibnamefont {Zong}}, \bibinfo {author} {\bibfnamefont {Y.}~\bibnamefont
  {Liu}},\ and\ \bibinfo {author} {\bibfnamefont {P.}~\bibnamefont {Yu}},\
  }\bibfield  {title} {\bibinfo {title} {Perspective on nanofluidic memristors:
  From mechanism to application},\ }\href@noop {} {\bibfield  {journal}
  {\bibinfo  {journal} {Chemistry--An Asian Journal}\ }\textbf {\bibinfo
  {volume} {17}},\ \bibinfo {pages} {e202200682} (\bibinfo {year}
  {2022})}\BibitemShut {NoStop}%
\bibitem [{\citenamefont {Noy}\ \emph {et~al.}(2023)\citenamefont {Noy},
  \citenamefont {Li},\ and\ \citenamefont {Darling}}]{noy2023fluid}%
  \BibitemOpen
  \bibfield  {author} {\bibinfo {author} {\bibfnamefont {A.}~\bibnamefont
  {Noy}}, \bibinfo {author} {\bibfnamefont {Z.}~\bibnamefont {Li}},\ and\
  \bibinfo {author} {\bibfnamefont {S.~B.}\ \bibnamefont {Darling}},\
  }\bibfield  {title} {\bibinfo {title} {Fluid learning: Mimicking brain
  computing with neuromorphic nanofluidic devices},\ }\href@noop {} {\bibfield
  {journal} {\bibinfo  {journal} {Nano Today}\ }\textbf {\bibinfo {volume}
  {53}},\ \bibinfo {pages} {102043} (\bibinfo {year} {2023})}\BibitemShut
  {NoStop}%
\bibitem [{\citenamefont {Chicca}\ and\ \citenamefont
  {Indiveri}(2020)}]{chicca2020recipe}%
  \BibitemOpen
  \bibfield  {author} {\bibinfo {author} {\bibfnamefont {E.}~\bibnamefont
  {Chicca}}\ and\ \bibinfo {author} {\bibfnamefont {G.}~\bibnamefont
  {Indiveri}},\ }\bibfield  {title} {\bibinfo {title} {A recipe for creating
  ideal hybrid memristive-cmos neuromorphic processing systems},\ }\href@noop
  {} {\bibfield  {journal} {\bibinfo  {journal} {Applied Physics Letters}\
  }\textbf {\bibinfo {volume} {116}},\ \bibinfo {pages} {120501} (\bibinfo
  {year} {2020})}\BibitemShut {NoStop}%
\bibitem [{\citenamefont {Tanaka}\ \emph {et~al.}(2019)\citenamefont {Tanaka},
  \citenamefont {Yamane}, \citenamefont {H{\'e}roux}, \citenamefont {Nakane},
  \citenamefont {Kanazawa}, \citenamefont {Takeda}, \citenamefont {Numata},
  \citenamefont {Nakano},\ and\ \citenamefont {Hirose}}]{tanaka2019recent}%
  \BibitemOpen
  \bibfield  {author} {\bibinfo {author} {\bibfnamefont {G.}~\bibnamefont
  {Tanaka}}, \bibinfo {author} {\bibfnamefont {T.}~\bibnamefont {Yamane}},
  \bibinfo {author} {\bibfnamefont {J.~B.}\ \bibnamefont {H{\'e}roux}},
  \bibinfo {author} {\bibfnamefont {R.}~\bibnamefont {Nakane}}, \bibinfo
  {author} {\bibfnamefont {N.}~\bibnamefont {Kanazawa}}, \bibinfo {author}
  {\bibfnamefont {S.}~\bibnamefont {Takeda}}, \bibinfo {author} {\bibfnamefont
  {H.}~\bibnamefont {Numata}}, \bibinfo {author} {\bibfnamefont
  {D.}~\bibnamefont {Nakano}},\ and\ \bibinfo {author} {\bibfnamefont
  {A.}~\bibnamefont {Hirose}},\ }\bibfield  {title} {\bibinfo {title} {Recent
  advances in physical reservoir computing: A review},\ }\href@noop {}
  {\bibfield  {journal} {\bibinfo  {journal} {Neural Networks}\ }\textbf
  {\bibinfo {volume} {115}},\ \bibinfo {pages} {100} (\bibinfo {year}
  {2019})}\BibitemShut {NoStop}%
\bibitem [{\citenamefont {Cucchi}\ \emph {et~al.}(2021)\citenamefont {Cucchi},
  \citenamefont {Gruener}, \citenamefont {Petrauskas}, \citenamefont {Steiner},
  \citenamefont {Tseng}, \citenamefont {Fischer}, \citenamefont {Penkovsky},
  \citenamefont {Matthus}, \citenamefont {Birkholz}, \citenamefont {Kleemann}
  \emph {et~al.}}]{cucchi2021reservoir}%
  \BibitemOpen
  \bibfield  {author} {\bibinfo {author} {\bibfnamefont {M.}~\bibnamefont
  {Cucchi}}, \bibinfo {author} {\bibfnamefont {C.}~\bibnamefont {Gruener}},
  \bibinfo {author} {\bibfnamefont {L.}~\bibnamefont {Petrauskas}}, \bibinfo
  {author} {\bibfnamefont {P.}~\bibnamefont {Steiner}}, \bibinfo {author}
  {\bibfnamefont {H.}~\bibnamefont {Tseng}}, \bibinfo {author} {\bibfnamefont
  {A.}~\bibnamefont {Fischer}}, \bibinfo {author} {\bibfnamefont
  {B.}~\bibnamefont {Penkovsky}}, \bibinfo {author} {\bibfnamefont
  {C.}~\bibnamefont {Matthus}}, \bibinfo {author} {\bibfnamefont
  {P.}~\bibnamefont {Birkholz}}, \bibinfo {author} {\bibfnamefont
  {H.}~\bibnamefont {Kleemann}}, \emph {et~al.},\ }\bibfield  {title} {\bibinfo
  {title} {Reservoir computing with biocompatible organic electrochemical
  networks for brain-inspired biosignal classification},\ }\href@noop {}
  {\bibfield  {journal} {\bibinfo  {journal} {Science Advances}\ }\textbf
  {\bibinfo {volume} {7}},\ \bibinfo {pages} {eabh0693} (\bibinfo {year}
  {2021})}\BibitemShut {NoStop}%
\bibitem [{\citenamefont {Cao}\ \emph {et~al.}(2022)\citenamefont {Cao},
  \citenamefont {Zhang}, \citenamefont {Cheng}, \citenamefont {Qiu},
  \citenamefont {Liu}, \citenamefont {Wang},\ and\ \citenamefont
  {Liu}}]{cao2022emerging}%
  \BibitemOpen
  \bibfield  {author} {\bibinfo {author} {\bibfnamefont {J.}~\bibnamefont
  {Cao}}, \bibinfo {author} {\bibfnamefont {X.}~\bibnamefont {Zhang}}, \bibinfo
  {author} {\bibfnamefont {H.}~\bibnamefont {Cheng}}, \bibinfo {author}
  {\bibfnamefont {J.}~\bibnamefont {Qiu}}, \bibinfo {author} {\bibfnamefont
  {X.}~\bibnamefont {Liu}}, \bibinfo {author} {\bibfnamefont {M.}~\bibnamefont
  {Wang}},\ and\ \bibinfo {author} {\bibfnamefont {Q.}~\bibnamefont {Liu}},\
  }\bibfield  {title} {\bibinfo {title} {Emerging dynamic memristors for
  neuromorphic reservoir computing},\ }\href@noop {} {\bibfield  {journal}
  {\bibinfo  {journal} {Nanoscale}\ }\textbf {\bibinfo {volume} {14}},\
  \bibinfo {pages} {289} (\bibinfo {year} {2022})}\BibitemShut {NoStop}%
\bibitem [{\citenamefont {Cucchi}\ \emph {et~al.}(2022)\citenamefont {Cucchi},
  \citenamefont {Abreu}, \citenamefont {Ciccone}, \citenamefont {Brunner},\
  and\ \citenamefont {Kleemann}}]{cucchi2022hands}%
  \BibitemOpen
  \bibfield  {author} {\bibinfo {author} {\bibfnamefont {M.}~\bibnamefont
  {Cucchi}}, \bibinfo {author} {\bibfnamefont {S.}~\bibnamefont {Abreu}},
  \bibinfo {author} {\bibfnamefont {G.}~\bibnamefont {Ciccone}}, \bibinfo
  {author} {\bibfnamefont {D.}~\bibnamefont {Brunner}},\ and\ \bibinfo {author}
  {\bibfnamefont {H.}~\bibnamefont {Kleemann}},\ }\bibfield  {title} {\bibinfo
  {title} {Hands-on reservoir computing: a tutorial for practical
  implementation},\ }\href@noop {} {\bibfield  {journal} {\bibinfo  {journal}
  {Neuromorphic Computing and Engineering}\ }\textbf {\bibinfo {volume} {2}},\
  \bibinfo {pages} {032002} (\bibinfo {year} {2022})}\BibitemShut {NoStop}%
\bibitem [{\citenamefont {Midya}\ \emph {et~al.}(2019)\citenamefont {Midya},
  \citenamefont {Wang}, \citenamefont {Asapu}, \citenamefont {Zhang},
  \citenamefont {Rao}, \citenamefont {Song}, \citenamefont {Zhuo},
  \citenamefont {Upadhyay}, \citenamefont {Xia},\ and\ \citenamefont
  {Yang}}]{midya2019reservoir}%
  \BibitemOpen
  \bibfield  {author} {\bibinfo {author} {\bibfnamefont {R.}~\bibnamefont
  {Midya}}, \bibinfo {author} {\bibfnamefont {Z.}~\bibnamefont {Wang}},
  \bibinfo {author} {\bibfnamefont {S.}~\bibnamefont {Asapu}}, \bibinfo
  {author} {\bibfnamefont {X.}~\bibnamefont {Zhang}}, \bibinfo {author}
  {\bibfnamefont {M.}~\bibnamefont {Rao}}, \bibinfo {author} {\bibfnamefont
  {W.}~\bibnamefont {Song}}, \bibinfo {author} {\bibfnamefont {Y.}~\bibnamefont
  {Zhuo}}, \bibinfo {author} {\bibfnamefont {N.}~\bibnamefont {Upadhyay}},
  \bibinfo {author} {\bibfnamefont {Q.}~\bibnamefont {Xia}},\ and\ \bibinfo
  {author} {\bibfnamefont {J.~J.}\ \bibnamefont {Yang}},\ }\bibfield  {title}
  {\bibinfo {title} {Reservoir computing using diffusive memristors},\
  }\href@noop {} {\bibfield  {journal} {\bibinfo  {journal} {Advanced
  Intelligent Systems}\ }\textbf {\bibinfo {volume} {1}},\ \bibinfo {pages}
  {1900084} (\bibinfo {year} {2019})}\BibitemShut {NoStop}%
\bibitem [{\citenamefont {Du}\ \emph {et~al.}(2017)\citenamefont {Du},
  \citenamefont {Cai}, \citenamefont {Zidan}, \citenamefont {Ma}, \citenamefont
  {Lee},\ and\ \citenamefont {Lu}}]{du2017reservoir}%
  \BibitemOpen
  \bibfield  {author} {\bibinfo {author} {\bibfnamefont {C.}~\bibnamefont
  {Du}}, \bibinfo {author} {\bibfnamefont {F.}~\bibnamefont {Cai}}, \bibinfo
  {author} {\bibfnamefont {M.~A.}\ \bibnamefont {Zidan}}, \bibinfo {author}
  {\bibfnamefont {W.}~\bibnamefont {Ma}}, \bibinfo {author} {\bibfnamefont
  {S.~H.}\ \bibnamefont {Lee}},\ and\ \bibinfo {author} {\bibfnamefont {W.~D.}\
  \bibnamefont {Lu}},\ }\bibfield  {title} {\bibinfo {title} {Reservoir
  computing using dynamic memristors for temporal information processing},\
  }\href@noop {} {\bibfield  {journal} {\bibinfo  {journal} {Nature
  communications}\ }\textbf {\bibinfo {volume} {8}},\ \bibinfo {pages} {1}
  (\bibinfo {year} {2017})}\BibitemShut {NoStop}%
\bibitem [{\citenamefont {Pyo}\ and\ \citenamefont {Kim}(2022)}]{pyo2022non}%
  \BibitemOpen
  \bibfield  {author} {\bibinfo {author} {\bibfnamefont {J.}~\bibnamefont
  {Pyo}}\ and\ \bibinfo {author} {\bibfnamefont {S.}~\bibnamefont {Kim}},\
  }\bibfield  {title} {\bibinfo {title} {Non-volatile and volatile switching
  behaviors determined by first reset in ag/taox/tin device for neuromorphic
  system},\ }\href@noop {} {\bibfield  {journal} {\bibinfo  {journal} {Journal
  of Alloys and Compounds}\ }\textbf {\bibinfo {volume} {896}},\ \bibinfo
  {pages} {163075} (\bibinfo {year} {2022})}\BibitemShut {NoStop}%
\bibitem [{\citenamefont {Kim}\ \emph {et~al.}(2022{\natexlab{a}})\citenamefont
  {Kim}, \citenamefont {Shin},\ and\ \citenamefont
  {Kim}}]{kim2022implementation}%
  \BibitemOpen
  \bibfield  {author} {\bibinfo {author} {\bibfnamefont {D.}~\bibnamefont
  {Kim}}, \bibinfo {author} {\bibfnamefont {J.}~\bibnamefont {Shin}},\ and\
  \bibinfo {author} {\bibfnamefont {S.}~\bibnamefont {Kim}},\ }\bibfield
  {title} {\bibinfo {title} {Implementation of reservoir computing using
  volatile wox-based memristor},\ }\href@noop {} {\bibfield  {journal}
  {\bibinfo  {journal} {Applied Surface Science}\ ,\ \bibinfo {pages} {153876}}
  (\bibinfo {year} {2022}{\natexlab{a}})}\BibitemShut {NoStop}%
\bibitem [{\citenamefont {Choi}\ \emph {et~al.}(2016)\citenamefont {Choi},
  \citenamefont {Wang}, \citenamefont {Chang},\ and\ \citenamefont
  {Park}}]{choi2016high}%
  \BibitemOpen
  \bibfield  {author} {\bibinfo {author} {\bibfnamefont {E.}~\bibnamefont
  {Choi}}, \bibinfo {author} {\bibfnamefont {C.}~\bibnamefont {Wang}}, \bibinfo
  {author} {\bibfnamefont {G.~T.}\ \bibnamefont {Chang}},\ and\ \bibinfo
  {author} {\bibfnamefont {J.}~\bibnamefont {Park}},\ }\bibfield  {title}
  {\bibinfo {title} {High current ionic diode using homogeneously charged
  asymmetric nanochannel network membrane},\ }\href@noop {} {\bibfield
  {journal} {\bibinfo  {journal} {Nano Letters}\ }\textbf {\bibinfo {volume}
  {16}},\ \bibinfo {pages} {2189} (\bibinfo {year} {2016})}\BibitemShut
  {NoStop}%
\bibitem [{\citenamefont {Kim}\ \emph {et~al.}(2022{\natexlab{b}})\citenamefont
  {Kim}, \citenamefont {Jeon}, \citenamefont {Wang}, \citenamefont {Chang},\
  and\ \citenamefont {Park}}]{kim2022asymmetric}%
  \BibitemOpen
  \bibfield  {author} {\bibinfo {author} {\bibfnamefont {J.}~\bibnamefont
  {Kim}}, \bibinfo {author} {\bibfnamefont {J.}~\bibnamefont {Jeon}}, \bibinfo
  {author} {\bibfnamefont {C.}~\bibnamefont {Wang}}, \bibinfo {author}
  {\bibfnamefont {G.~T.}\ \bibnamefont {Chang}},\ and\ \bibinfo {author}
  {\bibfnamefont {J.}~\bibnamefont {Park}},\ }\bibfield  {title} {\bibinfo
  {title} {Asymmetric nanochannel network-based bipolar ionic diode for
  enhanced heavy metal ion detection},\ }\href@noop {} {\bibfield  {journal}
  {\bibinfo  {journal} {ACS nano}\ }\textbf {\bibinfo {volume} {16}},\ \bibinfo
  {pages} {8253} (\bibinfo {year} {2022}{\natexlab{b}})}\BibitemShut {NoStop}%
\bibitem [{\citenamefont {Mani}\ \emph {et~al.}(2009)\citenamefont {Mani},
  \citenamefont {Zangle},\ and\ \citenamefont
  {Santiago}}]{mani2009propagation}%
  \BibitemOpen
  \bibfield  {author} {\bibinfo {author} {\bibfnamefont {A.}~\bibnamefont
  {Mani}}, \bibinfo {author} {\bibfnamefont {T.~A.}\ \bibnamefont {Zangle}},\
  and\ \bibinfo {author} {\bibfnamefont {J.~G.}\ \bibnamefont {Santiago}},\
  }\bibfield  {title} {\bibinfo {title} {On the propagation of concentration
  polarization from microchannel- nanochannel interfaces part i: analytical
  model and characteristic analysis},\ }\href@noop {} {\bibfield  {journal}
  {\bibinfo  {journal} {Langmuir}\ }\textbf {\bibinfo {volume} {25}},\ \bibinfo
  {pages} {3898} (\bibinfo {year} {2009})}\BibitemShut {NoStop}%
\bibitem [{\citenamefont {Mani}\ and\ \citenamefont
  {Bazant}(2011)}]{mani2011deionization}%
  \BibitemOpen
  \bibfield  {author} {\bibinfo {author} {\bibfnamefont {A.}~\bibnamefont
  {Mani}}\ and\ \bibinfo {author} {\bibfnamefont {M.~Z.}\ \bibnamefont
  {Bazant}},\ }\bibfield  {title} {\bibinfo {title} {Deionization shocks in
  microstructures},\ }\href@noop {} {\bibfield  {journal} {\bibinfo  {journal}
  {Physical Review E}\ }\textbf {\bibinfo {volume} {84}},\ \bibinfo {pages}
  {061504} (\bibinfo {year} {2011})}\BibitemShut {NoStop}%
\bibitem [{\citenamefont {Boon}\ \emph {et~al.}(2022)\citenamefont {Boon},
  \citenamefont {Veenstra}, \citenamefont {Dijkstra},\ and\ \citenamefont {van
  Roij}}]{boon2021nonlinear}%
  \BibitemOpen
  \bibfield  {author} {\bibinfo {author} {\bibfnamefont {W.~Q.}\ \bibnamefont
  {Boon}}, \bibinfo {author} {\bibfnamefont {T.~E.}\ \bibnamefont {Veenstra}},
  \bibinfo {author} {\bibfnamefont {M.}~\bibnamefont {Dijkstra}},\ and\
  \bibinfo {author} {\bibfnamefont {R.}~\bibnamefont {van Roij}},\ }\bibfield
  {title} {\bibinfo {title} {Pressure-sensitive ion conduction in a conical
  channel: optimal pressure and geometry},\ }\href@noop {} {\bibfield
  {journal} {\bibinfo  {journal} {Physics of Fluids}\ }\textbf {\bibinfo
  {volume} {34}},\ \bibinfo {pages} {101701} (\bibinfo {year}
  {2022})}\BibitemShut {NoStop}%
\bibitem [{\citenamefont {Kamsma}\ \emph
  {et~al.}(2023{\natexlab{b}})\citenamefont {Kamsma}, \citenamefont {Boon},
  \citenamefont {Spitoni},\ and\ \citenamefont {van
  Roij}}]{kamsma2023unveiling}%
  \BibitemOpen
  \bibfield  {author} {\bibinfo {author} {\bibfnamefont {T.~M.}\ \bibnamefont
  {Kamsma}}, \bibinfo {author} {\bibfnamefont {W.~Q.}\ \bibnamefont {Boon}},
  \bibinfo {author} {\bibfnamefont {C.}~\bibnamefont {Spitoni}},\ and\ \bibinfo
  {author} {\bibfnamefont {R.}~\bibnamefont {van Roij}},\ }\bibfield  {title}
  {\bibinfo {title} {Unveiling the capabilities of bipolar conical channels in
  neuromorphic iontronics},\ }\href@noop {} {\bibfield  {journal} {\bibinfo
  {journal} {Faraday Discussions}\ } (\bibinfo {year}
  {2023}{\natexlab{b}})}\BibitemShut {NoStop}%
\bibitem [{\citenamefont {Schmuck}\ and\ \citenamefont
  {Bazant}(2015)}]{schmuck2015homogenization}%
  \BibitemOpen
  \bibfield  {author} {\bibinfo {author} {\bibfnamefont {M.}~\bibnamefont
  {Schmuck}}\ and\ \bibinfo {author} {\bibfnamefont {M.~Z.}\ \bibnamefont
  {Bazant}},\ }\bibfield  {title} {\bibinfo {title} {Homogenization of the
  poisson--nernst--planck equations for ion transport in charged porous
  media},\ }\href@noop {} {\bibfield  {journal} {\bibinfo  {journal} {SIAM
  Journal on Applied Mathematics}\ }\textbf {\bibinfo {volume} {75}},\ \bibinfo
  {pages} {1369} (\bibinfo {year} {2015})}\BibitemShut {NoStop}%
\bibitem [{\citenamefont {Zangle}\ \emph {et~al.}(2009)\citenamefont {Zangle},
  \citenamefont {Mani},\ and\ \citenamefont
  {Santiago}}]{zangle2009propagation}%
  \BibitemOpen
  \bibfield  {author} {\bibinfo {author} {\bibfnamefont {T.~A.}\ \bibnamefont
  {Zangle}}, \bibinfo {author} {\bibfnamefont {A.}~\bibnamefont {Mani}},\ and\
  \bibinfo {author} {\bibfnamefont {J.~G.}\ \bibnamefont {Santiago}},\
  }\bibfield  {title} {\bibinfo {title} {On the propagation of concentration
  polarization from microchannel- nanochannel interfaces part ii: numerical and
  experimental study},\ }\href@noop {} {\bibfield  {journal} {\bibinfo
  {journal} {Langmuir}\ }\textbf {\bibinfo {volume} {25}},\ \bibinfo {pages}
  {3909} (\bibinfo {year} {2009})}\BibitemShut {NoStop}%
\bibitem [{\citenamefont {Jubin}\ \emph {et~al.}(2018)\citenamefont {Jubin},
  \citenamefont {Poggioli}, \citenamefont {Siria},\ and\ \citenamefont
  {Bocquet}}]{jubin2018dramatic}%
  \BibitemOpen
  \bibfield  {author} {\bibinfo {author} {\bibfnamefont {L.}~\bibnamefont
  {Jubin}}, \bibinfo {author} {\bibfnamefont {A.}~\bibnamefont {Poggioli}},
  \bibinfo {author} {\bibfnamefont {A.}~\bibnamefont {Siria}},\ and\ \bibinfo
  {author} {\bibfnamefont {L.}~\bibnamefont {Bocquet}},\ }\bibfield  {title}
  {\bibinfo {title} {Dramatic pressure-sensitive ion conduction in conical
  nanopores},\ }\href@noop {} {\bibfield  {journal} {\bibinfo  {journal}
  {Proceedings of the National Academy of Sciences}\ }\textbf {\bibinfo
  {volume} {115}},\ \bibinfo {pages} {4063} (\bibinfo {year}
  {2018})}\BibitemShut {NoStop}%
\bibitem [{\citenamefont {Markin}\ \emph {et~al.}(2014)\citenamefont {Markin},
  \citenamefont {Volkov},\ and\ \citenamefont {Chua}}]{markin2014analytical}%
  \BibitemOpen
  \bibfield  {author} {\bibinfo {author} {\bibfnamefont {V.~S.}\ \bibnamefont
  {Markin}}, \bibinfo {author} {\bibfnamefont {A.~G.}\ \bibnamefont {Volkov}},\
  and\ \bibinfo {author} {\bibfnamefont {L.}~\bibnamefont {Chua}},\ }\bibfield
  {title} {\bibinfo {title} {An analytical model of memristors in plants},\
  }\href@noop {} {\bibfield  {journal} {\bibinfo  {journal} {Plant signaling \&
  behavior}\ }\textbf {\bibinfo {volume} {9}},\ \bibinfo {pages} {e972887}
  (\bibinfo {year} {2014})}\BibitemShut {NoStop}%
\bibitem [{\citenamefont {Chua}(2014)}]{chua2014if}%
  \BibitemOpen
  \bibfield  {author} {\bibinfo {author} {\bibfnamefont {L.}~\bibnamefont
  {Chua}},\ }\bibfield  {title} {\bibinfo {title} {If it's pinched it's a
  memristor},\ }\href@noop {} {\bibfield  {journal} {\bibinfo  {journal}
  {Semicond. Sci. Technol.}\ }\textbf {\bibinfo {volume} {29}},\ \bibinfo
  {pages} {104001} (\bibinfo {year} {2014})}\BibitemShut {NoStop}%
\bibitem [{\citenamefont {Brown}\ \emph {et~al.}(2022)\citenamefont {Brown},
  \citenamefont {Kvetny}, \citenamefont {Yang},\ and\ \citenamefont
  {Wang}}]{brown2022selective}%
  \BibitemOpen
  \bibfield  {author} {\bibinfo {author} {\bibfnamefont {W.}~\bibnamefont
  {Brown}}, \bibinfo {author} {\bibfnamefont {M.}~\bibnamefont {Kvetny}},
  \bibinfo {author} {\bibfnamefont {R.}~\bibnamefont {Yang}},\ and\ \bibinfo
  {author} {\bibfnamefont {G.}~\bibnamefont {Wang}},\ }\bibfield  {title}
  {\bibinfo {title} {Selective ion enrichment and charge storage through
  transport hysteresis in conical nanopipettes},\ }\href@noop {} {\bibfield
  {journal} {\bibinfo  {journal} {The Journal of Physical Chemistry C}\
  }\textbf {\bibinfo {volume} {126}},\ \bibinfo {pages} {10872} (\bibinfo
  {year} {2022})}\BibitemShut {NoStop}%
\bibitem [{\citenamefont {Rotman}\ \emph {et~al.}(2011)\citenamefont {Rotman},
  \citenamefont {Deng},\ and\ \citenamefont {Klyachko}}]{rotman2011short}%
  \BibitemOpen
  \bibfield  {author} {\bibinfo {author} {\bibfnamefont {Z.}~\bibnamefont
  {Rotman}}, \bibinfo {author} {\bibfnamefont {P.-Y.}\ \bibnamefont {Deng}},\
  and\ \bibinfo {author} {\bibfnamefont {V.~A.}\ \bibnamefont {Klyachko}},\
  }\bibfield  {title} {\bibinfo {title} {Short-term plasticity optimizes
  synaptic information transmission},\ }\href@noop {} {\bibfield  {journal}
  {\bibinfo  {journal} {Journal of Neuroscience}\ }\textbf {\bibinfo {volume}
  {31}},\ \bibinfo {pages} {14800} (\bibinfo {year} {2011})}\BibitemShut
  {NoStop}%
\bibitem [{\citenamefont {Abbott}\ and\ \citenamefont
  {Regehr}(2004)}]{abbott2004synaptic}%
  \BibitemOpen
  \bibfield  {author} {\bibinfo {author} {\bibfnamefont {L.}~\bibnamefont
  {Abbott}}\ and\ \bibinfo {author} {\bibfnamefont {W.~G.}\ \bibnamefont
  {Regehr}},\ }\bibfield  {title} {\bibinfo {title} {Synaptic computation},\
  }\href@noop {} {\bibfield  {journal} {\bibinfo  {journal} {Nature}\ }\textbf
  {\bibinfo {volume} {431}},\ \bibinfo {pages} {796} (\bibinfo {year}
  {2004})}\BibitemShut {NoStop}%
\bibitem [{\citenamefont {Deng}\ and\ \citenamefont
  {Klyachko}(2011)}]{deng2011diverse}%
  \BibitemOpen
  \bibfield  {author} {\bibinfo {author} {\bibfnamefont {P.-Y.}\ \bibnamefont
  {Deng}}\ and\ \bibinfo {author} {\bibfnamefont {V.~A.}\ \bibnamefont
  {Klyachko}},\ }\bibfield  {title} {\bibinfo {title} {The diverse functions of
  short-term plasticity components in synaptic computations},\ }\href@noop {}
  {\bibfield  {journal} {\bibinfo  {journal} {Communicative \& Integrative
  Biology}\ }\textbf {\bibinfo {volume} {4}},\ \bibinfo {pages} {543} (\bibinfo
  {year} {2011})}\BibitemShut {NoStop}%
\bibitem [{\citenamefont {Kamsma}\ \emph
  {et~al.}(2024{\natexlab{b}})\citenamefont {Kamsma}, \citenamefont {van
  Roij},\ and\ \citenamefont {Spitoni}}]{kamsma2024simple}%
  \BibitemOpen
  \bibfield  {author} {\bibinfo {author} {\bibfnamefont {T.~M.}\ \bibnamefont
  {Kamsma}}, \bibinfo {author} {\bibfnamefont {R.}~\bibnamefont {van Roij}},\
  and\ \bibinfo {author} {\bibfnamefont {C.}~\bibnamefont {Spitoni}},\
  }\bibfield  {title} {\bibinfo {title} {A simple mathematical theory for
  simple volatile memristors and their spiking circuits},\ }\href@noop {}
  {\bibfield  {journal} {\bibinfo  {journal} {arXiv preprint arXiv:2404.08647}\
  } (\bibinfo {year} {2024}{\natexlab{b}})}\BibitemShut {NoStop}%
\bibitem [{\citenamefont {Deng}(2012)}]{deng2012mnist}%
  \BibitemOpen
  \bibfield  {author} {\bibinfo {author} {\bibfnamefont {L.}~\bibnamefont
  {Deng}},\ }\bibfield  {title} {\bibinfo {title} {The mnist database of
  handwritten digit images for machine learning research [best of the web]},\
  }\href@noop {} {\bibfield  {journal} {\bibinfo  {journal} {IEEE Signal
  Processing Magazine}\ }\textbf {\bibinfo {volume} {29}},\ \bibinfo {pages}
  {141} (\bibinfo {year} {2012})}\BibitemShut {NoStop}%
\end{thebibliography}
%

\end{document}


\title{Supplemental Information for: Brain-inspired computing with fluidic iontronic nanochannels}
\author{T. M. Kamsma}
\thanks{These two authors contributed equally to this work}
\affiliation{Institute for Theoretical Physics, Utrecht University,  Princetonplein 5, 3584 CC Utrecht, The Netherlands}
\affiliation{Mathematical Institute, Utrecht University, Budapestlaan 6, 3584 CD Utrecht, The Netherlands}
\author{\normalfont\textsuperscript{,$\dag$}\;J. Kim}
\thanks{These two authors contributed equally to this work}
\affiliation{Department of Mechanical Engineering, Sogang University, 35 Baekbeom-ro (Sinsu-dong), Mapo-gu, Seoul 04107, Republic of Korea}
\author{K. Kim}
\affiliation{Department of Mechanical Engineering, Sogang University, 35 Baekbeom-ro (Sinsu-dong), Mapo-gu, Seoul 04107, Republic of Korea}
\author{W. Q. Boon}
\affiliation{Institute for Theoretical Physics, Utrecht University,  Princetonplein 5, 3584 CC Utrecht, The Netherlands}
\author{C. Spitoni}
\affiliation{Mathematical Institute, Utrecht University, Budapestlaan 6, 3584 CD Utrecht, The Netherlands}
\author{J. Park}
\thanks{Corresponding author}
\affiliation{Department of Mechanical Engineering, Sogang University, 35 Baekbeom-ro (Sinsu-dong), Mapo-gu, Seoul 04107, Republic of Korea}
\author{R. van Roij}
\thanks{Corresponding author}
\affiliation{Institute for Theoretical Physics, Utrecht University,  Princetonplein 5, 3584 CC Utrecht, The Netherlands}

\date{\today}

\maketitle

\onecolumngrid\

\section{Theory}\label{sec:theory}
\subsection{Poisson-Nernst-Planck equations}
We consider ionic transport through a triangular tapered channel of uniform height $H$, length $L$, and widths $2R_{\ch{t}}$ and $2R_{\ch{b}}$ at the (smaller) tip and the (larger) base, i.e.\ $R_{\ch{t}}<R_{\ch{b}}$, as illustrated in Fig.~1(a) of the main text. In our experiments we have $L=150\text{ }\mu$m (unless stated otherwise), $R_{\ch{t}}=5\text{ }\mu$m and $R_{\ch{b}}=100\text{ }\mu$m. We introduce the axial coordinate $x$, with $x=0$ at the base and $x=L$ at the tip, and the cartesian width-coordinate $y\in[0,R(x)]$, with $y=0$ the symmetry axis and $y=R(x)$ the half width of the channel given by $R(x)=R_{\ch{b}} - (R_{\ch{b}}-R_{\ch{t}})x/L$ for $x\in[0,L]$. In the channel the height coordinate $z$ lies in the interval $z\in[-H/2,H/2]$. The channel is filled with a rigid close-packed fcc crystal of charged colloidal spheres (in the experiments with packing fraction $\eta\simeq 0.74$, radius $a=100$ nm, and zeta potential $\psi_0\approx-39$ mV). The channel connects two large and deep aqueous reservoirs containing a 1:1 electrolyte at room temperature with salt bulk concentration $\rho_{\ch{b}}$ (in the experiments $\rho_{\ch{b}}=10$ mM KCl with a Debye length $\lambda_{\ch{D}}=3.1$ nm). The ionic transport takes place through the electrolyte that fills the space between the colloids in the colloidal crystal and is driven by a time-dependent voltage $V(t)$. We define the applied voltage $V=V(L)-V(0)$ to be the voltage at the tip minus the voltage at the base, such that conductance enhancement always occurs when $V>0$, regardless of which side is grounded. In the main text we have a grounded tip so we apply $V_{\text{app}}(t)=-V(t)$ at the base. The transport is described in terms of the Poisson-Nernst-Planck (PNP) equations, that relate the electrostatic potential $\Psi(x,y,z,t)$ and the cationic and anionic concentrations $\rho_+(x,y,z,t)$ and $\rho_-(x,y,z,t)$, respectively, to the ionic fluxes ${\bf j}_\pm(x,y,z,t)$. In the space between the colloidal spheres we write

\begin{gather}
	\nabla^2\Psi=-\frac{e}{\epsilon}(\rho_+-\rho_-),\label{eq:poisson}\\
	\dfrac{\partial\rho_{\pm}}{\partial t}+\nabla\cdot\mathbf{j}_{\pm}=0,\label{eq:ce}\\
 \mathbf{j}_{\pm}=-D\left(\nabla\rho_{\pm}\pm\rho_{\pm}\frac{e\nabla \Psi}{k_{\mathrm{B}}T}\right),\label{eq:NP}
\end{gather}
where $e$ denotes the elementary charge, $\epsilon=80.23\epsilon_0$ the dielectric constant of water at room temperature $T$, $k_{\ch{B}}$ the Boltzmann constant, and $D$ the ionic diffusion coefficient that we take equal for the cations and the anions. The electrostatics is accounted for by the Poisson equation (\ref{eq:poisson}), the conservation of ions by the continuity equation (\ref{eq:ce}), and the combination of Fickian diffusion and Ohmic conduction by the Nernst-Planck equation (\ref{eq:NP}). We neglect electro-osmotic fluid flow, which we expect to be relatively small due to the narrow constrictions of the geometry. The system of equations (\ref{eq:poisson})-(\ref{eq:NP}) is closed upon imposing blocking boundary conditions on all solid walls, $\mathbf{n}\cdot\mathbf{j}_{\pm}=0$,  with ${\mathbf n}$ the (inward) normal on the walls of the channel and the colloids, together with  Gauss' law $\mathbf{n}\cdot\nabla\Psi=-e\sigma/\epsilon$ with $\sigma$ the surface charge density (on the wall and on the colloidal surfaces). We also impose that $\rho_\pm(x,y,z,t)$ equals the bulk concentration $\rho_{\ch{b}}$ at the far end of either reservoir, that $\Psi(0,y,z,t)=0$ and $\Psi(L,y,z,t)=V(t)$ to account for the applied potentials. 

The resulting closed set of PNP equations and boundary conditions can in principle be solved numerically by finite-element methods. The geometry of a colloidal crystal in a tapered channel, however, is computationally challenging as it requires a spatial resolution on the nm length scale of the electric double layer, on the 1-100 nm length scale of the pore structure in between the colloidal particles, and on the 10-100 $\mu$m length scale of the channel dimensions. By treating the complex porous structure homogeneous medium, these equations were modified and successfully solved numerically to describe the simplified physics inside the type of channel of interest here \cite{choi2016high}, however at a considerable computational cost and no analytic insights. Instead of a computationally costly numerical approach, we will derive analytical results straight from the Nernst-Planck equation (\ref{eq:NP}) to obtain an analytic approximation for the channel dynamics. This will yield a computationally significantly cheaper theoretical model which can be treated analytically to investigate the origin of the ion current rectifying properties of the channel, and to predict features such as its conductance memory properties.

\subsection{Slab-averaged electric field, space charge, and salt concentration}\label{sec:coarse}
Our theoretical approach is based upon a methodology that we successfully developed and applied recently to quantitatively explain the steady-state and dynamic conductance properties of simpler channels filled with a homogeneous aqueous electrolyte \cite{boon2021nonlinear,kamsma2023iontronic,kamsma2023unveiling}. Here we show that this methodology can be extended to the nanoporous channel network that characterizes the channel we study in this work. The colloidal structure within the channel forms a (nearly) close-packed face centered cubic (fcc) crystal at a volume fraction $\eta\simeq 0.74$ as we saw before. With a colloid radius $a=100\text{ nm}$,  this means that the pores through which ions can be transported have diameters as large as several tens of nm in the octahedral and tetrahedral holes of the fcc-lattice \cite{soloveva2021evaluation,heidig2017influence}, which is much larger than the Debye length of $\lambda_{\ch{D}}\approx 3.1$ nm that characterises the thickness of the electric double layers. In other words, the channels are mostly in the regime of non-overlapping and hence fully developed thin electric double layers, bringing us into the scope of area-averaging techniques \cite{mani2009propagation,zangle2009propagation,boon2021nonlinear,kamsma2023unveiling}. Specifically, this justifies the same underlying assumption as in Refs.~\cite{boon2021nonlinear,kamsma2023iontronic,kamsma2023unveiling} that the local and voltage-dependent total salt concentration $\rho_{\ch{s}}\equiv\rho_+ + \rho_-$, the total ionic space charge density $\rho_e=\rho_+-\rho_-$, and the local electric potential $\Psi$ can faithfully be represented by the $y-z$-slab-averaged functions $\overline{\rho}_{\ch{s}}(x,V)$, $\overline{\rho_e}(x)$, and $\overline{\Psi}(x)$, respectively, where we recall that the lateral coordinate $x\in[0,L]$ runs from base to tip. Here we explicitly denote the dependence of $\overline{\rho}_{\ch{s}}$ on $V$, while refraining from denoting the explicit (linear) dependence of $\overline{\Psi}$ on $V$ below for notational convenience. While we expect also a $V$-dependence of $\overline{\rho_e}$ in a full calculation, we will restrict ourselves to a $V$-independent form below.

If the slab-averaged electric field lines cannot escape the tapered channel, a realistic assumption in our experiments as the dielectric constant of water is much higher than that of the wall-material, then the slab-averaged electric field component $-\partial_x\overline{\Psi}(x)$ must be proportional to $1/R(x)$ on the basis of charge neutrality on the length scale beyond the Debye length. Since we define $V=V(L)-V(0)$, i.e.\ the tip minus the base voltage, the applied voltage also satisfies $\int_{0}^{L}\partial_x \overline{\Psi}(x)\dd x=V$. Combining the scaling with this property, we find with $\Delta R=R_{\ch{b}}-R_{\ch{t}}$ that

\begin{equation}\label{eq:Efield}
    \partial_x\overline{\Psi}(x)=\frac{\Delta RV}{L\ln\left(\frac{R_{\rm{b}}}{R_{\rm{t}}}\right)R(x)}.
\end{equation}
This slab-averaged electric field in the channel is therefore proportional to the applied field $V/L$ and gets progressively stronger closer to the tip.  
The total steady-state salt flux $\mathbf{j}_{\ch{s}}(x,y,z)=\mathbf{j}_{+}(x,y,z,t)+\mathbf{j}_{-}(x,y,z)$ can now be integrated over slabs in the $y$ and $z$ direction to obtain for the $x$-component of the total salt flux $J_{x}(x)=\int_{-R(x)}^{R(x)}\int_{-H/2}^{H/2}\mathbf{j}_{\ch{s}}(x,y,z)\cdot\hat{\mathbf{x}}\dd y\dd z$ through the channel

\begin{align}\label{eq:saltflux}
    J_x(x)=-D\epsilon_{\ch{fcc}}\left(2R(x)H\partial_x\overline{\rho}_{\ch{s}}(x,V)+2R(x)H\overline{\rho_{e}}(x)\frac{e\partial_x\overline{\Psi}(x)}{k_{\rm{B}}T}\right).
\end{align}
Here we introduced the porosity $\epsilon_{\ch{fcc}}=1-\eta$ to take the volume into account that is excluded to the electrolyte by the colloidal fcc crystal, where we assume the colloidal particles to be impenetrable to the electrolyte. As we take slab averages one would expect an area term, instead of the porosity. However, microscopically, the available electrolyte area through which the ions can diffuse in the slab has a periodicity in $x$ that is dictated by the lattice spacing, which is much smaller than the channel length $L$ and can hence be ignored in our slab-averaged description. Therefore we consider each slab to have the same available surface area for ions, which in this simplified one-dimensional view is the porosity $\epsilon_{\ch{fcc}}$ \cite{mani2011deionization}.

Eq.~(\ref{eq:saltflux}) depends on the slab-averaged ionic space charge density $\overline{\rho_{\ch{e}}}(x)$, that picks up contributions from the electric double layers (EDLs) around the charged colloidal spheres. Given that EDLs have spatial extensions as small as the Debye length (here $\lambda_{\ch{D}}=3.1\text{ nm}$) around the colloidal spheres (here of radius $a=100\text{ nm}$), one might expect the slab-averaged space charge density to be a constant on the much larger length scale of the channel, with a magnitude $\propto \eta Z$ with $Z$ the charge of a colloid.  Interestingly, however, earlier measurements presented in  Ref.~\cite{choi2016high} on channels nearly identical to the ones we study here convincingly showed a heterogeneous rather than a homogeneous space charge density that could well be fitted by the monotonic functional form

\begin{align}\label{eq:rhoe}
    \overline{\rho_{\ch{e}}}(x)=\frac{\rho_{\ch{e,t}}}{1-(1-\rho_{\ch{e,t}}/\rho_{\ch{e,b}})(L-x)/L},
\end{align}
where $\rho_{\ch{e,t}}$ and $\rho_{\ch{e,b}}$ are the space charge density at the tip and the base, respectively, which may differ from each other even at zero applied voltage \cite{choi2016high}. In this work we have $\rho_{\rm{e,t}}=8.85\text{ mM}$ and $\rho_{\rm{e,b}}=8.00\text{ mM}$ for the steady-state results of Fig.~1(b) and $\rho_{\rm{e,b}}=6.3\text{ mM}$ for the other (dynamic) results, the choice for these values is further discussed in Sec.~\ref{sec:params}. Whereas we take the functional form of Eq.(\ref{eq:rhoe}) as experimental input from now on, the reason for this ionic charge heterogeneity on length scales of the channel length remains an interesting open question. Our hypothesis, that we underpin with standard Poisson-Boltzmann calculations of a single colloid in a spherical Wigner-Seitz cell in Sec.~\ref{sec:cellmodel}, shows that a reduction of the colloidal packing fraction from 0.74 to 0.73 can already increase the colloidal charge density (and hence the average ionic charge in the EDLs)  by as much as $\sim20$\% at a fixed zeta potential. Thus, a small heterogeneity of the colloidal packing fraction from tip to base could provide a microscopic explanation for the spatial dependence of $\overline{\rho_{\ch{e}}}(x)$ by assuming macroscopic electroneutrality \cite{mani2009propagation,mani2011deionization}. Given the device fabrication (see Sec.~\ref{sec:devicefabrication}) such a small asymmetry of $\eta$ between tip and base is quite well possible although such a small deviation in packing fraction is difficult to observe experimentally.

With Eqs.~(\ref{eq:Efield}) and (\ref{eq:rhoe}) every component of Eq.~(\ref{eq:saltflux}) is known except for $\overline{\rho}_{\ch{s}}(x,V)$. We can use Eq.~(\ref{eq:saltflux}) for $J_x(x)$ to find the steady-state salt concentration $\overline{\rho_{\ch{s}}}(x,V)$ explicitly by imposing the steady-state condition $\partial_xJ_x(x)=0$. This yields a differential equation for $\overline{\rho_{\ch{s}}}(x,V)$, that can be solved analytically after inserting Eqs.(\ref{eq:Efield}) and (\ref{eq:rhos}), yielding 

\begin{align}\label{eq:rhos}
	\overline{\rho_{\ch{s}}}(x,V)=2\rho_{\ch{b}}-\rho_{\ch{e,b}}\frac{eV}{k_{\rm{B}} T }\frac{\Delta R}{R_{\rm{b}}} \frac{\ln\left(\frac{R_{\rm{b}}}{R_{\rm{t}}}\right)\ln\left(\frac{\rho_{\ch{e,t}}}{\rho_{\ch{e,b}}}\frac{L-x}{L}+x/L\right)+\ln\left(\frac{R(x)}{R_{\ch{t}}}\right)\ln\left(\frac{\rho_{\ch{e,b}}}{\rho_{\ch{e,t}}}\right)}{\ln^2\left(\frac{R_{\rm{b}}}{R_{\rm{t}}}\right) \left(\rho_{\ch{e,b}}/\rho_{\ch{e,t}}- R_{\rm{t}}/R_{\rm{b}}\right)}.
\end{align}
We note that $H$, $D$, and $\epsilon_{\ch{fcc}}$ do not appear in Eq.~(\ref{eq:rhos}) since these drop out of the underlying differential equation $\partial_xJ_x(x)=0$. More importantly, $\overline{\rho_{\ch{s}}}(x,V)$ is voltage-dependent, resulting in the voltage-dependent channel conductance we derive next.
\begin{figure}[h]
\centering
     \includegraphics[width=0.6\textwidth]{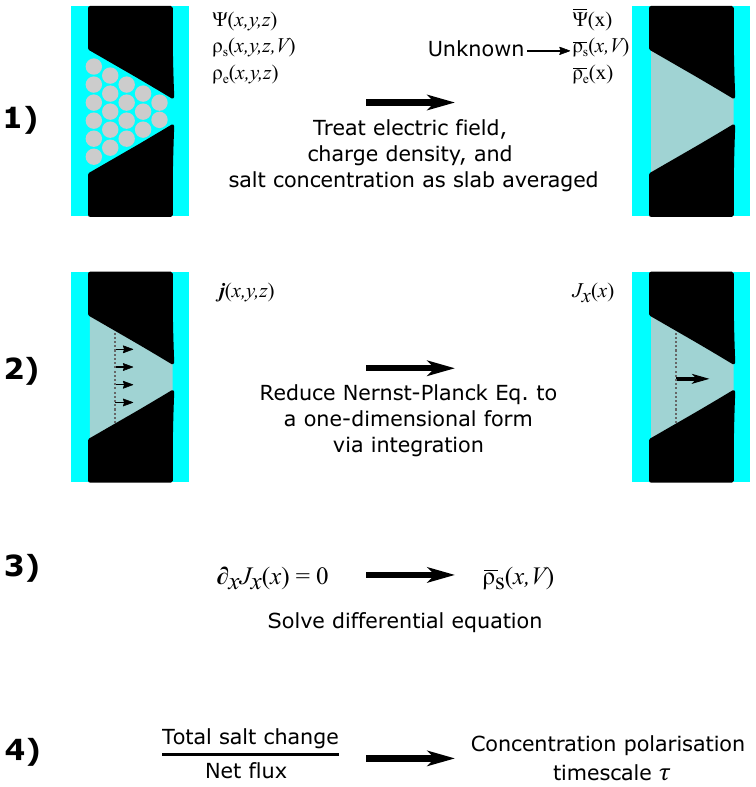}
        \caption{Schematic depiction of the steps taken in our theoretical analysis, involving 1) the slab-averaging assumption, 2) the integration of the salt flux over each slab to reduce the full three-dimensional Nernst-Planck equation to its one-dimensional slab-averaged form as in Eq.~(\ref{eq:saltflux}), and 3) the solution of the steady-state condition of a divergence-free flux to obtain Eq.~(\ref{eq:rhos}). Lastly, 4) the combination of the salt concentration and the salt flux to extract a timescale for the salt concentration polarisation as detailed in Sec.~\ref{sec:ts}.}
        \label{fig:Theory_Fig}
\end{figure}

\subsection{Static channel conductance}\label{sec:cond}
The tapered microchannels of our interest are well known to exhibit ionic rectification properties characterised by a static conductance $g_{\infty}\equiv I(V)/V$ that is a nontrivial function of the applied static voltage $V$. By viewing the slabs of thickness $\dd x$ at $x\in[0,L]$ as a series of resistors with resistivities $\propto \dd x/\rho_{\ch{s}}(x)$, one can write the static conductance of the channels of present interest as

\begin{equation}
    g_{\infty}(V)=g_0\int_0^L \overline{\rho}_{\mathrm{s}}(x,V) \dd x/(2\rho_{\rm{b}}L),\label{eq:cond}
\end{equation}
where we made an approximation compared to the more accurate dependence on $L/\int_{0}^{L}(\overline{\rho}_{\mathrm{s}}(x,V))^{-1} \dd x$, which reduces computational complexity and yields mostly the same results \cite{boon2021nonlinear,kamsma2023iontronic,kamsma2023unveiling}. The $V$-dependence stems from the salt concentration dependence on $V$ as given by Eq.(\ref{eq:rhos}). As in Ref.~\cite{kamsma2023unveiling}, we replace $\overline{\rho}_{\ch{s}}(x,V)$ by $\max\left[0.2\rho_{\ch{b}},\overline{\rho}_{\ch{s}}(x,V)\right]$ in the actual (numerical) evaluations of Eq.~(\ref{eq:cond}) in order to account for the possibility of unphysical negative concentrations  that could follow from Eq.(\ref{eq:rhos}) at strongly negative voltages in part of the density profiles. (In this regime the underlying assumption that the Debye length is much larger than than the channel dimensions breaks down). 

The reference (zero-voltage) conductance $g_0$ of the channel can, in direct analogy with recent results  
from Refs.~\cite{boon2021nonlinear,kamsma2023iontronic,kamsma2023unveiling}, be written as  

\begin{align}\label{eq:g0}
    g_0=&2\rho_{\rm{b}}\epsilon_{\ch{fcc}}eD\frac{2\Delta RH}{L\ln(\frac{R_{\ch{b}}}{R_{\ch{t}}})}\frac{e}{k_{\rm{B}}T}+g_{\rm{s}}=2\rho_{\rm{b}}\frac{e^2D}{k_{\ch{B}}T}\epsilon_{\ch{fcc}}\frac{2\Delta RH}{L\ln(\frac{R_{\ch{b}}}{R_{\ch{t}}})}\left[1+\frac{4\lambda_{\ch{D}}}{R_{\ch{pore}}}\left(\cosh\left(\frac{e\psi_{\rm{0}}}{2k_{\rm{B}}T}\right)-1\right)\right],
\end{align}
which includes the volumetric contribution $\propto\epsilon_{\ch{fcc}}$ that depends on the channel-geometry parameters calculated with the total charge flux $\int_{-R(x)}^{R(x)}\int_{-H/2}^{H/2}\left[\mathbf{j}_{+}(x,y,z)-\mathbf{j}_{-}(x,y,z)\right]\cdot\hat{\mathbf{x}}\dd y\dd z$. Additionally we also consider a surface contribution $\propto\lambda_{\ch{D}}/R_{\ch{pore}}$, with $R_{\ch{pore}}$ the effective radius of the pores embedded in the fcc crystal, which accounts for the excess conductivity due to the excess salt concentration in the colloidal EDLs \cite{aarts2022ion} and which we determine below. This surface term is of direct relevance here due to the large internal surface in the channel as a result of the colloidal structure. However, the internal structure of the fcc crystal is complex with pores of varying sizes and shapes, and regions with fully developed EDLs in the pores with the size of several tens of $\lambda_{\ch{D}}$ neighboured by regions where the EDLs are not fully developed since the shortest distance between colloids is smaller than a few $\lambda_{\ch{D}}$. Therefore, the surface conductance of the channel is a highly non-trivial property and investigating this in full detail falls outside the scope of this work. Instead we will employ an effective method where we treat the internal pores as a channel of radius $R_{\ch{pore}}$, with a total slab surface area $2\epsilon_{\ch{fcc}}R(x)H$. This approach then yields Eq.~(\ref{eq:g0}) \cite{aarts2022ion}. We find good agreement with the experiments if we take $R_{\ch{pore}}=2\lambda_{\ch{D}}=0.06a$. Although this seems somewhat small compared to the radii of up to $\sim10\lambda_{\ch{D}}$ for the larger pores, it is still in a reasonable regime given the fact that many regions in a close-packed fcc crystal feature much smaller distances between colloids. For the parameters used in the experiment (see Sec.~\ref{sec:params}) we have that $g_{\ch{s}}/g_0\approx 0.38$, so the conductance is still mostly dictated by bulk conductance, as expected with the relatively large pores compared to the Debye length.

The combination of Eq.~(\ref{eq:cond}) and Eq.~(\ref{eq:g0}) provides us with the steady-state conductance function $g_{\infty}(V)$ used in the main text. 

\subsubsection{Space charge density inhomogeneity is a salt source or sink term}\label{sec:sourcesink}
The methodology we employ in Sec.~\ref{sec:coarse} to derive salt concentration polarisation can provide insights into designing new devices that also exhibit salt accumulation and depletion. As shown in Sec.~\ref{sec:cond}, the presence of salt concentration polarisation affects the channel conductance and will hence lead to current rectification. Our theoretical model explains current rectification in the colloid-filled channels we present here, but other types of channels also exhibit current rectification as a result of an inhomogeneous space charge \cite{kim2022asymmetric,sabbagh2023designing}. Our theoretical description might provide insights into such other systems as well, provided some assumptions are satisfied. (i) The underlying PNPS equations only describe continuum transport, so sub-nm or atomic scale systems \cite{robin2021principles,robin2023long} will fall outside the scope of this theoretical framework. (ii) The radial dependence of the potential and the concentrations must be relatively weak and/or short ranged, such that the slab-averaged salt-concentration $\overline{\rho}_{\ch{s}}(x,V)$ and electric field $-\partial_x\overline{\Psi}(x)$ are fair approximations.  For instance, a single channel with strongly overlapping electric double layers throughout the entire channel will exhibit a significant salt concentration profile in the radial direction of the channel, possibly resulting in physical phenomena that a slab average $\overline{\rho}_{\ch{s}}(x,V)$ does not take into account.

Eq.~[\ref{eq:saltflux}] can be generalised to
\begin{align}\label{eq:saltfluxGeneral}
    J_x(x)=-Da\left(A(x)\partial_x\overline{\rho}_{\ch{s}}(x,V)+Q_{e}(x)\frac{e\partial_x\overline{\Psi}(x)}{k_{\rm{B}}T}\right)=-Da\left(A(x)\partial_x\overline{\rho}_{\ch{s}}(x,V)+A(x)\overline{\rho_{e}}(x)\frac{e\partial_x\overline{\Psi}(x)}{k_{\rm{B}}T}\right),
\end{align}
with $Q_{e}(x)\dd x$ the total ionic charge in a slab of infinitesimal thickness $\dd x$ with volume $A(x)\dd x$ at location $x$, and where we define $\overline{\rho_{e}}(x)\equiv Q_{e}(x)/A(x)$ the slab-averaged ionic charge density. Eq.~(\ref{eq:saltfluxGeneral}) describes a channel with cross-sectional area $A(x)$ and the average area per slab available for ions $a$ (which would be 1 for materials without any solid obstructions for the ions), where we remark that in the simplified one-dimensional view of treating each slab to feature the same available area one can use the porosity as we did in Eq.~(\ref{eq:saltflux}) \cite{mani2011deionization}. In the present channel geometry we have $A(x)=2HR(x)$, but Eq.~(\ref{eq:saltfluxGeneral}) could also apply to hourglass shaped channels \cite{sabbagh2023designing}, T-shaped channels \cite{kim2022asymmetric}, or conical channels \cite{boon2021nonlinear,kamsma2023unveiling}. If macroscopic charge neutrality is ensured within the channel, e.g. in the case of thin electric double layers, and if the electric field lines cannot leave the channel because of a weakly polarising outside medium, then the relation $-\partial_x\overline{\Psi}(x)\propto 1/A(x)$ must hold. A fully analytical solution as in Eq.~(\ref{eq:Efield}) might not always be possible, however the inverse proportionality with the channel area $A(x)$ suffices for the qualitative mechanistic understanding discussed here. Consider, namely, a channel without any applied voltage such that the salt concentration $\overline{\rho}_{\ch{s}}(x,V)$ is constant within the channel. Upon applying a voltage, the electric field $-\partial_x\overline{\Psi}(x)$ will form quasi-instantaneously, so a short time after the voltage is applied we have a fully formed electric field, but still a constant salt concentration $\overline{\rho}_{\ch{s}}(x,V)$. In this case the diffusion term in Eq.~(\ref{eq:saltfluxGeneral}) vanishes and due to the aforementioned proportionality $-\partial_x\overline{\Psi}(x)\propto 1/A(x)$, the only $x$-dependence that remains in Eq.~(\ref{eq:saltfluxGeneral}) is $\overline{\rho_{e}}(x)$. Therefore, if we take the divergence $\partial_xJ_x(x)$ of the total salt flux in Eq.~(\ref{eq:saltfluxGeneral}) and apply the continuity Eq.~[\ref{eq:ce}] we find 
\begin{align}
    \dfrac{\dd \overline{\rho}_{\ch{s}}(x,V)}{\dd t}=-\dfrac{\dd J_x(x)}{\dd x}\propto \overline{\rho_{e}}(x).\label{eq:sourcesink}
\end{align}
This shows how any inhomogeneous ionic space charge density forms a source or sink for salt term upon applying a voltage, thereby inducing salt concentration polarisation and consequently current rectification. An important understanding is that the ionic charge \textit{density} must be inhomogeneous, and not just the \textit{total} ionic charge in the slab. The reason is that even though the total charge $Q_{e}(x)\equiv \overline{\rho_{e}}(x)A(x)$ in the slab for a constant space charge density $\overline{\rho_{e}}(x)=\overline{\rho_{e}}$ could still be $x$-dependent due to its scaling with $A(x)$, this dependence would cancel out with the $1/A(x)$ dependence of the electric field.

The insight that an inhomogeneous charge density forms a source-sink term previously explained how a constant surface charge density $\sigma$ in a conical geometry could induce current rectification \cite{boon2021nonlinear} as the total (surface) charge is in this case given by $Q_{e}(x)=2\pi R(x)\sigma\dd x$. Since in this geometry $A(x)=\pi R(x)^2$ we see that $Q_{e}(x)\propto \sqrt{A(x)}$ and therefore $\overline{\rho_{e}}(x)\equiv Q_{e}(x)/A(x)\propto 1/\sqrt{A(x)}$, exhibiting the required $x$-dependence. Additionally Eq.~(\ref{eq:sourcesink}) shows why merely a geometric inhomogeneity, such as a tapered geometry, is not enough to induce current rectification; it must go coupled with a spatially varying slab-averaged ionic charge density. The insight of Eq.~(\ref{eq:sourcesink}) could not only explain current rectification in channels with a space charge density step-function as in recent polyelectrolyte channels \cite{sabbagh2023designing} and NCNM channels with colloids of opposing charge \cite{kim2022asymmetric}, it may also provide specific guidance to design current rectification properties in future iontronics.

\subsection{Typical conductance memory retention time}\label{sec:ts}
As detailed in Refs.~\cite{kamsma2023iontronic,kamsma2023unveiling}, the process of ion accumulation and depletion is not instantaneous. To investigate this timescale for the channel of interest here we can apply the same approach as in Refs.~\cite{kamsma2023iontronic,kamsma2023unveiling}. We consider two quantities, the total number of ions $N=\int_0^L2R(x)H\epsilon_{\ch{fcc}}\overline{\rho}_{\mathrm{s}}(x,V)\dd x$ in the channel and the net salt flux $J_{x}(0)-J_{x}(L)$ into the channel. The change of $N$ given by Eq.~(\ref{eq:rhos}) upon a small voltage perturbation $V^{\prime}$ around $V=0$ yields

\begin{equation}\label{eq:alpha}
    \begin{aligned}
    \left.\dfrac{\partial N}{\partial V}\right|_{V=0} V^{\prime}=&\frac{\epsilon_{\ch{fcc}}e H L \rho_{\ch{e,t}} \rho_{\ch{e,b}}\Delta RV^{\prime}}{2 k_{\ch{B}} T \ln ^2\left(\frac{R_{\ch{b}}}{R_{\ch{t}}}\right) (\rho_{\ch{e,t}} R_{\ch{t}}-\rho_{\ch{e,b}} R_{\ch{b}})}\left((R_{\ch{b}}+R_{\ch{t}}) \ln \left(\frac{\rho_{\ch{e,t}}}{\rho_{\ch{e,b}}}\right)\right.\\
    &\left.+\frac{\ln \left(\frac{R_{\ch{b}}}{R_{\ch{t}}}\right) \left(-(\rho_{\ch{e,t}}-\rho_{\ch{e,b}}) \Delta R (\rho_{\ch{e,t}} (R_{\ch{b}}+3 R_{\ch{t}})-\rho_{\ch{e,b}} (3 R_{\ch{b}}+R_{\ch{t}}))-2 \ln \left(\frac{\rho_{\ch{e,t}}}{\rho_{\ch{e,b}}}\right) (\rho_{\ch{e,t}} R_{\ch{t}}-\rho_{\ch{e,b}}
   R_{\ch{b}})^2\right)}{(\rho_{\ch{e,t}}-\rho_{\ch{e,b}})^2 \Delta R}\right)\\
   \equiv& \alpha V^{\prime},
    \end{aligned}
\end{equation}
where $\alpha>0$ for our parameters, in agreement with the enhanced (reduced) conductance of a positive (negative) potential $V^{\prime}$ found in the experiments and as can be seen in Fig.~1(b).

At $V=0$ the concentration profile is at equilibrium, so for a small voltage perturbation $V^{\prime}$ we can assume $\bar{\rho}_{\rm{s}}(x)=2\rho_{\rm{b}}$. With this assumption the first term in Eq.~(\ref{eq:saltflux}) vanishes. The net salt flux into the channel, $J_x(0)-J_x(L)$, is then determined by the remaining conductive terms

\begin{align}\label{eq:netflux}
    J_x(0)-J_x(L)=2D\epsilon_{\ch{fcc}}\frac{\Delta R H}{L\ln\left(\frac{R_{\ch{b}}}{R_{\ch{t}}}\right)}(\rho_{\ch{e,t}}-\rho_{\ch{e,b}})\frac{e}{k_{\ch{B}}T}V^{\prime}\equiv \gamma V^{\prime},
\end{align}
where $\gamma>0$ for our parameter choices, again in agreement with the enhanced (reduced) conductance of a positive (negative) potential $V^{\prime}$ found in the experiments and as can be seen in Fig.~1(b). We see that the net flux is determined by the difference in space charge density between the tip and the base $(\rho_{\ch{e,t}}-\rho_{\ch{e,b}})$. This is consistent with Eq.~(\ref{eq:saltflux}), where we saw that the total salt flux is proportional to the charge density in the absence of a diffusion term. Therefore, this reconfirms the underlying mechanistic insight that a inhomogeneous charge density is what drives a net salt flux and consequent salt concentration polarisation, as also argued in Sec.~\ref{sec:sourcesink}.

\begin{figure}[h]
\centering
     \includegraphics[width=0.8\textwidth]{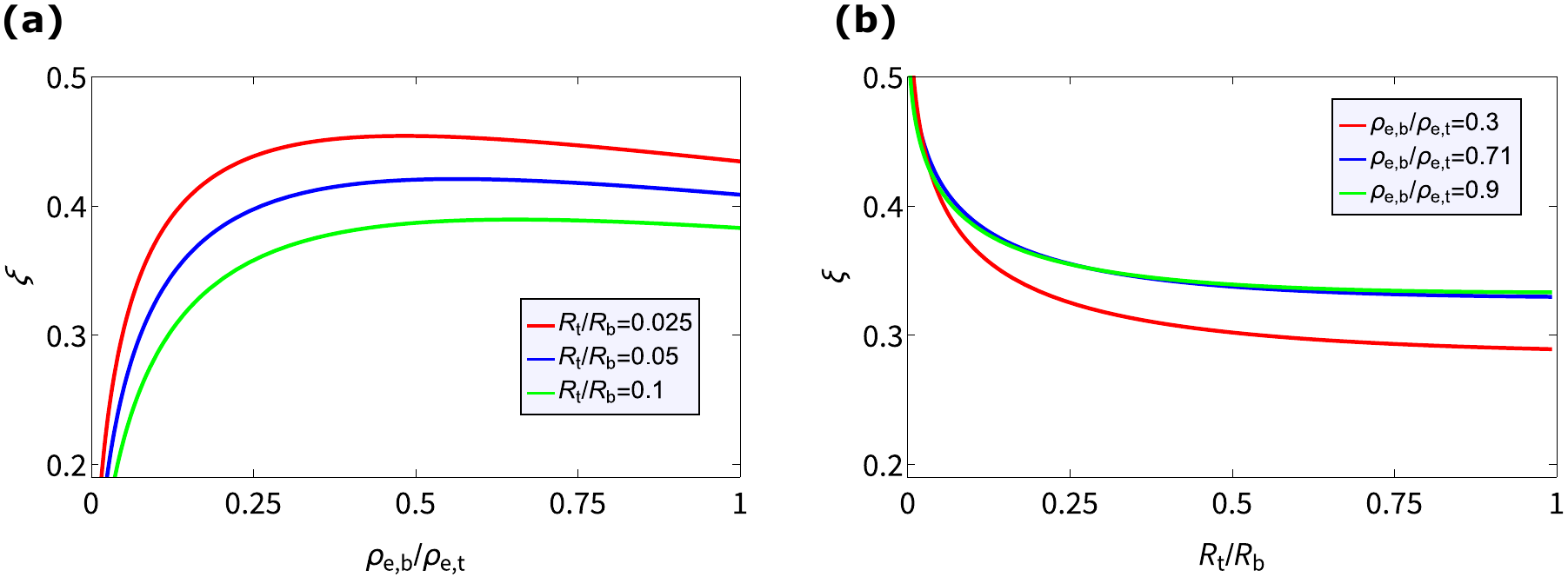}
        \caption{The memory time proportionality function $\xi$ as a function the internal space charge ratio $\rho_{\ch{e,b}}/\rho_{\ch{e,t}}$ in \textbf{(a)} and the tip and base radii ratio $R_{\ch{t}}/R_{\ch{b}}$ in \textbf{(b)}, for three different values of $R_{\ch{t}}/R_{\ch{b}}$ and $\rho_{\ch{e,b}}/\rho_{\ch{e,t}}$, respectively. In both \textbf{(a)} and \textbf{(b)} the blue graph represents $R_{\ch{t}}/R_{\ch{b}}=0.05$ and $\rho_{\ch{e,b}}/\rho_{\ch{e,t}}=0.71$, respectively, the system parameters used in the main text, except for Fig.~1(b) which uses $\rho_{\ch{e,b}}/\rho_{\ch{e,t}}=0.90$.}
        \label{fig:hPlots}
\end{figure}

The typical time it takes for ion depletion or accumulation, and thus the typical memory retention timescale, is then approximated by $\tau=\alpha/\gamma$. This yields

\begin{align*}
    \tau=&\frac{L^2 \rho_{\ch{e,t}} \rho_{\ch{e,b}} \left((R_{\ch{b}}-R_{\ch{t}}) (R_{\ch{b}}+R_{\ch{t}}) \ln \left(\frac{\rho_{\ch{e,t}}}{\rho_{\ch{e,b}}}\right)+\frac{\ln \left(\frac{R_{\ch{b}}}{R_{\ch{t}}}\right) \left(-(\rho_{\ch{e,t}}-\rho_{\ch{e,b}}) (R_{\ch{b}}-R_{\ch{t}}) (\rho_{\ch{e,t}} (R_{\ch{b}}+3 R_{\ch{t}})-\rho_{\ch{e,b}} (3 R_{\ch{b}}+R_{\ch{t}}))-2 \ln \left(\frac{\rho_{\ch{e,t}}}{\rho_{\ch{e,b}}}\right) (\rho_{\ch{e,t}} R_{\ch{t}}-\rho_{\ch{e,b}}
   R_{\ch{b}})^2\right)}{(\rho_{\ch{e,t}}-\rho_{\ch{e,b}})^2}\right)}{4 D (\rho_{\ch{e,t}}-\rho_{\ch{e,b}}) (R_{\ch{b}}-R_{\ch{t}}) \ln \left(\frac{R_{\ch{b}}}{R_{\ch{t}}}\right) (\rho_{\ch{e,t}} R_{\ch{t}}-\rho_{\ch{e,b}} R_{\ch{b}})}\\
   =&\frac{L^2}{4D}\rho_{\ch{e,t}} \rho_{\ch{e,b}}\left(\frac{(R_{\ch{b}}+R_{\ch{t}}) \ln \left(\frac{\rho_{\ch{e,t}}}{\rho_{\ch{e,b}}}\right)}{(\rho_{\ch{e,t}}-\rho_{\ch{e,b}}) \ln \left(\frac{R_{\ch{b}}}{R_{\ch{t}}}\right) (\rho_{\ch{e,t}} R_{\ch{t}}-\rho_{\ch{e,b}} R_{\ch{b}})}-\frac{\rho_{\ch{e,t}}R_{\ch{b}}-\rho_{\ch{e,b}}R_{\ch{t}}}{(\rho_{\ch{e,t}}-\rho_{\ch{e,b}})^2(\rho_{\ch{e,t}} R_{\ch{t}}-\rho_{\ch{e,b}} R_{\ch{b}})}\right.\\
   &\left.-\frac{3}{(\rho_{\ch{e,t}}-\rho_{\ch{e,b}})^2}-\frac{2\ln
   \left(\frac{\rho_{\ch{e,t}}}{\rho_{\ch{e,b}}}\right) (\rho_{\ch{e,t}} R_{\ch{t}}-\rho_{\ch{e,b}} R_{\ch{b}})}{(\rho_{\ch{e,t}}-\rho_{\ch{e,b}})^3 \Delta R}\right)\\
   =&\frac{L^2}{4D}\left(\frac{(1+R_{\ch{t}}/R_{\ch{b}}) \ln \left(\frac{\rho_{\ch{e,t}}}{\rho_{\ch{e,b}}}\right)}{(1-\rho_{\ch{e,b}}/\rho_{\ch{e,t}}) \ln \left(\frac{R_{\ch{b}}}{R_{\ch{t}}}\right) (\rho_{\ch{e,t}} R_{\ch{t}}/(\rho_{\ch{e,b}}R_{\ch{b}})-1)}-\frac{1-\rho_{\ch{e,b}}R_{\ch{t}}/(R_{\ch{b}}\rho_{\ch{e,t}})}{(\rho_{\ch{e,t}}/\rho_{\ch{e,b}}+\rho_{\ch{e,b}}/\rho_{\ch{e,t}}-2)(R_{\ch{t}}/R_{\ch{b}}-\rho_{\ch{e,b}}/\rho_{\ch{e,t}})}\right.\\
   &\left.-\frac{3}{\rho_{\ch{e,t}}/\rho_{\ch{e,b}}+\rho_{\ch{e,b}}/\rho_{\ch{e,t}}-2}-\frac{2\ln
   \left(\frac{\rho_{\ch{e,t}}}{\rho_{\ch{e,b}}}\right) ( R_{\ch{t}}/R_{\ch{b}}-\rho_{\ch{e,b}}/\rho_{\ch{e,t}})}{(1-\rho_{\ch{e,b}}/\rho_{\ch{e,t}})(\rho_{\ch{e,t}}/\rho_{\ch{e,b}}+\rho_{\ch{e,b}}/\rho_{\ch{e,t}}-2) \Delta R/R_{\ch{b}}}\right).
\end{align*}
From the above calculation it has now become clear that $\tau\propto L^2/(4D)$, which corresponds to the time it takes to diffuse over distances of the order of the channel length $L$. The remaining involved terms form a dimensionless number 

\begin{align*}
   \xi=&\frac{(1+R_{\ch{t}}/R_{\ch{b}}) \ln \left(\frac{\rho_{\ch{e,t}}}{\rho_{\ch{e,b}}}\right)}{(1-\rho_{\ch{e,b}}/\rho_{\ch{e,t}}) \ln \left(\frac{R_{\ch{b}}}{R_{\ch{t}}}\right) (\rho_{\ch{e,t}} R_{\ch{t}}/(\rho_{\ch{e,b}}R_{\ch{b}})-1)}-\frac{1-\rho_{\ch{e,b}}R_{\ch{t}}/(R_{\ch{b}}\rho_{\ch{e,t}})}{(\rho_{\ch{e,t}}/\rho_{\ch{e,b}}+\rho_{\ch{e,b}}/\rho_{\ch{e,t}}-2)(R_{\ch{t}}/R_{\ch{b}}-\rho_{\ch{e,b}}/\rho_{\ch{e,t}})}\\
   &-\frac{3}{\rho_{\ch{e,t}}/\rho_{\ch{e,b}}+\rho_{\ch{e,b}}/\rho_{\ch{e,t}}-2}-\frac{2\ln
   \left(\frac{\rho_{\ch{e,t}}}{\rho_{\ch{e,b}}}\right) ( R_{\ch{t}}/R_{\ch{b}}-\rho_{\ch{e,b}}/\rho_{\ch{e,t}})}{(1-\rho_{\ch{e,b}}/\rho_{\ch{e,t}})(\rho_{\ch{e,t}}/\rho_{\ch{e,b}}+\rho_{\ch{e,b}}/\rho_{\ch{e,t}}-2)(1-R_{\ch{t}}/R_{\ch{b}})}
\end{align*}
of order $\mathcal{O}(10^{-1})$ depending on the channel geometry $R_{\ch{t}}/R_{\ch{b}}$ and internal space charge distribution $\rho_{\ch{e,b}}/\rho_{\ch{e,t}}$. Fig.~\ref{fig:hPlots} shows the dependence of $\xi$ for typical tip-to-base ratios of the space charge concentrations in Fig.~\ref{fig:hPlots}(a) and the channel widths in Fig.~\ref{fig:hPlots}(b). For our standard parameters we find $\xi\approx 0.42$. This simplifies our notation considerably such that we arrive at the final concise expression Eq.~(\ref{eq:tauSI}) for $\tau$ which can be found in the main text as Eq.~[1], 

\begin{align}\label{eq:tauSI}
   \tau=&\frac{L^2}{4D}\xi.
\end{align}
For our standard parameter set as laid out in Sec.~\ref{sec:params}, we have $\tau\approx 1.62\text{ s}$.\\

To then arrive at Eq.~[2] of the main text, we make the natural assumption that the time-derivative of the dynamic conductance $\partial_tg(t)$ depends on the difference with the corresponding steady-state conductance, i.e.\ $\partial_tg(t)=f\big(g_\infty(V(t))-g(t)\big)$ for some function $f$, where $f(0)$ must vanish based on stability arguments. By expanding $f$ up to first order we find $\partial_tg(t)\propto g_\infty( V(t))-g(t)$ with the proportionality constant naturally given by the typical timescale $\tau$ of the underlying salt concentration polarisation that drives the conductance change, given by Eq.~(\ref{eq:tauSI}) (Eq.~[1] in the main text).

\subsection{Cell model}\label{sec:cellmodel}
In Sec.~\ref{sec:coarse} we showed how the inhomogeneous charge distribution leads to salt concentration polarisation and thereby to ion current rectification. However, Eq.~(\ref{eq:rhoe}) presented in Ref.~\cite{choi2016high}, which describes the inhomogeneous space charge, follows from empirical measurements without a microscopic explanation of the actual origin of this important feature of the channel. Here we will leverage the well-established Poisson-Boltzmann cell-model \cite{von2001gas,trizac2003alexander} to propose a tentative explanation of the emergence of the observed inhomogeneous space charge.

We consider a dispersion of charged colloidal spheres of radius $a$ at packing fraction $\eta$ in a 1:1 electrolyte of Debye length $\lambda_{\ch{D}}$, such that the volume per particle equals $(4\pi/3)b^3$ with $b\equiv a\eta^{-1/3}>a$. The environment of each particle fluctuates due to colloidal Brownian motion, and hence the calculation of the profile of the electric potential $\psi({\bf r};\{{\bf R}\})$ is a complicated many-body problem that depends on the instantaneous configuration $\{{\bf R}\}$ of the colloidal particles. This problem can be reduced tremendously, however, if we assume each sphere to be at the center of an electrically neutral and spherically symmetric Wigner-Seitz cell of radius $b$. The dimensionless electrostatic potential $\phi(r)\equiv e\psi(r)/k_{\ch{B}}T$ is then the same in each cell, and can for $r\in[a,b]$ be described by the Poisson-Boltzmann equation with boundary conditions

\begin{align}\label{eq:PBcell}
    \phi^{\prime\prime}(r)+\frac{2}{r}\phi^{\prime}(r)=&\lambda_{\ch{D}}^{-2}\sinh{\phi(r)},\\
    \phi^{\prime}(b)=&0,\\
    \phi(a)=&\phi_0,
\end{align}
where a prime denotes a radial derivative and where $\phi_0$ is the fixed dimensionless zeta potential in units of $k_{\ch{B}}T/e\simeq 25\text{ mV}$. 
This is a closed system of equations that can easily be solved numerically for fixed $a$, $b$, $\lambda_{\ch{D}}$, and $\psi_0$ or rather for fixed dimensionless $\eta$, $a/\lambda_{\ch{D}}$ and $\phi_0$. The resulting surface charge density $e\sigma$ of the spheres is then from Gauss's law given by $\sigma=-\phi^{\prime}(a)/4\pi\lambda_B$ with $\lambda_{\rm{B}}=e^2/4\pi\epsilon k_{\rm{B}}T=0.72\text{ nm}$ the Bjerrum length of water. 

\begin{figure}[h]
\centering
     \includegraphics[width=0.5\textwidth]{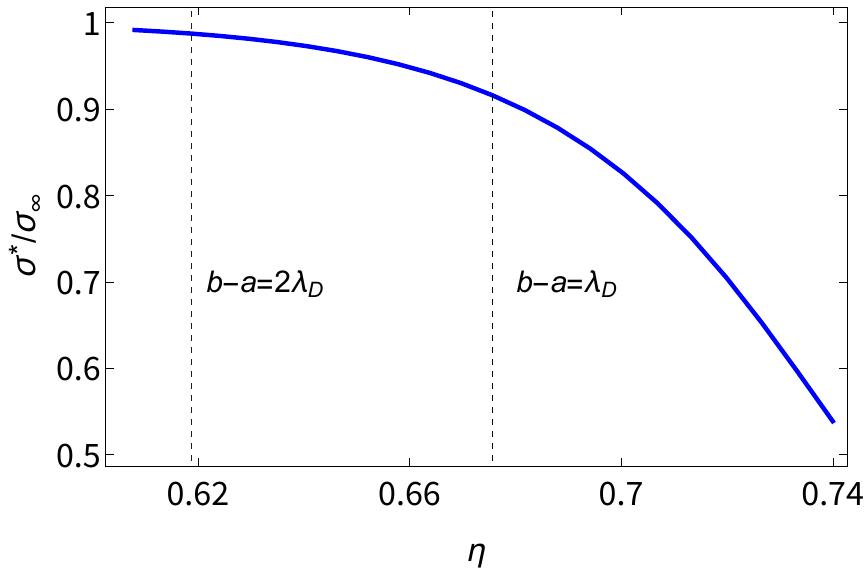}
        \caption{Dependence on the colloidal packing fraction $\eta$ of the effective surface charge density $\sigma^{*}$ of a colloid in an fcc crystal as a fraction of the surface charge density $\sigma_\infty=\sigma_{\ch{C}}=0.01\text{ C/m}^2$ of a free isolated colloid. An increase of up to $\sim100\%$ in $\sigma^{*}$ is already visible when the cell radius $b$ in the direction of the nearest neighbour just extends one or two Debye lengths $\lambda_{\rm{D}}$ beyond the colloid of radius $a$, i.e.\ when the nearest neighbour distance $b-a\in [0,2\lambda_{\rm{D}}]$.}
        \label{fig:ZPlot}
\end{figure}

For dispersions that are dilute enough that $b-a\gg \lambda_{\ch{D}}$, the EDLs described by Eqs.(\ref{eq:PBcell}) are fully developed. In this regime it is well known that the colloidal particles obtain their maximum (absolute) surface charge \cite{smallenburg2011phase}. At high packing fraction where $b-a\simeq \lambda_{\ch{D}}$,  the colloidal surfaces discharge and loose their charge completely in the (unphysical) limit $b=a$. The situation for the colloidal dispersion of our system is intricate, since for $a=100\text{ nm}$ and $\eta=0.74$ we have $b/a\simeq1.105$ and hence $b-a\simeq 3\lambda_{\ch{D}}$. In this regime the assumption of spherical symmetry of the EDLs is highly questionable, because the surface-surface distances between a central particle in a close-packed fcc crystal at $\eta=0.74$ vary between smaller than $\lambda_{\ch{D}}$ in the 12 directions of its nearest neighbours to about $10 \lambda_{\ch{D}}$ in the directions of the 8 tetrahedral and the 6 octahedral holes of the fcc crystal. We can therefore expect that the particles only significantly discharge in the vicinity of nearest-neigbour contact. Within the spherical-cell approximation, we mimic this anisotropy of the colloidal surface charge by averaging over these $12+8+6=26$ directions, with equal weight for simplicity, to define the effective surface charge density $\sigma^*=(12\sigma_2 + 14\sigma_\infty)/26$, with the surface charges $\sigma_2$ and $\sigma_\infty$ of the spherical cell with $b$ half the nearest-neighbour distance and $b\gg a$ the dilute limit, respectively. In Fig.~\ref{fig:ZPlot} we show the dependence of $\sigma^*/\sigma_{\infty}$ on the packing fraction $\eta$ for the present system parameter with $\phi_0=-1.53$, which is such that $\sigma_C=\sigma_\infty=0.01\text{ C/m}^2$. It shows that decreasing the packing fraction from $\eta=0.74$ to $0.73$, which corresponds to reducing $b$ by as little as 0.5 nm (i.e. by $\simeq\lambda_{\ch{D}}/6$), the colloidal charge can increase by as much as $\sim20$\%. A reduction to $\eta\simeq0.67$ and $\eta\simeq0.63$ (which corresponds to $b-a=\lambda_{\ch{D}}$ and $2\lambda_{\ch{D}}$) increases the surface charge by about 80\% and 100\%, respectively, compared to that at $\eta=0.74$. On the basis of these simple estimates, we argue that the slab-averaged space charge in the channel (that compensates the colloidal surface charge) can therefore also exhibit the same relative change between tip and base. The variation we use of $\rho_{\ch{e,b}}/\rho_{\ch{e,t}}=0.90$ for the steady-state results of Fig.~1(b) and $\rho_{\ch{e,b}}/\rho_{\ch{e,t}}=0.71$ for the other (dynamic) results falls well within this reasonable range of change predicted by the explanation we offer here. We leave an additional estimate on an alleged spatial variation of $\phi_0$ in the channel to explain the heterogeneous space charge as future work.    

\section{System parameters}\label{sec:params}
The channel has a base radius of $R_{\ch{b}}=100 \mu\text{m}$, a tip radius of $R_{\ch{t}}=5 \mu\text{m}$ and a height of $H=5 \mu\text{m}$. The channel connects two reservoirs with a bulk concentration $\rho_{\ch{b}}=10 \text{ mM}$ of aqueous KCl electrolyte, yielding a Debye length of $\lambda_{\ch{D}}\approx 3.1\text{ nm}$. The colloids have a radius of $a=100\text{ nm}$ and carry a uniform charge density of $\sigma_{\rm{C}}=-0.01\text{ C/m}^2$, squeezed together during the device fabrication to form a face centered cubic crystal with a close-packed packing fraction $\eta=1-\epsilon_{\ch{fcc}}\approx 0.74$, where $\epsilon_{\ch{fcc}}$ is the porosity. The maximum space charge at the tip is determined in the same way as in Ref.~\cite{choi2016high}, i.e.\ $\overline{\rho_{\ch{e}}}(L)=\rho_{\rm{e,t}}=4\pi a^2\sigma_{\rm{C}}n/(e\epsilon_{\rm{fcc}}\Omega)=8.85\text{ mM}$ with $n=\frac{3}{4}\eta\Omega/(\pi a^3)$ the number of colloids in the channel and $\Omega$ the overall channel volume. The space charge concentration at the base is assumed to be $\overline{\rho_{\ch{e}}}(0)=\rho_{\ch{e,b}}=8\text{ mM}$ in the steady-state calculations for Fig.~1(b) and $\rho_{\ch{e,b}}=6.3\text{ mM}$ in the rest of the manuscript for the time-dependent calculations. These values were chosen such that (i) the change in space charge density does not exceed the reasonable regime predicted by our tentative explanation for the the inhomogeneous space charge in Sec.~\ref{sec:cellmodel}, i.e.\ the difference between $\rho_{\ch{e,b}}$ and $\rho_{\ch{e,b}}$ is not more than $\sim40\%$, and (ii) a good agreement is found with the experiments. Although the space charge density $\overline{\rho_{\ch{e}}}(x)$ of Eq.~(\ref{eq:rhoe}) has a clear physical meaning, its precise form and parameters values are not entirely clear and may contain a voltage-dependence that we ignore in the present study and leave for future investigations. We stress, however, that our present overall results do not depend strongly on the detailed value of $\rho_{\ch{e,b}}$ and $\rho_{\ch{e,t}}$, as it only marginally changes the conductance properties in the parameter regime we consider here and the overall behaviour relevant to reservoir computing remains. However, we do note that our theory predicts that some difference between the space charge density at the tip and at the base is necessary for any current rectification, so even though the precise values of $\rho_{\ch{e,b}}$ and $\rho_{\ch{e,t}}$ are not of major importance for our overall results, our theory still predicts it is crucial that there is some difference between the two. Lastly, the effective diffusion coefficient is also taken to be similar to Ref.~\cite{choi2016high} with $D_{\rm{eff}}=\epsilon_{\rm{fcc}}D=0.38\text{ }\mu\text{m}^2\text{ms}^{-1}$.

\section{Device fabrication}\label{sec:devicefabrication}
\begin{figure}[h]
\centering
     \includegraphics[width=0.65\textwidth]{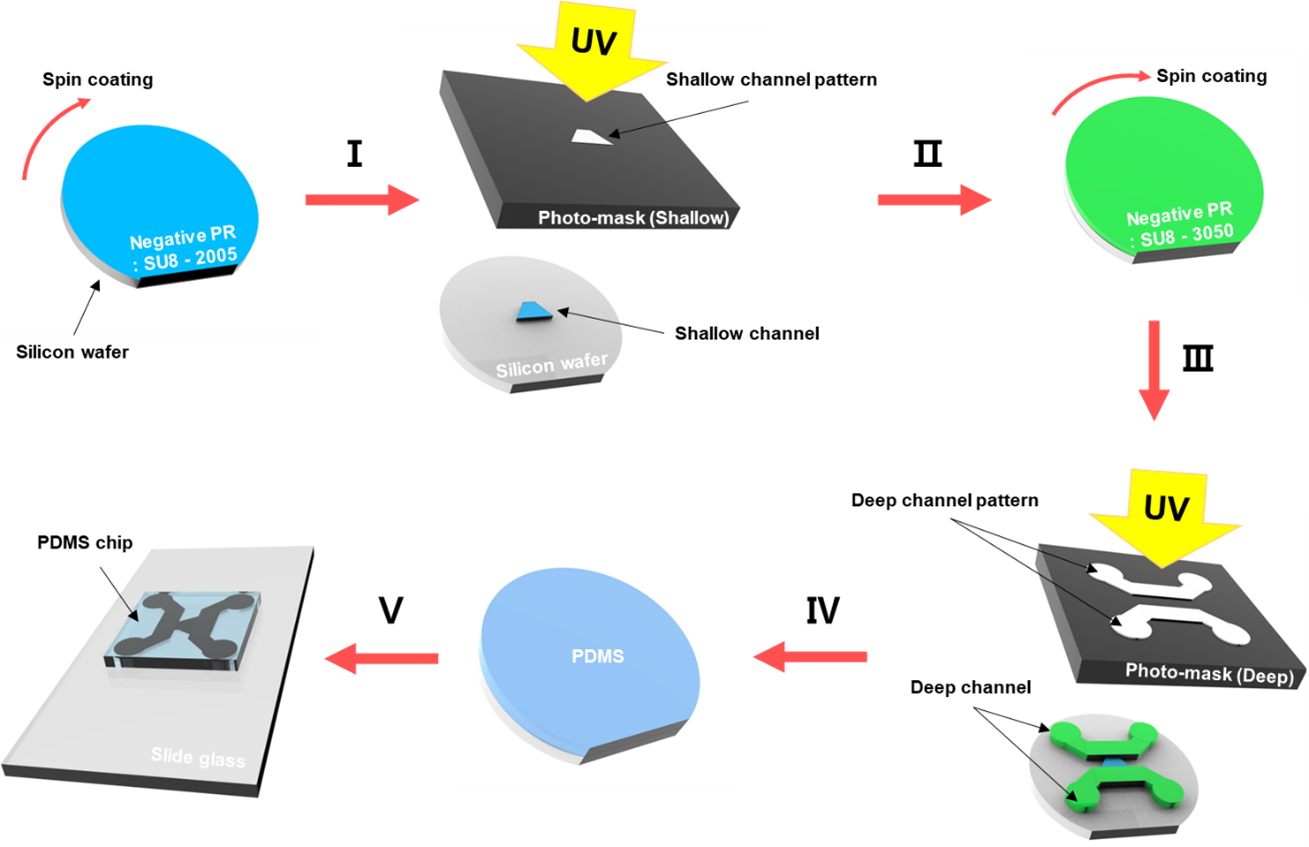}
        \caption{Soft lithography process to construct our microfluidic memristor.}
        \label{fig:FabricationS1}
\end{figure}
Fig.~\ref{fig:FabricationS1} shows schematic diagrams describing the fabrication procedures for microfluidic devices using soft lithography: A negative photoresist (PR) was applied to the silicon wafer (SU-8 2005; Microchem Co., Westborough, Massachusetts, USA) using a spin coater and then soft-baked. The PR was then exposed to UV to create a shallow channel and the wafer was hard baked. The unexposed PR was removed to form the shallow $5\text{ }\mu$m channel. A deep $100\text{ }\mu$m channel was then formed using a negative photoresist, SU8-3050 (Microchem Co., Westborough, Massachusetts, USA), in the same process as above. After completion of the master mold, the surface was treated with (3,3,3-trifluoropropyl)silane (452807; Sigma-Aldrich, St. Louis, Missouri, USA). Polydimethylsiloxane (PDMS; Sylgard; Dow Corning Korea Ltd., Gwangju-si, Gyeonggi-do, Republic of Korea) was then poured over the master mold and heated on a hot plate at 95 $ ^{\circ}$C for 1 hour. The reservoir of the PDMS device was punched out with a 1.5 mm medical punch. The surface of the PDMS and the slide glass were treated using a plasma device (Cute-MP; Femto Science, Hwaseong-si, Gyeonggi-do, Republic of Korea) and bonded together.
\begin{figure}[h]
\centering
     \includegraphics[width=0.65\textwidth]{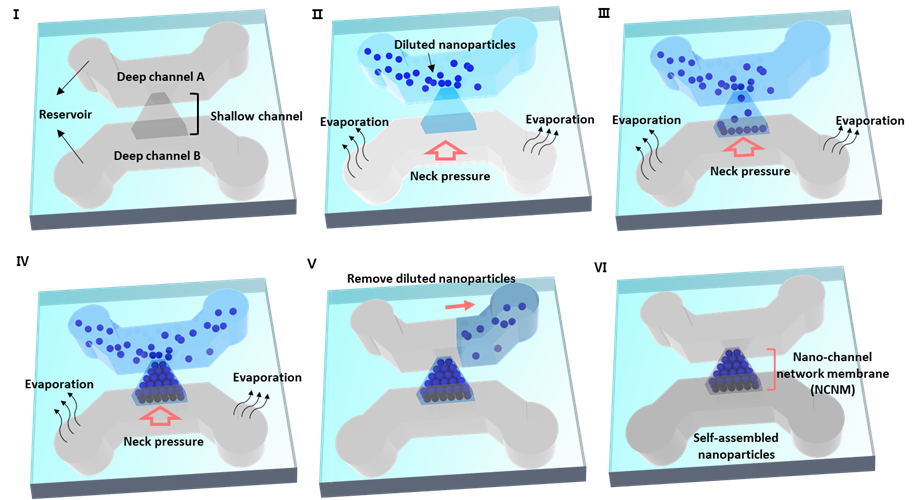}
        \caption{Schematic depiction of in situ nanoparticle assembly in the microchannel.}
        \label{fig:FabricationS2}
\end{figure}

As shown in Fig.~\ref{fig:FabricationS2}, the diluted nanoparticles (with a carboxylic (-COOH) end groups on the surface and with radius $a=100\text{ nm}$) dispersed in a $2\text{ }\mu$L 70\% ethanol solution were injected into the deep channel, such that the solution including the nanoparticles fills the shallow channel through the capillary. Due to the neck pressure at the interface between the shallow and deep channels, the flow of the particle dispersion stopped, after which evaporation of the solvent (70\% of the ethanol) was induced through the deep channel. This way additional nanoparticles were transported toward the neck by the convective flow that compensates for the loss of solvent by evaporation. This influx of particles promotes the growth (and the compression) of the ordered fcc lattice in the shallow channel. When the self-assembled nanoparticles completely filled the shallow channel with a wet close-packed fcc lattice, the remaining diluted nanoparticles were removed from the deep channel by suction. The finished device was dried for one day and then used for the experiment.

\section{Additional experimental results}
\subsection{Voltage pulse measurements}
\begin{figure}[h]
\centering
     \includegraphics[width=1\textwidth]{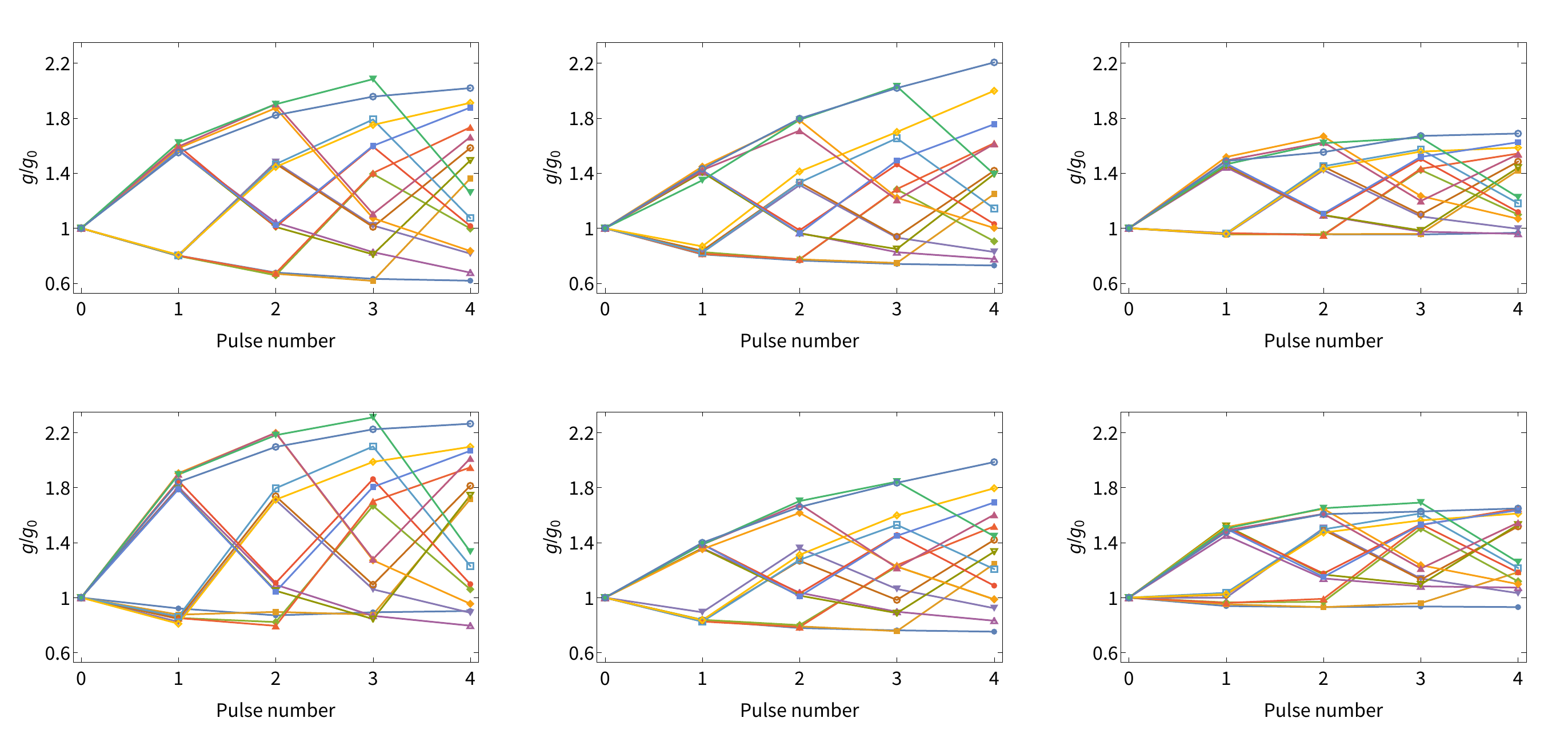}
        \caption{Six measurements of the normalized channel conductance for all $2^4=16$ different voltage pulse trains, obtained from three devices. Each column of two figures corresponds to a device. The average for each different pulse train is used to create Fig.~4(a). Standard deviations for the conductance measurement after the fourth (last) pulse were also calculated with these six measurements and shown in Table~\ref{tab:stdev}.}
        \label{fig:pulseTrainPlots}
\end{figure}

To obtain Fig.~4(a) we applied all $2^4=16$ different voltage trains of four pulses in sequence with 30 s between each pulse train. This was performed on three different devices with two cycles per device, resulting in 6 independent measurements per voltage train, all individually shown in Fig.~\ref{fig:pulseTrainPlots}. The average of these measurement is shown in Fig.~4(a), the standard deviations calculated with the six conductance values after the fourth (last) pulse for each pulse train are shown in Table \ref{tab:stdev}. These values are used in the main text to take the measured (device-to-device) variability into account. Therefore we stress that all (device-to-device) variability visible in Fig.~\ref{fig:pulseTrainPlots} is incorporated in the main text.

\begin{figure}[h]
\centering
     \includegraphics[width=1\textwidth]{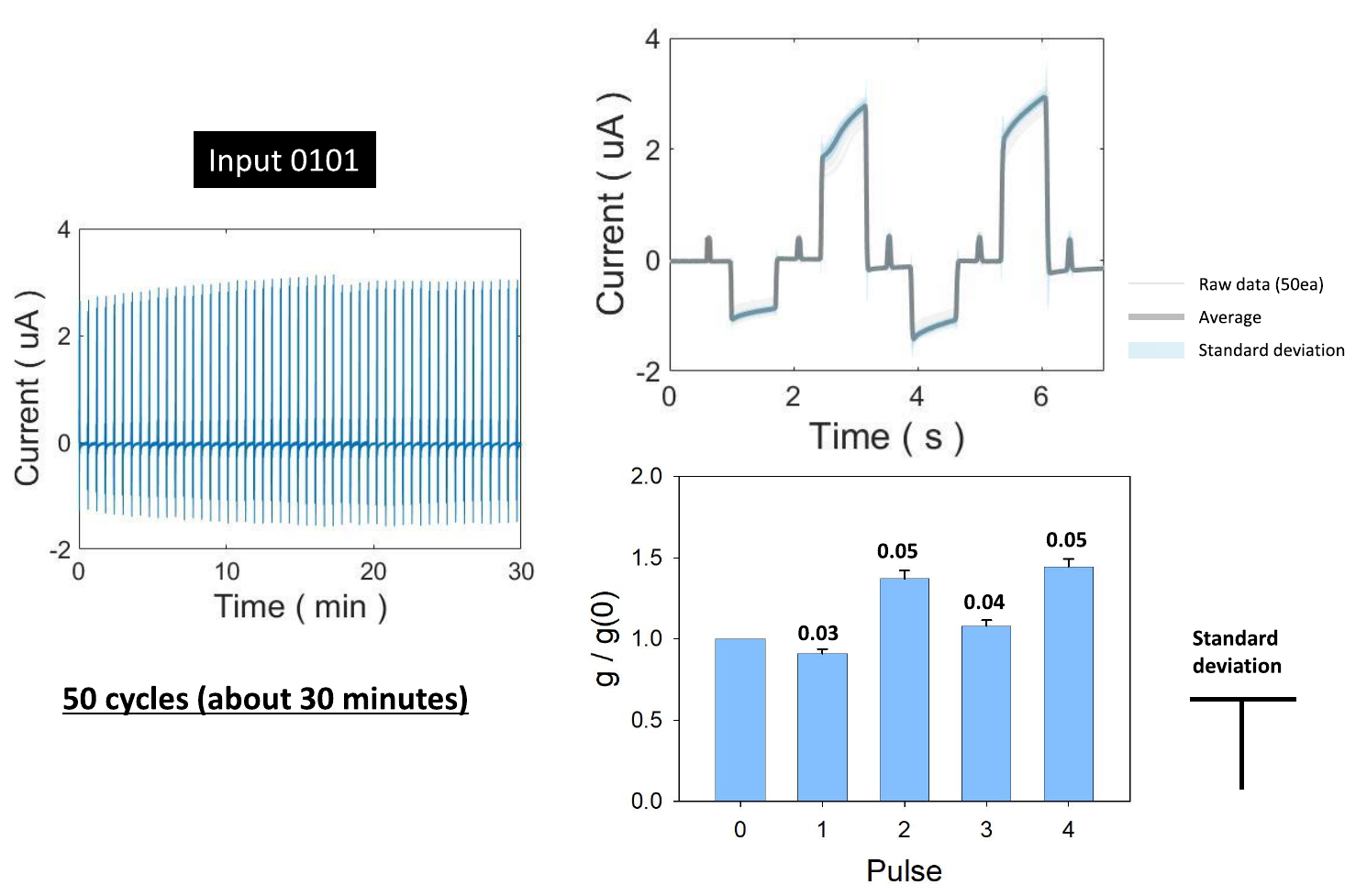}
        \caption{A 50-cycle repeat of the 0101 voltage pulse train. With (left) the current measurement during all cycles, (top right) all cycles overlaid with the light grey spread the raw overlaid data, the dark grey line the average of the measurements, and the variability characterised by the standard deviation in light blue. (bottom right) The resulting normalized conductances determined by the five read-pulses with their respective standard deviations.}
        \label{fig:0101}
\end{figure}

\begin{figure}[h]
\centering
     \includegraphics[width=1\textwidth]{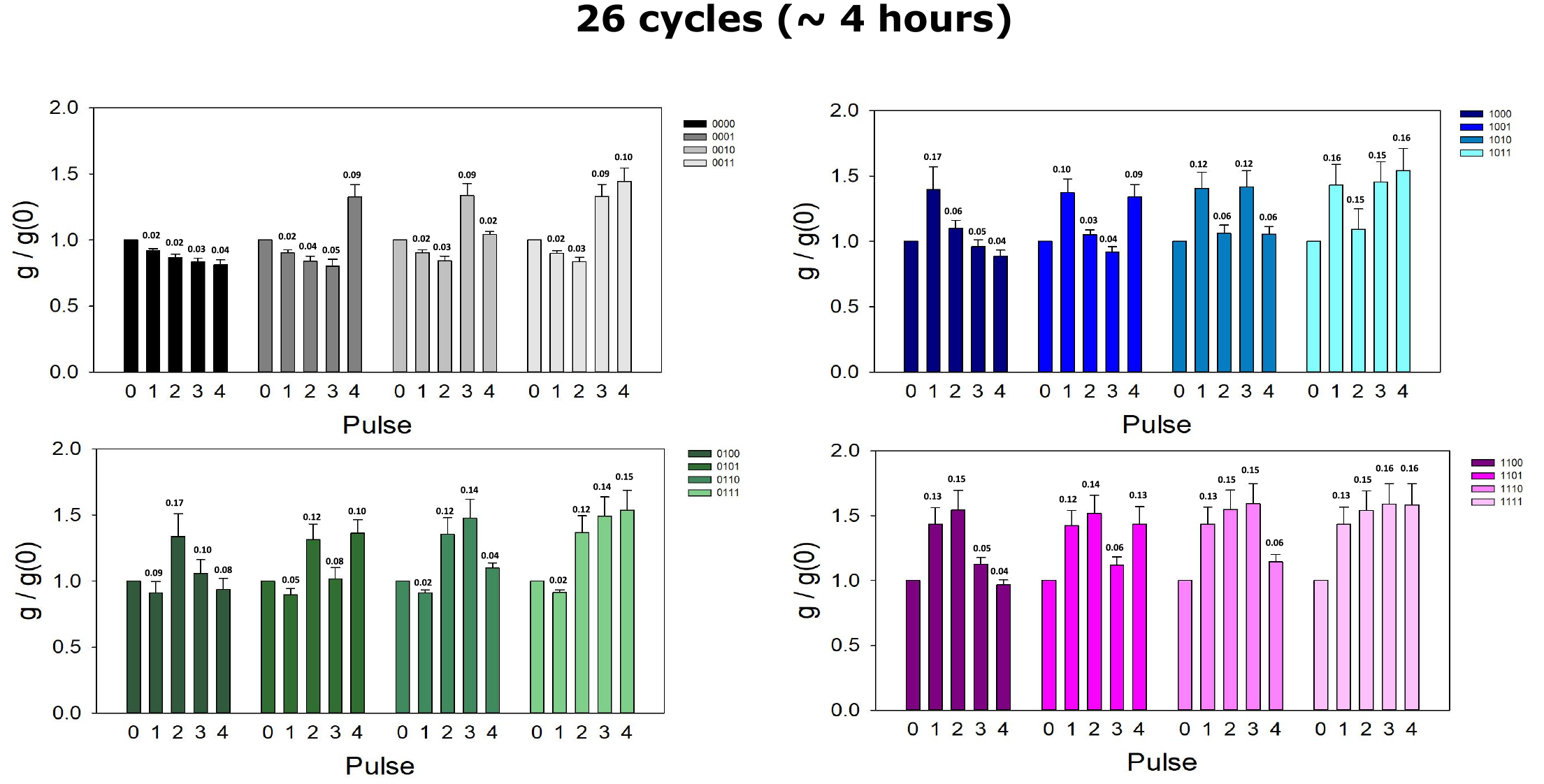}
        \caption{Normalized conductances and their respective standard deviations obtained by cycling all 16 voltage pulse trains for a total of 50 times, taking roughly 4 hours to complete.}
        \label{fig:0000_1111}
\end{figure}

To ensure that our devices also remain reliable and stable over longer periods of time, we performed various additional cycles of pulse trains over a single device. In Fig.~~\ref{fig:0101} we show a 50 cycle repeat, which lasted about 30 minutes, of the pulse train corresponding to the bitstring 0101. We see a remarkable stability, with essentially the same current response each cycle (which we also show as Fig.~2 in the main text), yielding a narrow spread for each pulse and reliably measured altered conductances with conductance standard deviations of $g/g_0\sim0.03-0.05$. Moreover, we cycled through all 16 different bitstrings for a total of 26 cycles, lasting around 4 hours, and found essentially the same current response for each cycle, even after the device had been cycling for 4 hours. The resulting average normalized conductances per pulse, per cycle, are shown in Fig.~\ref{fig:0000_1111}, with the corresponding standard deviations in the range $g/g_0\sim0.02-0.15$.

\begin{figure}[h]
\centering
     \includegraphics[width=0.65\textwidth]{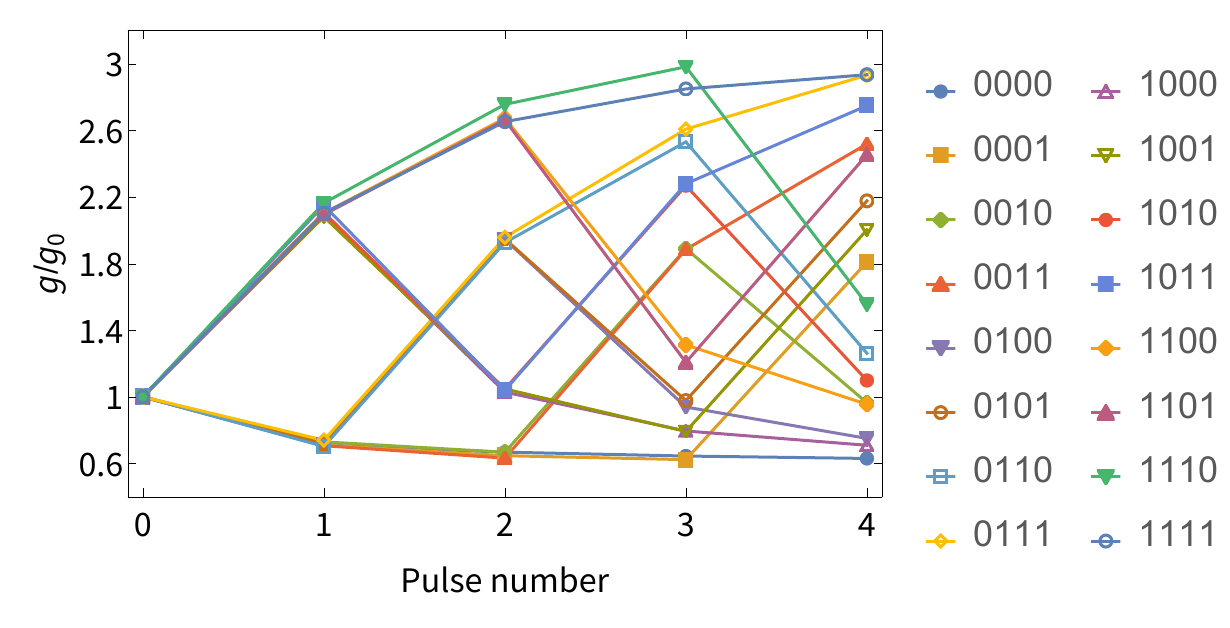}
        \caption{Measurements of the normalized channel conductance for all $2^4=16$ different voltage pulse trains, but here the channel base is grounded and the voltage is applied at the tip.}
        \label{fig:pulseTrainPlotstipGround}
\end{figure}

\begin{figure}[h]
\centering
     \includegraphics[width=0.75\textwidth]{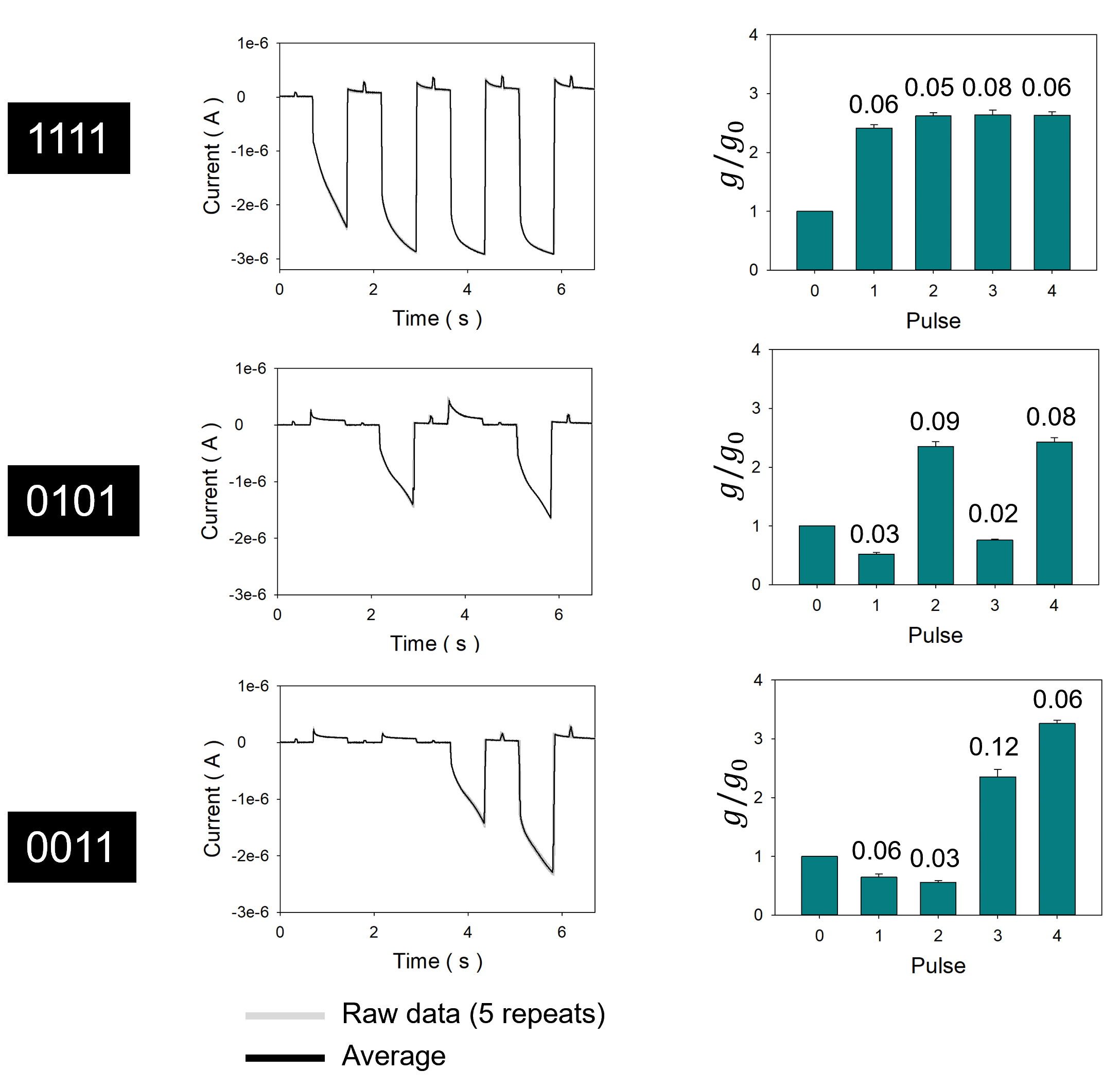}
        \caption{Voltage pulse train measurements with the ground at the tip and the applied voltage at the channel base. A ``0" corresponds to a pulse of 2 V, while a ``1" corresponds to a pulse of -5 V. Pulse duration and interval are 0.75 s, with read pulses of 1 V and 50 ms duration. Voltage pulse trains corresponding to the bit-strings 1111, 0101 and 0011 were repeated 5 times, where we show the 5 individual measurements (light grey), the average of the measured current (black) and the calculated normalized conductances in the bar plots, averaged over the 5 measurements. The error bars depict the measured standard deviations.}
        \label{fig:repeatPulses}
\end{figure}

We repeated a similar protocol, but now with the ground at the base and the applied voltage at the channel tip. In this instance, a ``0" corresponds to a pulse of 2 V, while a ``1" corresponds to a pulse of -5 V. Pulse duration and interval are still 0.75 s, the read pulses are 1 V of 50 ms duration. The result is shown in Fig.~\ref{fig:pulseTrainPlotstipGround}, showing a similar clear separation of the different bit-strings. To again ensure the device stability, the bit-strings 1111, 0101 and 0011 were repeated 5 times, which we present in Fig.~\ref{fig:repeatPulses}. Here we show the 5 individual measurements (light grey), the average of the measured current (black) and the calculated normalized conductances in the bar plots, averaged over the 5 measurements. The error bars depict the measured standard deviations, where we find good reproducibility for each voltage pulse train. 

\begin{figure}[h]
\centering
     \includegraphics[width=0.75\textwidth]{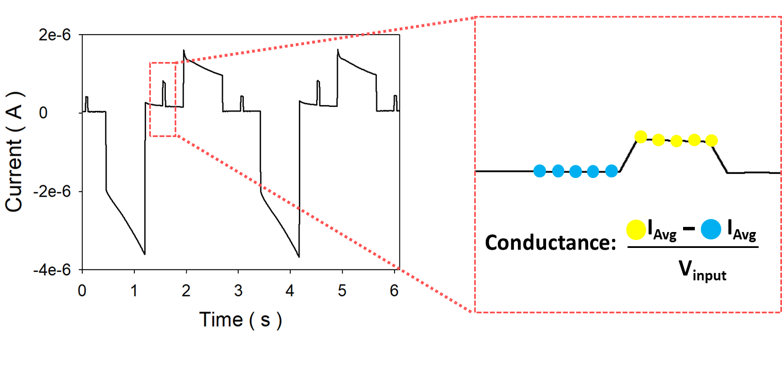}
        \caption{Schematic depiction of how the read-pulses are used to calculate the channel conductance. The current measurements during the read pulse are averaged. The difference with the average current just before the read pulse then yields the channel conductance after dividing by the applied voltage. Before each voltage pulse train, a read pulse is applied to obtain the base conductance $g_0$, which is used to normalise the measurements.}
        \label{fig:condMeasurement}
\end{figure}

In Fig.~\ref{fig:condMeasurement} we schematically show how we use the read pulses to calculate the channel conductance. We calculate the difference in (average) current during a read pulse and the measured current just before the read pulse to calculate the conductance.

\begin{figure}[h]
\centering
     \includegraphics[width=0.75\textwidth]{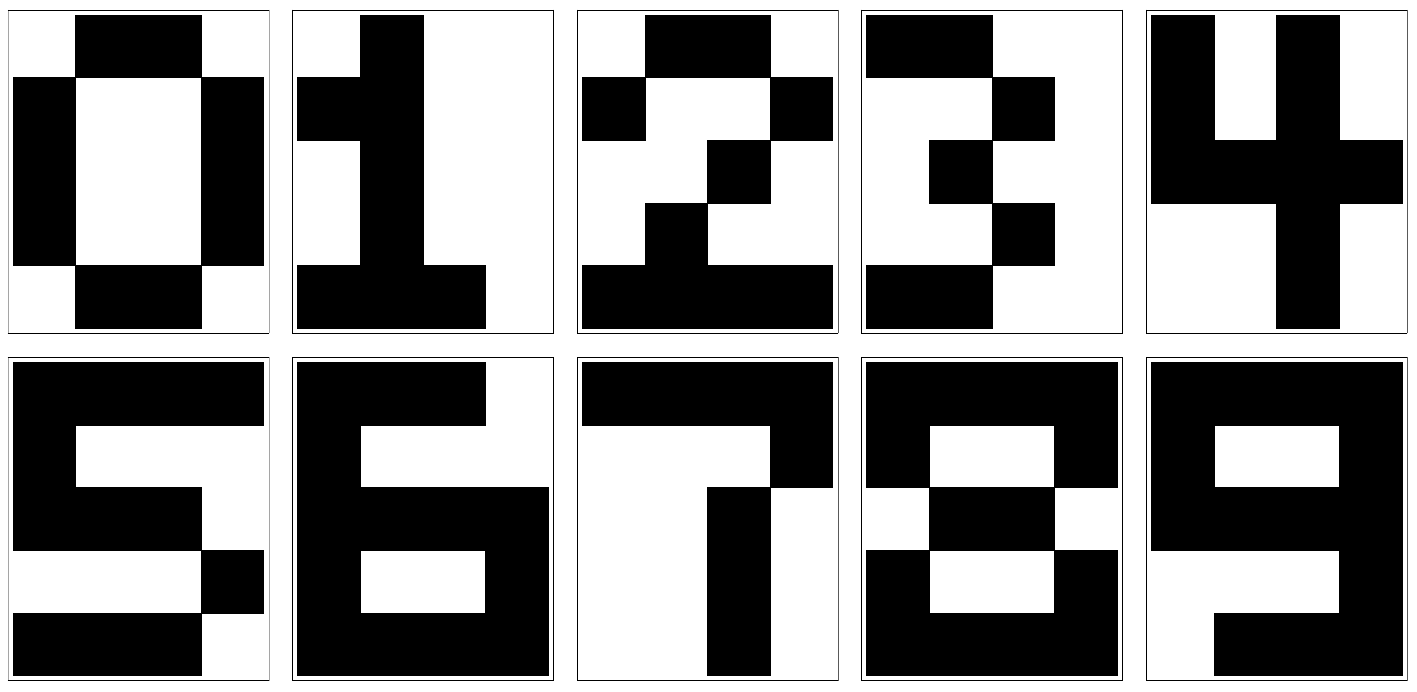}
        \caption{Simple single digit numbers used for classification in the main text as shown in Fig.~3(b) and Fig.~3(c).}
        \label{fig:simpleNumGrid}
\end{figure}

In the main text we classified simple single-digit images consisting of $4\times 5$ black and white pixels. In Fig.~3(b) we only showed the ``2'' as an example, the other digits used are depicted in Fig.~\ref{fig:simpleNumGrid}.
\subsection{Timescale measurements}
\begin{figure}[h]
\centering
     \includegraphics[width=1\textwidth]{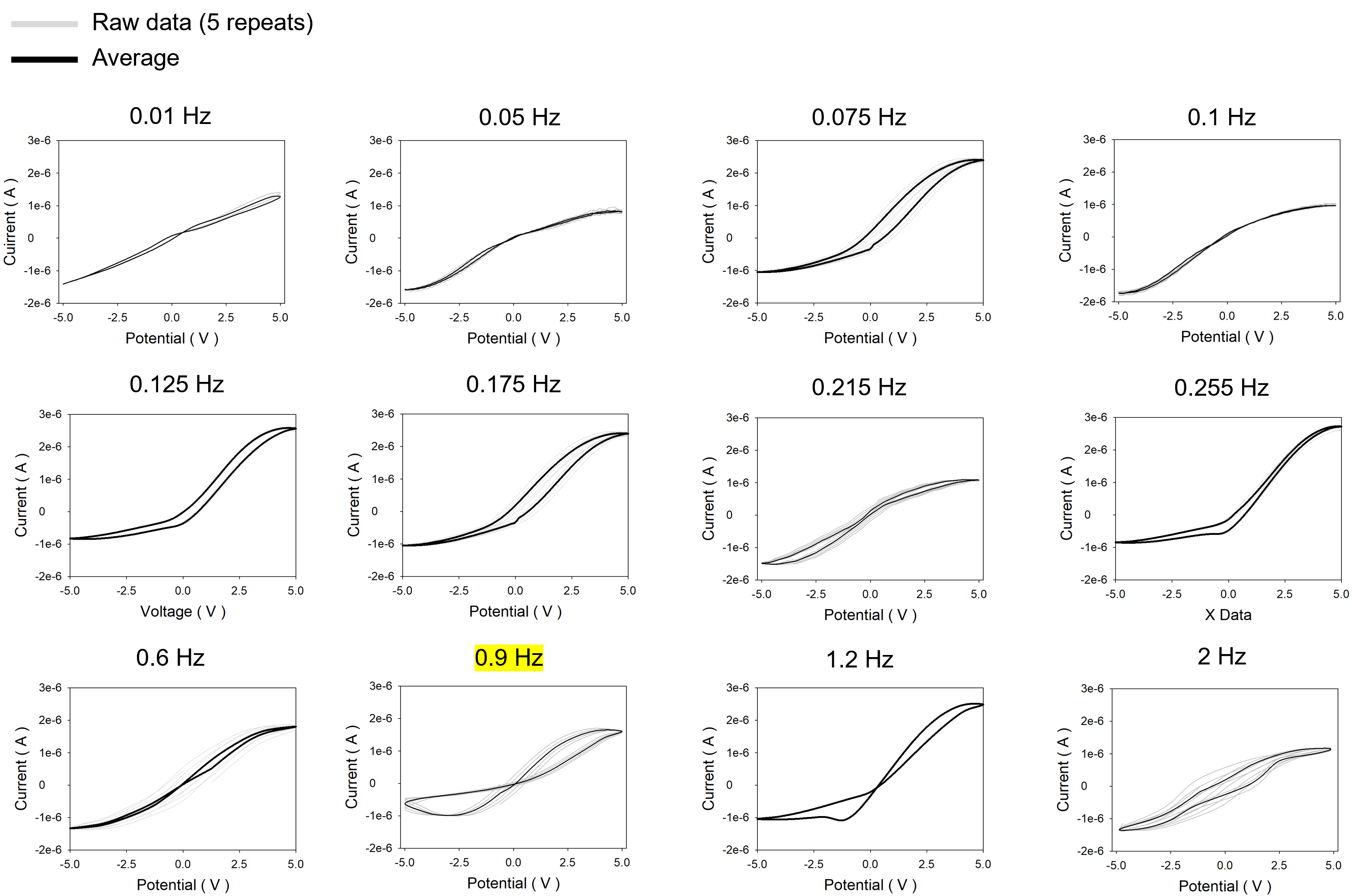}
        \caption{Current-voltage hysteresis loops for the $50\text{ }\mu$m channel as a result of a sinusoidal voltage over the channel of amplitude 5 V for the various frequencies shown. Measurements were gathered during five periods, depicted as the light grey graphs, where the average of the measurements is shown as a black graph. Enclosed areas were calculated with $f_{\ch{max}}=0.9$ Hz (highlighted yellow) exhibiting the most open hysteresis loop.}
        \label{fig:hysteresisLoops50um}
\end{figure}

\begin{figure}[h]
\centering
     \includegraphics[width=1\textwidth]{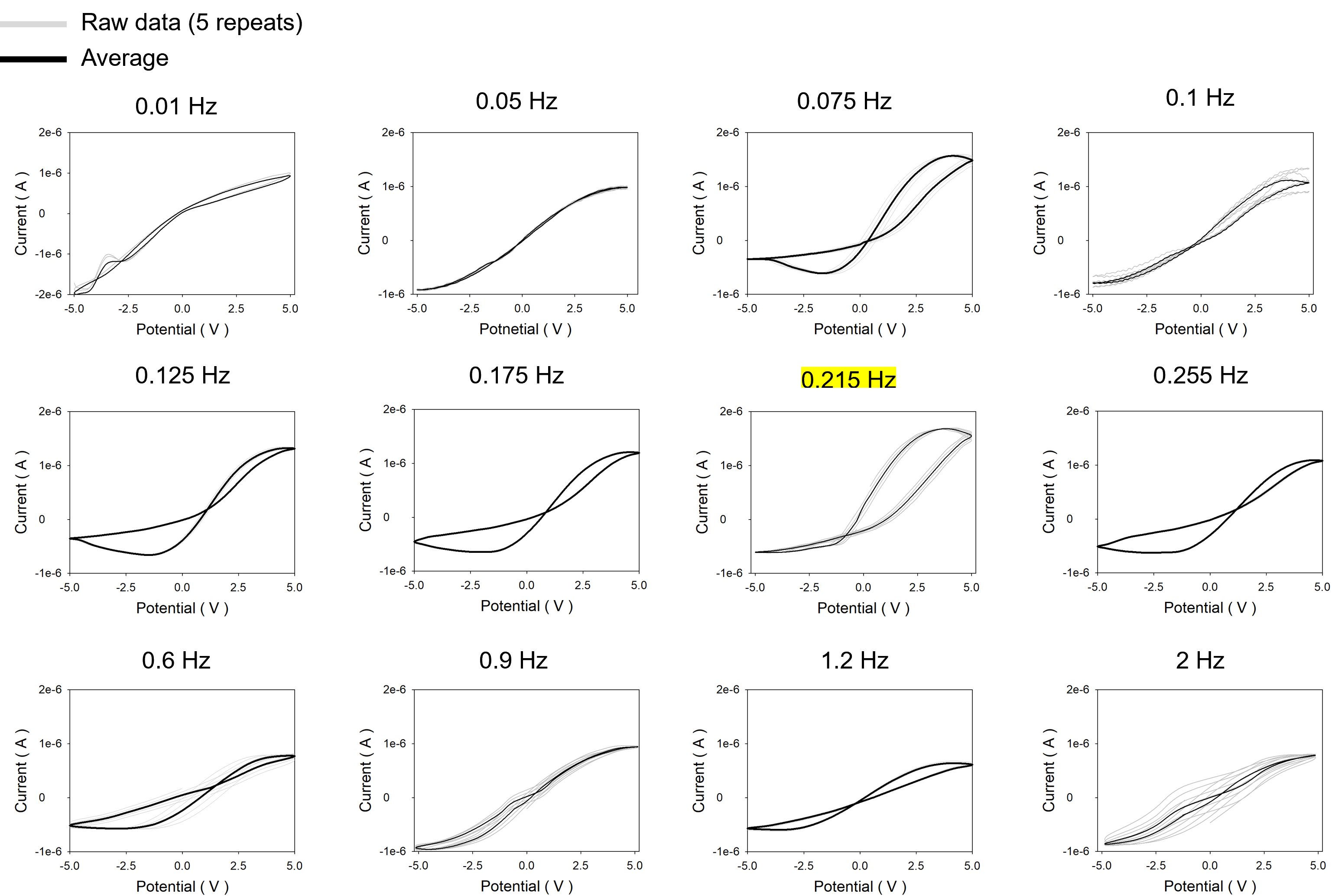}
        \caption{Current-voltage hysteresis loops for the $100\text{ }\mu$m channel as a result of a sinusoidal voltage over the channel of amplitude 5 V for the various frequencies shown. Measurements were gathered during five periods, depicted as the light grey graphs, where the average of the measurements is shown as a black graph. Enclosed areas were calculated with $f_{\ch{max}}=0.215$ Hz (highlighted yellow) exhibiting the most open hysteresis loop.}
        \label{fig:hysteresisLoops100um}
\end{figure}

\begin{figure}[h]
\centering
     \includegraphics[width=1\textwidth]{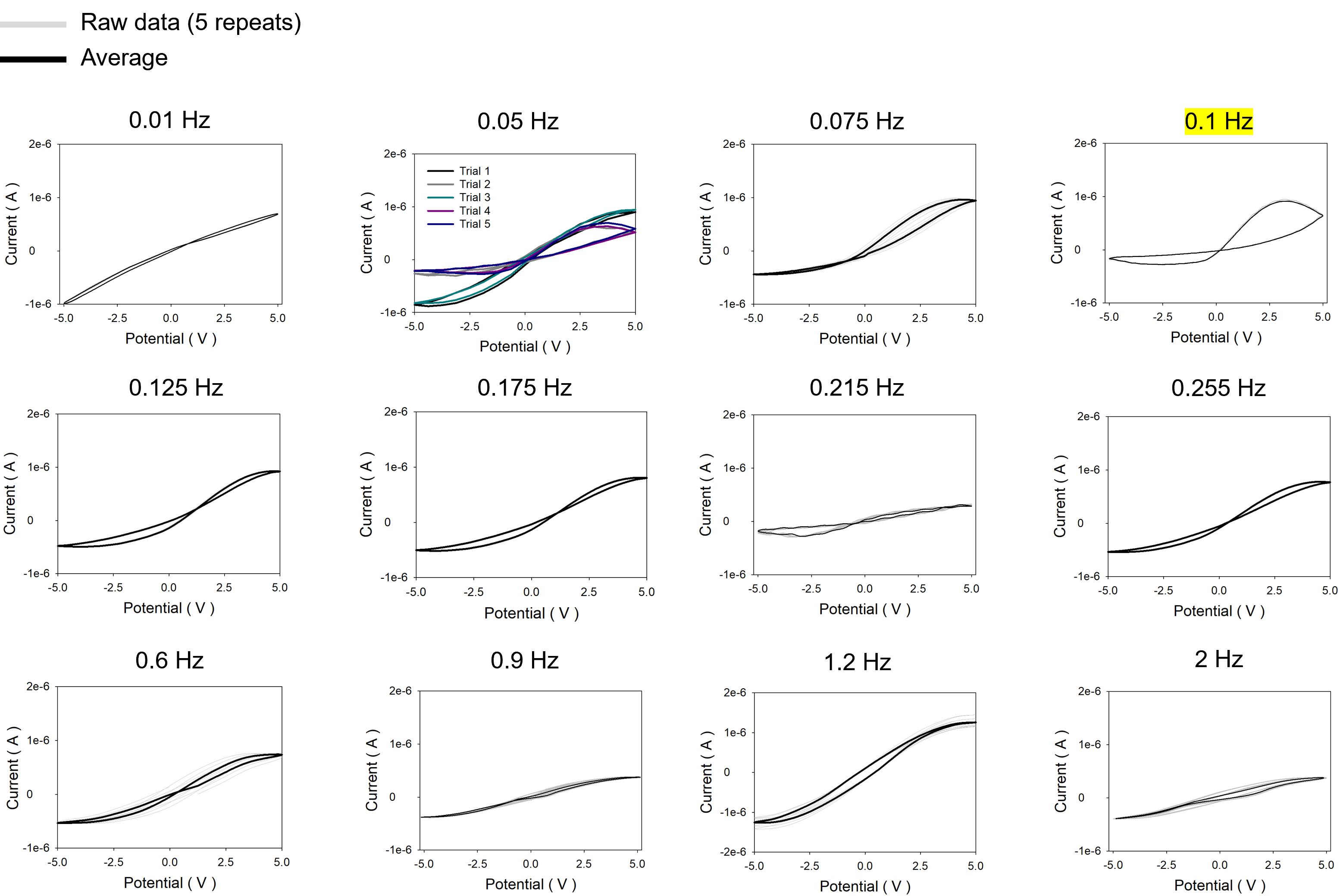}
        \caption{Current-voltage hysteresis loops for the $150\text{ }\mu$m channel as a result of a sinusoidal voltage over the channel of amplitude 5 V for the various frequencies shown. Measurements were gathered during five periods, depicted as the light grey graphs, where the average of the measurements is shown as a black graph. Enclosed areas were calculated with $f_{\ch{max}}=0.1$ Hz (highlighted yellow) exhibiting the most open hysteresis loop. In the first run of measurements, 0.05 Hz yielded a hysteresis loop with an enclosed area very close to 0.1 Hz. Both frequencies have been tested various more times, where the 0.1 Hz loops were consistently more open, one of these experiments was used for Fig.~1(c). The other 0.05 Hz measurements are shown here.}
        \label{fig:hysteresisLoops150um}
\end{figure}

In Fig.~2 we show measured hysteresis loops for 3 different frequencies for channels of lengths $50\text{ }\mu$m, $100\text{ }\mu$m and $150\text{ }\mu$m. Additional measurements were conducted for intermediate frequencies in order to find upper and lower bound for the frequency which exhibits the most open hysteresis loop. All loops are shown in Fig.~\ref{fig:hysteresisLoops50um}, Fig.~\ref{fig:hysteresisLoops100um} and Fig.~\ref{fig:hysteresisLoops150um} for channel lengths of $50\text{ }\mu$m, $100\text{ }\mu$m and $150\text{ }\mu$m respectively.

\clearpage

\begin{table}[h]
    \centering
    \begin{tabular}{c|c c c |c}
       \textbf{Bit-string } & \textbf{ Standard deviation}& \quad\quad\quad &\textbf{Bit-string }  & \textbf{ Standard deviation}\\
       0000  & 0.137 & \quad\quad\quad & 1000 & 0.142\\
       0001  & 0.193 & \quad\quad\quad & 1001 & 0.143\\
       0010  & 0.080 & \quad\quad\quad & 1010 & 0.061\\
       0011  & 0.157 & \quad\quad\quad & 1011 & 0.170\\
       0100  & 0.088 & \quad\quad\quad & 1100 & 0.093\\
       0101  & 0.148 & \quad\quad\quad & 1101 & 0.178\\
       0110  & 0.058 & \quad\quad\quad & 1110 & 0.087\\
       0111  & 0.207 & \quad\quad\quad & 1111 & 0.256\\
    \end{tabular}
    \caption{The standard deviations of the normalized channel conductance $g/g_0$ for each different bit-string, determined through the six measurements per bit-string shown in Fig.~\ref{fig:pulseTrainPlots}.}
    \label{tab:stdev}
\end{table}


\clearpage

%